\documentclass[9pt,a4paper,twoside,twocolumn]{article}
\usepackage{changepage} 
\usepackage{graphicx}
\usepackage{verbatim}
\usepackage{gensymb}
\usepackage{amsmath,amssymb,amsfonts} 
\usepackage{bm}

\usepackage[usenames,dvipsnames]{color}
\usepackage[table]{xcolor}
\definecolor{light-gray}{gray}{0.90}

\usepackage[T1]{fontenc}
\usepackage{lmodern}
\usepackage{courier} 
\normalfont

\usepackage{titlesec}

\titleformat{\section}{\normalfont\Large\color{black}}{\bf\thesection}{1em}{}

\titleformat{\subsection}{\normalfont\large\color{black}}{\bf\thesubsection}{1em}{}

\usepackage{fancyhdr}
\fancypagestyle{plain}{ %
  \fancyhf{} 
  
}

\usepackage{epstopdf}
\DeclareGraphicsExtensions{.png,.jpg,.jpeg,.eps}
\graphicspath{ {./Figures/} }
\usepackage{float}
\usepackage{caption}
\usepackage{subcaption}

\usepackage{booktabs}

\usepackage[square,comma,numbers,sort&compress]{natbib} 
\usepackage{url} 
\usepackage[colorlinks]{hyperref}
\hypersetup{linkcolor=blue, citecolor = blue}

\usepackage{siunitx}
\usepackage{authblk}

\usepackage{upgreek}
\newcommand{\chithree}{ \ensuremath{\chi^{(3)}} }
\newcommand{\ket}[1]{\ensuremath{|#1\rangle\mkern-1mu}}
\newcommand{\bra}[1]{\ensuremath{\mkern-1mu\langle#1|}}
\newcommand{\ketbra}[2]{\ensuremath{|#1\rangle\mkern-5mu\langle#2|}}
\newcommand{\braket}[2]{\ensuremath{|#1\langle\mkern-5mu\rangle#2|}}

\newcommand{\Tr}[1]{\mathrm{Tr}#1}

\newcommand{\ad}[1]{\textsuperscript{#1}\kern-2pt}

\usepackage{array}
\newcolumntype{P}[1]{>{\arraybackslash}p{#1}}
\newcolumntype{Q}[1]{>{\centering\arraybackslash}p{#1}}

\usepackage{mathtools}

\title{Chip-to-chip quantum teleportation and multi-photon entanglement in silicon}

\author[$^{\dagger}$,1]{Daniel Llewellyn} 
\author[$^{\dagger}$,2,3]{Yunhong Ding}
\author[$^{\dagger}$,1]{Imad I. Faruque}
\author[1]{Stefano Paesani}
\author[2,3]{Davide Bacco} 
\author[1]{Raffaele Santagati}
\author[4]{Yan-Jun Qian} 
\author[4]{Yan Li} 
\author[4,5]{Yun-Feng Xiao} 
\author[6]{Marcus Huber} 
\author[7]{Mehul Malik} 
\author[1]{Gary F. Sinclair} 
\author[8]{Xiaoqi Zhou} 
\author[2,3]{Karsten Rottwitt}
\author[1]{Jeremy L. O\textquoteright Brien} 
\author[1]{John G. Rarity}
\author[4,5]{Qihuang Gong} 
\author[2,3]{Leif K. Oxenlowe}
\author[1,4,5,*]{Jianwei Wang} 
\author[1]{Mark G. Thompson}

\affil[1]{Quantum Engineering Technology Labs, H. H. Wills Physics Laboratory \& Department of Electrical and Electronic Engineering, University of Bristol, Merchant Venturers Building, Woodland Road, Bristol BS8 1UB, United Kingdom.}
\affil[2]{Department of Photonics Engineering, Technical University of Denmark, 2800 Kgs. Lyngby, Denmark}
\affil[3]{Center for Silicon Photonics for Optical Communication (SPOC), Technical University of Denmark, 2800 Kgs. Lyngby, Denmark}
\affil[4]{State Key Laboratory for Mesoscopic Physics  and  Collaborative Innovation Center of Quantum Matter, School of Physics,  Peking University, Beijing 100871, China}
\affil[5]{Beijing Academy of Quantum Information Sciences, West Bld.3,No.10 Xibeiwang East Rd., Haidian District, Beijing 100193,China}
\affil[6]{ Institute for Quantum Optics and Quantum Information (IQOQI), Austrian Academy of Sciences, Vienna, Austria}
\affil[7]{Institute of Photonics and Quantum Sciences (IPaQS), Heriot-Watt University, Edinburgh, UK}
\affil[8]{State Key Laboratory of Optoelectronic Materials and Technologies and School of Physics, Sun Yat-sen University, Guangzhou, China}
\affil[$^{\dagger}$]{These authors contributed equally to this work.}
\affil[*]{ Email: jianwei.wang@pku.edu.cn}

\date{}                     

\begin{document}
\maketitle
\begin{figure*} 
\centering
\includegraphics[width=0.99\textwidth]{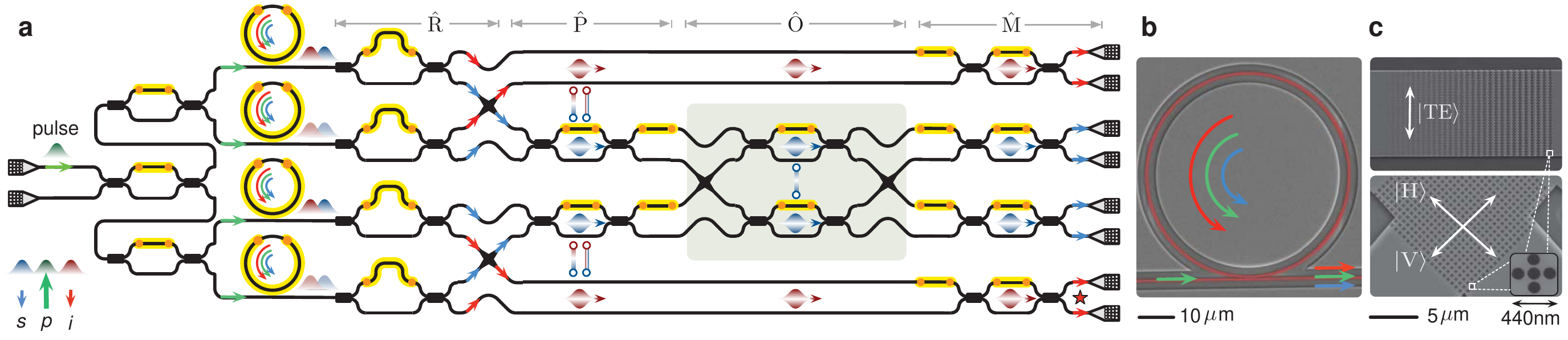} 
\captionof{figure}{\textbf{A microresonator-enhanced multiphoton quantum processor in silicon.} 
\textbf{a,} Schematic. A network of nonlinear single-photon sources and linear-optic multiqubit circuits are integrated in a single silicon chip.  
Two pairs of nondegenerate photons (red idler, blue signal) are generated in an array of four MRR single-photon sources. The MRRs strongly enhance the SFWM nonlinearity and allow bright, pure, and indistinguishable photon generation while also suppressing background noise from all waveguides and linear circuits. 
A linear-optic quantum circuit ($\hat{\text{O}}$) is programmed to work as a bosonic Bell operator and a fusion entangling operator on the two blue photons. 
The four photons are demultiplexed by asymmetric MZIs and routed via waveguide-crossers ($\hat{\text{R}}$). 
An array of MZIs and phase-shifters allow the preparation ($\hat{\text{P}}$) and projective measurement ($\hat{\text{M}}$) of multiqubit states. 
Yellow parts refer to electronically controllable thermal-optic phase-shifters. 
\textbf{b}, The scanning electron microscope (SEM) images of a MRR single-photon source coupled to a bus-waveguide (pseudo red colour), and \textbf{c,}  metasurface-assisted low-loss subwavelength grating couplers (SGC) with bonded aluminium reflectors. 
Top: 1D SGC for fiber-chip interface, bottom: 2D SGC for path-polarisation conversion, and inset: zoom-in view of a metasurface cell. White arrows denote the polarised state of photon. 
Star "$\star$" in \textbf{a} refers to a switchable router (not shown) for either single-chip (via 1D SGCs) or chip-to-chip (via 2D SGCs)  experiments. 
}
\label{fig:DeviceSchematics} 
\end{figure*}    

\noindent\textbf{Exploiting semiconductor fabrication techniques, natural carriers of quantum information such as atoms, electrons, and photons can be embedded in scalable integrated devices~\cite{MonroeIon,SilconCNOT,JeremyPhotons}. 
Integrated optics provides a versatile platform for large-scale quantum information processing and transceiving with photons~\cite{JeremyPhotons, OxfordChip,Spring798,RomeBS,Sparrow2018, Kumar2014, Mittal2018, MorandottiSource, USTCSource, Graphstate, Wang16d,Wang:QHL,Harris, TangSNSPD}. 
Scaling up the integrated devices for quantum applications requires high-performance single-photon generation and photonic qubit-qubit entangling operations~\cite{KLM,One-way,12photons,Telereview}. 
However, previous demonstrations report major challenges in producing multiple bright, pure and identical single-photons~\cite{Kumar2014, Mittal2018, MorandottiSource, USTCSource,Graphstate}, and entangling multiple photonic qubits with high fidelity~\cite{Harris, Wang16d,Wang:QHL}. 
Another notable challenge is to noiselessly interface multiphoton sources and multiqubit operators in a single device~\cite{JeremyPhotons, OxfordChip, Spring798,
RomeBS,Sparrow2018, Kumar2014, Mittal2018, USTCSource, Graphstate, MorandottiSource, Harris, Wang16d,Wang:QHL}. 
Here we demonstrate on-chip genuine multipartite entanglement and quantum teleportation in silicon, by coherently controlling an integrated network of microresonator nonlinear single-photon sources and linear-optic multiqubit entangling circuits. 
The microresonators are engineered to locally enhance the nonlinearity, producing multiple frequency-uncorrelated and indistinguishable single-photons, without requiring any spectral filtering.  
The multiqubit states are processed in a programmable linear circuit facilitating Bell-projection and fusion-operation in a measurement-based manner. 
We benchmark key functionalities, such as intra-/inter-chip teleportation of quantum states, and generation of four-photon Greenberger-Horne-Zeilinger entangled states. 
The production, control, and transceiving of states are all achieved in micrometer-scale silicon chips, fabricated by complementary metal-oxide-semiconductor processes. Our work lays the groundwork for scalable on-chip multiphoton  technologies for quantum computing and communication.
}

\noindent
Most systems in physics, chemistry, and engineering are inherently nonlinear, therefore engineering and controlling nonlinear devices presents a powerful tool to study those systems. 
In quantum physics and information science, the coherent control of  quantum nonlinear optic devices is of great significance, e.g, having enabled the generation of entangled photons~\cite{PhysRevLett.75.4337} and squeezed light~\cite{Squeezed} in parametric nonlinear processes, and aided our understanding of quantum correlation and measurement. 
Continued development in controlling photons in nonlinear and linear devices promise unprecedented capabilities in performing information tasks. 
For examples, universal linear-optic quantum computing is enabled by the measurement of large entangled cluster states~\cite{One-way, 12photons}, while the preparation of these states requires nonlinear single-photon sources and qubit-qubit entangling circuits. The transmission and teleportation of quantum states promises secure quantum communication networks~\cite{Telereview,satellite:tele}. 
Scattering photons in a specific linear-optic circuit allows the computation of boson sampling problems~\cite{RomeBS,Spring798} and the simulation of molecular dynamics~\cite{Sparrow2018}. In general, interacting linear and nonlinear quantum components in a sophisticated connected network would offer fundamentally new quantum photonic devices~\cite{Gu2018}. 

Integrated quantum photonics promises a scalable platform for large-scale integration of linear  and nonlinear quantum devices. 
Silicon quantum photonics is of particular interest, having recently shown large integration of two-photon sources and circuits~\cite{Wang16d}, precise control of photon states~\cite{Wang:QHL,Harris}, and efficient single-photon detection~\cite{TangSNSPD}. 
However, the scalability has been constrained by the inability to produce multiple high-performance single-photons, to coherently entangle multiple qubits with high fidelity, and to seamlessly interface the nonlinear and linear  devices. 
Previously, silicon waveguide sources with a $cm$-length~\cite{Wang16d,Wang:QHL,USTCSource,Graphstate}, have been used to create photon-pairs via the spontaneous four-wave mixing (SFWM) process. 
These photon-pairs however are highly correlated in frequency~\cite{Grice2001}, and improving their spectral purity requires narrowband spectral filtering --- causing a significant reduction of photon counts and heralding efficiency. 
In fact, single-photons are typically generated uniformly in waveguide sources and linear circuits~\cite{Wang16d,Wang:QHL,USTCSource,Graphstate}, thus noises are induced when performing quantum operations and error corrections. 
Moreover, on-chip implementations of quantum applications require photonic qubit-qubit interactions in linear-optic circuits, which rely on quantum interference and success of measurements~\cite{KLM}, having shown the controlled-Z gate in waveguides~\cite{JeremyPhotons, OxfordChip}.  
A broad range of multiqubit entangling operations on chip, such as fusion operation and Bell projection~\cite{PanReview}, is in high demand for chip-based photonic quantum computing~\cite{One-way} and quantum communication~\cite{Telereview}. In general, a seamless interface and coherent control of multiphoton nonlinear sources and multiqubit linear circuits remains a great challenge, due to the uniformity of waveguide sources and circuits and the lack of an efficient architecture. 

A solution is to locally enhance the nonlinearity using optical microresonator~\cite{Vahala2003} that can greatly strengthen the light-matter interaction by storing light inside the resonator. The generation of photons outside of the microresonators can thus be greatly suppressed by using a weak pump light. 
More importantly, the microresonators allow the generation of single-photons with high spectral purity~\cite{Vernon:17}, and high heralding efficiency by removing the requirement of spectral filtering~\cite{Vernon2016}. 
Composite microresonators also allow the further control of photon spectra~\cite{Kumar2014} and emission of topologically protected photons~\cite{Mittal2018}. 
The versatility of microresonator sources has thus far been demonstrated only for two-photon implementations. 
Recently, a single microresonator has been exploited to produce multiphoton states~\cite{MorandottiSource}; however, for scalable  quantum devices it is essential to verify the multiphoton indistinguishability and to perform multiqubit operations on the generated photons. 

Here we report a multiqubit quantum processor enabled by coherently controlling a network of microring resonators (MRRs) for near-optimal photon generation and linear-optic circuits for high-fidelity multiqubit operation. All the nonlinear and linear quantum devices are monolithically integrated in silicon, and can be individually programmed. 
We benchmark key protocols in quantum applications, including intra-/inter-chip teleportation of single-qubits and Bell states, and on-chip generation of three- and four-photon Greenberger-Horne-Zeilinger  (GHZ) genuine entangled states. 

Figure 1a shows the MRR-enhanced multiqubit processor that is fabricated on the silicon-on-insulator platform.  
The Si MRR-sources array (see SEM image in Fig.~1b) can produce two pairs of signal ($\lambda_s$) \& idler ($\lambda_i$) photons via the SFWM. 
Four dual-rail qubits are encoded in the four generated photons. 
Each qubit is represented in the logical basis \{$|0\rangle_{k}$, $|1\rangle_{k}$\} ($k=1,2,3,4$) and can be prepared ($\hat{\text{P}}$) and measured ($\hat{\text{M}}$) by a network of Mach-Zehnder interferometers (MZI) and phase-shifters (see Fig.~1a). 
Another key device is the programmable two-qubit operator ($\hat{\text{O}}$) that is able to entangle two qubits (previously never interacted) in two different manners, i.e, Bell-projection and fusion-operation. 
The MRR, qubit-generator, entangling-operator, and qubit-analyser are all individually controllable and fully programmable.  
The chip is coupled to optical fibers via an array of low-loss (0.8 dB) 1D SGC~\cite{Ding2014:SGC} (Fig.~1c), and photons are detected off-chip by 8 superconducting single-photon detectors ($\sim$0.85 efficiency). Details on device and setup are provided in Supplementary Information (SI) Sec.~\ref{sec:DesignSetup}.  

\begin{figure*}[ht!] 
\centering
\includegraphics[width=0.95\textwidth]{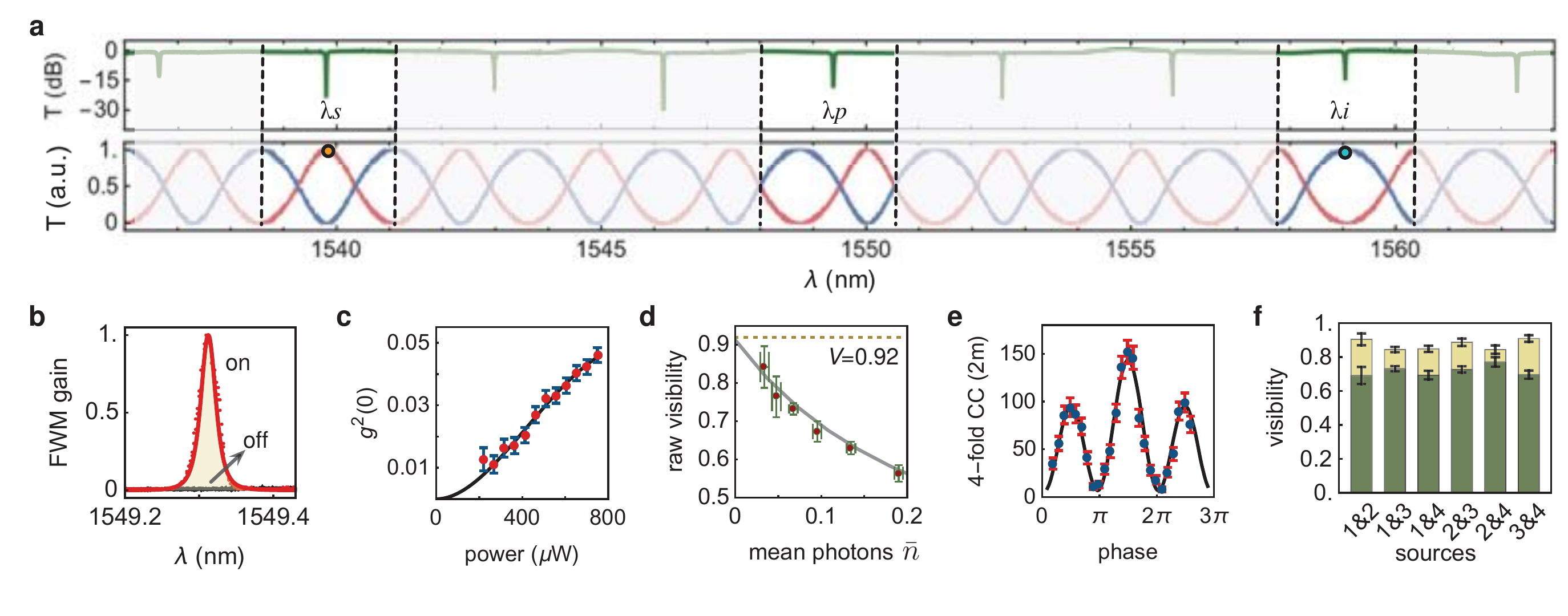} 
\caption{\textbf{Near-optimal photon-pair generation in an array of MRR-enhanced nonlinear sources. 
}    
\textbf{a,} The measured transmission spectra for an MRR (top) and asymmetric MZI (bottom), where FSR\textsubscript{MRR}$= 400$ GHz and  FSR\textsubscript{AMZI}$= 320$ GHz. 
Signal photons are created at $\lambda$\textsubscript{s}= 1539.758 {nm} and idler photons at $\lambda$\textsubscript{i}= 1559.015 {nm}, at the resonances of MRR. 
Asymmetric MZIs can demultiplex the $\lambda$\textsubscript{s} and $\lambda$\textsubscript{i} photons. Residual pump photons are removed by off-chip filters with a $\sim$1.1 nm bandwidth, much wider than the linewidth of MRRs $37.7\pm 1.9$ pm.  
\textbf{b,} FWM enhancement in MRR when on resonance. Background noise in the whole device is efficiently suppressed (off-resonance). 
\textbf{c,} Tests of photon number purity by measuring the heralded $g^{(2)}(0)$.  
\textbf{d,}  Measured raw visibility of the heralded quantum interference, as a function of mean photon number $\Bar{n}$ per pulse. The V = 0.92 dotted line is the maximum achievable visibility for our MRR designs. 
\textbf{e,} Tests of spectral indistinguishability, with a subtraction of multi-pair events. Visibility of $90.99\pm 3.91\%$  are obtained, agreeing with the theoretical limit. 
\textbf{f,} Measured visibility of quantum interference between pairwise MRRs in the array. Mean visibilities of $87.3\pm 1.9\%$ and $71.9\pm 2.4\%$ are measured, with and without multi-pair corrections. 
Points are all experimental data, while lines in \textbf{d} are theoretical values, and lines in \textbf{b},\textbf{c}, and \textbf{e} are fittings. 
All error bars refer to $\pm$1 standard deviation (s.d.) estimated from Poissonian photon-counting statistics.}
\label{fig:SourcesResults} 
\end{figure*}    

The MRRs each have a quality-factor of greater than $10^4$, yielding a strong SFWM enhancement in the cavity. As shown in Fig.~\ref{fig:SourcesResults}b, the generated two-photon rate is enhanced by a factor of $\sim$43 when MRR is on/off resonance. 
For each MRR, a raw rate of $\sim$20 kcts/s at a coincidences-to-accidentals ratio 
$\sim$50 was detected, using a pump laser with 15 ps pulse width with 500 MHz repetition rate at 800 $\mu$W power.  
Since the MRR nonlinear sources only require weak pump, negligible photons are created in surrounding waveguides and circuits, greatly suppressing noise there. That means the nonlinear and linear quantum photonic devices can now noiselessly interface each other.  
We further measured the photon-number purity of the single-photons, by performing the Hanbury-Brown-Twiss measurement of photon to obtain the heralded second-order correlation g\textsuperscript{(2)}(0). At the same power, we observed g\textsuperscript{(2)}(0) = 0.05 corresponding to 95\% photon-number purity (Fig.~\ref{fig:SourcesResults}c). All four MRRs have a high heralding efficiency of $\sim$50\% after the resonators (see SI Sec.~\ref{sec:MRRsource}), matching the theoretical limit well. 

The four MRRs are designed to be identical. The high-yield fabrication enables nearly identical free-spectral-ranges (FSRs$\sim$400 GHz, Fig.~\ref{fig:SourcesResults}a) and high spectral overlap at resonances (Fig.~\ref{fig:FigSMRRSources}). Each MRR can be individually tuned and frequency-locked (see SI Sec.~\ref{sec:MRRsource}), ensuring photon wavefunctions that are highly overlapped in spectral mode. 
The photon indistinguishability is estimated by the visibility of heralded two-photon quantum interference. We interfere two signal photons on a MZI (heralding two idler photons), which are emitted from two independent MRRs. 
Figure~\ref{fig:SourcesResults}e reports a quantum interference fringe having multi-pair corrected visibility of 90.99$\pm$ 3.91\%, which agrees within error of the fundamental limit of 92\% spectral purity for our MRR design (Sec.~\ref{sec:MRRsource}). 
In Fig.~\ref{fig:SourcesResults}d, the raw uncorrected visibility as a function of the mean photon number per pulse ($\bar{n}$) is also measured, e.g, $\sim$84\% raw visibility is achieved at $\bar{n}$ = 0.05. Compared to the spectral impurity, the photon-number impurity drastically affects the visibility~\cite{Faruque2018}. 
Figure~\ref{fig:SourcesResults}f  shows the pairwise indistinguishability of the four MRR sources, having a mean raw visibility of $71.9\pm2.4\%$ in a high brightness configuration, which if multi-pair corrected, reaches to $87.3\pm1.9$ mean visibility. We thus demonstrate the generation of highly identical photons in multiple MRRs. Remarkably, in all of our indistinguishability measurements, we have not used any spectral filtering to improve the spectral purity. 

\begin{figure}[!] 
\centering
\includegraphics[width=0.47\textwidth]{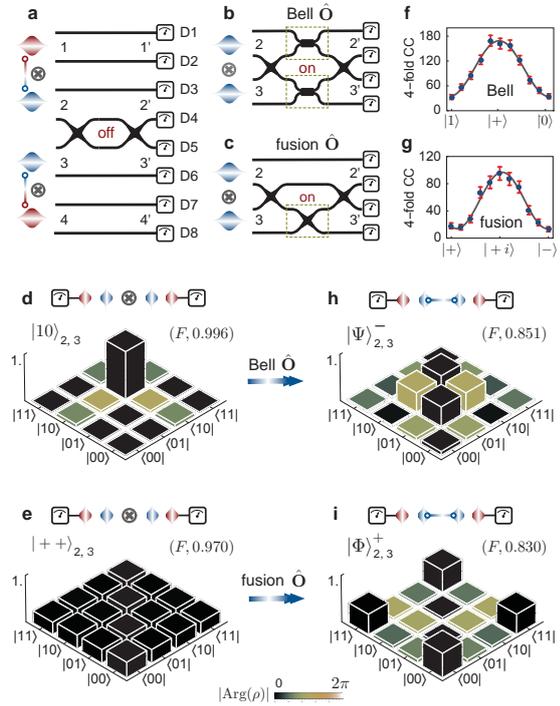} 
\caption{\textbf{Programmable linear-optic quantum circuits for Bell projection and fusion operation.} 
\textbf{a,} Diagram for a general operator, in this case \^{O} = \^{I}$\otimes$\^{I}  on qubits 2, 3, i.e, \^{O} is off. Each pair of lines refers to the dual-rail logical state $|0\rangle$ and $|1\rangle$. 
\textbf{b} (\textbf{c})\textbf{,} Diagram for Bell operator \^{O}\textsubscript{Bell} (fusion operator \^{O}\textsubscript{fusion}) in the spatial mode. 
For \^{O}\textsubscript{Bell}, a photon has 50\% probability of being measured in either 2' or 3', regardless of its input in 2 or 3. The \^{O}\textsubscript{fusion} transmits $|0\rangle$ and swaps $|1\rangle$ mode. In the dashed boxes the circuits are reconfigured as a Hadamard-like or swap-like operator.  
The generated states in the source array can be programmed as either  bipartite entangled (see SI), or separable \textbf{d,} $|10\rangle_{2,3}$ and \textbf{e,} $|++\rangle_{2,3}$ (two red photons are in herald).  The values in (*) refer to the measured state fidelity ($F$) to their ideal states. 
\textbf{f} (\textbf{g})\textbf{,} Quantum interference when the two blue qubits $|10\rangle_{2,3}$ ($|++\rangle _{2,3}$) meet at the \^{O}\textsubscript{Bell} (\^{O}\textsubscript{fusion}) and rotate qubit 2 
along the $\hat{\sigma}_y$ axis (
along the $\hat{\sigma}_z$ axis). Points are experimental data measured in the 
$\hat{\sigma}_x\hat{\sigma}_x$ basis, and lines are fitted with a sinusoidal function. 
\textbf{h} (\textbf{i})\textbf{,} Reconstructed density matrices for entangled states $\rho_{2,3}$, when performing \^{O}\textsubscript{Bell} (\^{O}\textsubscript{fusion}) on the separable states in \textbf{d} (\textbf{e}). 
The {$\rho$} are reconstructed by performing full QST on-chip. 
Column heights represent $|\rho|$ while colors represent $|\text{Arg}(\rho)|$. 
All error bars refer to $\pm$1 s.d. estimated from Poissonian photon-counting statistics. 
}
\label{fig:OperatorResults} 
\end{figure}    

The MRR-source array is programmed to create either entangled or separable bipartite states, by controlling the pump excitation in MRRs and re-configuring asymmetric MZIs. 
When the operator $\hat{\text{O}}$ between qubits 2,3 is turned off (see Fig.~\ref{fig:OperatorResults}a) the original bipartite states from the MRRs array are measured. 
We implement quantum state tomography (QST) to reconstruct the density matrix $\rho$. As examples, Figures~\ref{fig:OperatorResults}d,e show the measured $\rho$ for separable states $\ket{10}_{2,3}$ and $\ket{++}_{2,3}$ with fidelities of $0.964\pm0.072$ and $0.966 \pm 0.024$, respectively.  
The fidelity is defined as $F=\bra{ \psi_{0}}\rho\ket{\psi_{0}} $, where $\ket{\psi_{0}}$ represents the ideal state. 
In addition, the complete set of Bell states $\ket{\Phi}^{\pm}=({\ket{00} \pm \ket{11}})/{\sqrt{2}}$,  $\ket{\Psi}^{\pm}=({\ket{01} \pm \ket{10}})/{\sqrt{2}} $ can be generated, see $\rho$ data in Fig.~\ref{fig:FullBell}. 
We simultaneously prepared and measured two Bell pairs $\ket{\Phi}^{+}_{1,2}$ and $\ket{\Phi}^{+}_{3,4}$ from the four MRRs, having fidelities of $0.917\pm0.002$ and $0.915\pm0.003$, respectively. 

We then exploit a single programmable circuit to implement two key multiqubit operations in quantum applications, i.e., entangling initially separable qubits and measuring qubits in the Bell basis. 
Figure~\ref{fig:OperatorResults}b,c shows the diagrams for the bosonic Bell projector \^{O}\textsubscript{Bell} and fusion operator \^{O}\textsubscript{fusion} that are devised for dual-rail qubits. 
We here studied a case where the state $\ket{0}^{\otimes 4}$ is initially prepared, then $\rho_{2,3}$ is processed with the herald of photons 1, 4. 
Note the lithography-defined device ensures path-matching within a sub-wavelength accuracy for the photons 2, 3 and ensures their simultaneous arrival at \^{O}.  
The \^{O}\textsubscript{Bell} is capable of distinguishing the Bell states $\ket{\Psi}^\pm$ from the others. Here we distinguish $\ket{\Psi}^+$ when observing joint clicks in \{D3, D4\} or \{D5, D6\}  (Fig.~\ref{fig:OperatorResults}b). 
The \^{O}\textsubscript{fusion} transmits $\ket{0}$ and swaps $\ket{1}$ mode, able to fuse a two-qubit separable state into an entangled state when detecting only one photon in \{D3, D4\} and another in \{D5, D6\}  (Fig.~\ref{fig:OperatorResults}c). 
To verify these new on-chip building-blocks, heralded quantum interference and Bell state generation are performed. 
Figure~\ref{fig:OperatorResults}f reports the two-qubit ($\ket{10}_{2,3}$) interference, gradually rotating the qubit 2 around the $\hat{\sigma}_y$ axis. 
The observed {$80.5\pm 3.2 \%$} visibility confirms high quality interference of two bosons at \^{O}\textsubscript{Bell}. 
Figure~\ref{fig:OperatorResults}g shows the two-qubit (input $\ket{++}_{2,3}$) interference at \^{O}\textsubscript{fusion} having a {$85.8\pm 4.4\%$} visibility, when rotating the qubit 2 around the $\hat{\sigma}_z$ axis.  
In general, performing \^{O}\textsubscript{Bell} and \^{O}\textsubscript{fusion} enables the generation of entangled states, 
transforming the state $\ket{10}_{2,3}$ (Fig.~\ref{fig:OperatorResults}d) to $\ket{\Psi}^-_{2,3}$  (Fig.~\ref{fig:OperatorResults}h), and $\ket{++}_{2,3}$  (Fig.~\ref{fig:OperatorResults}e) to $\ket{\Phi}^+_{2,3}$  (Fig.~\ref{fig:OperatorResults}i).  
We obtained entangled state having fidelities of 0.$851\pm 0.040$ and $0.830\pm 0.032$, respectively. More details are provided in SI Sec.~\ref{sec:opeators}. 

\begin{figure*}[!] 
\centering
\includegraphics[width=0.98\textwidth]{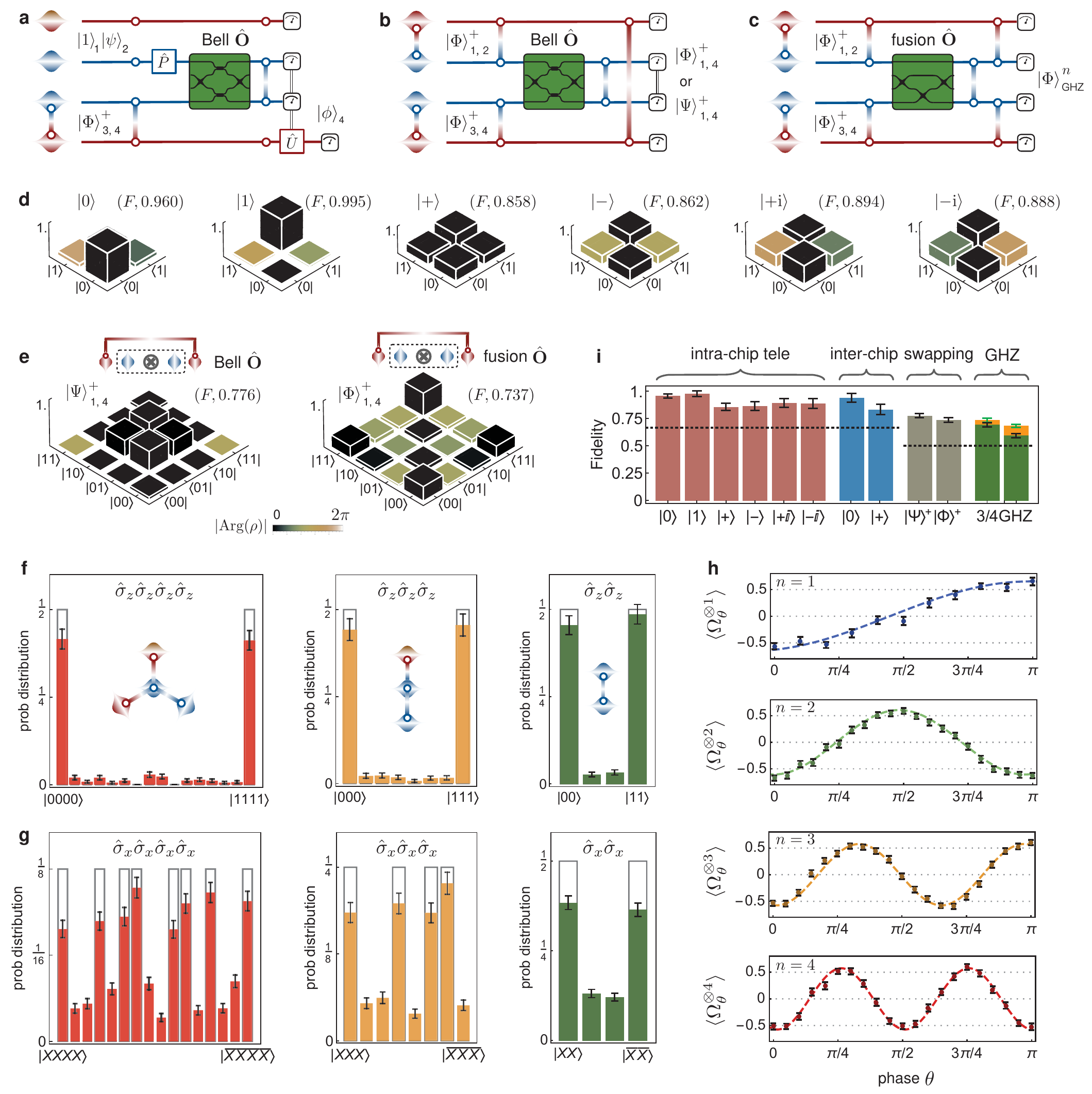} 
\caption{\textbf{On-chip multiphoton entanglement and intra-/inter-chip teleportation by programming nonlinear microresonators and linear circuits.} 
Quantum circuit diagrams for: \textbf{a,} teleportation of arbitrary single-qubit states from $|\psi\rangle_2$ to $|\phi\rangle_4$, by performing Bell measurement of qubits 2, 3; \textbf{b,} teleportation (entanglement swapping) of two-qubit entangled states from two Bell pairs \{$\ket{\Phi}^+ _{1,2}$ , $\ket{\Phi}^+ _{3,4}$ \} to $\ket{\Psi}^+ _{1,4}$ or $\ket{\Phi}^+ _{1,4}$; \textbf{c,} generation of three-photon and four-photon GHZ entangled states $\ket{\Phi}^{n}_\text{GHZ}$,  by fusing $\ket{\Phi}^+ _{1,2}$ and $\ket{\Phi}^+ _{3,4}$. 
Each red (blue) line denotes the evolution of one qubit in the logical representation. 
\textbf{d,} Experimental results for intra-chip single-qubit teleportation. 
The $\rho$ of six teleported states 
are reconstructed by full QST. A local unitary $\hat{U}$ has been applied on $\rho$.  
An inter-chip teleportation of \{$| {0}\rangle $, $| {+}\rangle $\} from one chip to another chip reports fidelities of \{0.940,0.832\}. 
 \textbf{e,} Experimental results for two-qubit entanglement swapping. Performing \^{O}\textsubscript{Bell} (\^{O}\textsubscript{fusion}) on $\ket{\Phi}^+ _{1,2}$ and $\ket{\Phi}^+ _{3,4}$ results in swapped entanglement $\ket{\Psi}^+ _{1,4}$ ($\ket{\Phi}^+ _{1,4}$) between photons that have not interacted. 
\textbf{f}-\textbf{h,} Verification and quantification of GHZ genuine entanglement.   
Measured four-fold coincidences (normalised) in the (\textbf{f})  $\hat{\sigma}_{z}^{\otimes {n} }$ basis and (\textbf{g})  $\hat{\sigma}_{x}^{\otimes {n}} $ basis for ${n}=2,3,4$ photons. Grey boxes are theoretical probability distributions. \{$\ket{X},\ket{\bar{X}}$\} refers to the $\mathrm{|0\rangle} \pm |1\rangle$ states. 
\textbf{i,} Expectation values of the coherence term $\hat{\Omega}_\theta^{\otimes {n}}$ for  ${n}=1,2,3,4$. Each point is derived from a set of 16 four-fold coincidences in the general $(\text{cos}\theta\hat{\sigma}_{x}+\text{sin}\theta\hat{\sigma}_{y})^{\otimes {n}} $ basis. 
The fringe in $\theta \in [0,\pi]$ is fitted with a sinusoidal function. 
\textbf{i,} Summary of measured fidelities for the teleported states and GHZ entangled states. 
Dotted line refers to the classical bound for teleported states $F=2/3$ (GHZ states $F=1/2$). To verify GHZ, entanglement witness (orange bars) and two-basis  measurement (green bars) are implemented. 
All error bars refer to $\pm$1 s.d. and are estimated from  Poissonian photon-counting statistics. 
}
\label{fig:TeleGHZResults} 
\end{figure*}    

We then control the nonlinear MRRs and linear operators to implement important quantum information tasks such as teleportation and multipartite entanglement. 
In the teleportation protocol, an unknown quantum state can be transmitted to another location, by locally collapsing the state and remotely reconstructing it~\cite{Telereview}. This requires access to Bell states and Bell measurements. 
The intra-chip teleportation experiment was initially implemented (Fig.~\ref{fig:TeleGHZResults}a), where we prepared an arbitrary single-qubit state $\ket{\psi} _2$ in photon 2 ("B") via a unitary $\hat {\text{P}}$, and a Bell pair $\ket {\Phi}^+_{3,4}$ in photon 3 ("C") and photon 4 ("D"). Photon 1 ("A") was used as a trigger.  
The \^{O}\textsubscript{Bell} measurement was performed at B and C, projecting the state into $\ket {\Psi}^+_{2,3}$ basis. 
This process allows the teleportation of B's state $\ket{\psi}_2$ to D up to a local rotation $\hat{\sigma}_x$. 
We prepared six different $\ket{\psi}_2$ states at B and reconstructed at D obtaining $\ket{\phi}_4$. 
Full QST was implemented to reconstruct density matrices for the six $\ket{\phi}_4$ states, see experimental data in Fig.~\ref{fig:TeleGHZResults}d. Each QST requires 3 measurement settings.  
We obtained high-fidelity teleported states with a mean $\bar{F}=0.906\pm0.014$, due to the high-quality control of nonlinear MRRs and linear circuits.  

A chip-to-chip teleportation of single-qubits is implemented as a proof-of-concept demonstration of a quantum transceiver system.  
To preserve coherent teleportation between the transmitter and receiver chips, we exploit a polarisation-rail conversion technique relying on the 2D SGCs (see SEM in Fig.~1c). 
In our experiments, the $\ket {\Phi}^+_{3,4}$ state was created on the transmitter in Fig.~1a (2D SGC not  shown), and qubit 4 was distributed to another receiver chip (see circuit in Fig.~\ref{fig:CTC}) via a 2m-long optical fiber. The states after distribution remain highly entangled.  
The \^{O}\textsubscript{Bell} was carried out on the transmitter.  
As examples, $\ket{0}_2$ and $\ket{+}_2$ were coherently teleported from the transmitter to the receiver (see data in Fig.~\ref{fig:CTCteledata}). The teleported states $\ket{\phi}_4$ are recovered on the receiver, and QST reports fidelities of $0.940\pm0.041$ and $0.832\pm0.048$, respectively. More details are provided in SI Sec.~\ref{sec:TELE}. 

Teleportation of entanglement (i.e, entanglement swapping) is a protocol whereby the sets of entanglement, say \{A, B\} and \{C, D\}, can be swapped by conducting a Bell projection on two unentangled qubits~\cite{Telereview}. This means that collapsing \{B, C\} onto any particular Bell state will result in entanglement of \{A,  D\}, which have never interacted with one another. 
Figure~\ref{fig:TeleGHZResults}b shows the circuit for entanglement swapping, where two Bell pairs $\ket{\Phi}^+_{1,2} \otimes \ket{\Phi}^+_{3,4}$ are created in the MRRs-array and \^{O}\textsubscript{Bell} is performed on the qubits 2, 3,  projecting the qubits 1, 4 into the entangled state $\ket{\Psi}^+_{1,4}$. 
Additionally, performing \^{O}\textsubscript{fusion} and measuring qubits 2, 3 in the $\hat{\sigma}_x\hat{\sigma}_x$ basis produces the entangled state $\ket{\Phi}^+_{1,4}$. 
We performed QST on qubits 1, 4 and reconstructed $\ket{\Psi}^+_{1,4}$ and $\ket{\Phi}^+_{1,4}$, each from 9 global measurement settings (36 four-fold coincidence measurements). Figure~\ref{fig:TeleGHZResults}e reports the measured $\rho$ with fidelities of $0.776 \pm 0.018$ and $0.737 \pm 0.019$, respectively, demonstrating successful on-chip entanglement swapping. 

We generated three- and four-photon GHZ entangled states on chip~\cite{GHZstate}. 
Performing \^{O}\textsubscript{fusion} on the two Bell pairs $\ket{\Phi}^+_{1,2} \otimes \ket{\Phi}^+_{3,4}$ enables the fusion between qubits 2 and 3, thus yielding the four-qubit GHZ state $\ket{\Phi}^n_{\text{GHZ}}=(\ket{0}^{\otimes {n}}+\ket{1}^{\otimes {n}})/\sqrt{2}$, $n=4$. 
This entangling process succeeds with a 0.5 probability, when detecting only one photon in each of \{D1, D2\}, \{D3, D4\}, \{D5, D6\} and \{D7, D8\} (Fig.~\ref{fig:TeleGHZResults}c). 
The four-photon coincidence events arise from one of the two cases, either all photons are in the $\ket{0}$ mode or in the $\ket{1}$ mode, which are quantum mechanically indistinguishable in the superposition basis and thus results in coherent entanglement. The $\ket{\Phi}^\text{3}_{\text{GHZ}}$ and $\ket{\Phi}^\text{2}_{\text{GHZ}}$ (i.e, $\ket{\Phi}^+_{1,4}$) states can be produced by locally measuring the remaining qubits in the $\hat{\sigma}_x$ basis. 

To verify genuine multipartite entanglement (GME) for the states where all subsystems are genuinely entangled, we measured an entanglement witness operator $\hat{W_n} = \hat{{I}}/2 - \ketbra{\Phi_\text{GHZ}^n}{\Phi_\text{GHZ}^n}$~\cite{
Friis19}. 
When the expectation value $\langle \hat{W}_n \rangle$ is negative, the presence of GME can be verified. For the GHZ states, $\langle \hat{W}_n\rangle$ can be estimated by $\Tr(\rho \hat{W} ) = 0.5- F'$, where $F'$ is the fidelity measured in global product bases without QST.  
Figures~\ref{fig:TeleGHZResults}f-h report the probability distribution in the basis $\hat{\sigma}_z ^{\otimes n}$, $\hat{\sigma}_x ^{\otimes n}$, and the general basis $\hat{\Omega}_{\theta}^{\otimes n}= (\cos{\theta} \hat{\sigma}_x + \sin{\theta} \hat{\sigma}_y)^{\otimes n}$, where \{$\hat{\sigma}_x$, $\hat{\sigma}_y$, $\hat{\sigma}_z$\} are the Pauli operators. 
In Fig.~\ref{fig:TeleGHZResults}h, the fringes of $\langle \hat{\Omega}_{\theta}^{\otimes n} \rangle $ show the measured coherence of the GHZ states, and the variable oscillatory frequency confirms the correlation nature of $n$-photon GHZ entanglement. 
The fringe is observed by simultaneously rotating the four projectors $\hat{\text{M}}$ along the axis $\mathrm{|0\rangle} + e^{i \theta} |1\rangle$. 
Using the data in the $\hat{\sigma}_z ^{\otimes n}$ basis and $\hat{\Omega}_{\theta}^{\otimes n}$ basis ($\theta=k\pi /n$, $k=0,1,...n-1$), with a total $n+1$ measurement settings, we calculated the fidelities for $\ket{\Phi}_\text{GHZ}^n$ (see results in Fig.~\ref{fig:TeleGHZResults}i). The measured $\langle \hat{W}\rangle_{n=3,4} $ values are $-0.235 \pm 0.017$ and $-0.183 \pm 0.014$, certifying the genuine entanglement of the three- and four-photon GHZ states with at least $13\sigma$. 

We further quantified GME of the generated GHZ states without QST or other assumptions on the state itself. 
A suitable quantifier is given by the GME-concurrence ($C_\text{GME}$) \cite{GMEconcurrence}, which has computable bounds that turn out to be exact for GHZ-diagonal states. 
To showcase the quality and scalability of such an evaluation, we used an efficient framework to lower bound the GME-concurrence from only two global measurement settings (SI). 
Using only the outcome statistics of two measurements, $\hat{\sigma}_z^{\otimes n}$ in Fig.~\ref{fig:TeleGHZResults}f and $\hat{\sigma}_x^{\otimes n}$ in  Fig.~\ref{fig:TeleGHZResults}g, we arrive at a value of $C_\text{GME}\leq 0.390\pm0.04$ and $C_\text{GME}\leq 0.192\pm0.039$ for three and four-photon GHZ states respectively, thus efficiently quantifying genuine multipartite entanglement. The two-basis measurements can also be used to efficiently lower bound the state fidelity \cite{OAMNatPhys}, and results are provided in  Fig.~\ref{fig:TeleGHZResults}i. See SI Sec.~\ref{sec:GHZ} for more details on the generation, verification, and quantification of GME. 

We have presented silicon-photonic quantum devices able to generate, process, and transceive  multiqubit states. 
The nonlinear multiphoton sources and linear multiqubit circuits are naturally interfaced with low noise and coherently controlled in a single system, where each part is individually programmable. 
The multiple MRR-enhanced photon sources are approaching optimal levels of purity, indistinguishability, and heralding efficiency. We have performed fundamental protocols in quantum computing and communication, showcasing the abilities of our devices for multiqubit entanglement and teleportation. 
Unprecedented high-fidelity quantum operations are achieved, e.g, teleportation with a $\sim$0.90 fidelity is among the highest ones (see Ref.~\cite{Telereview}).  
In future, the heralding efficiency of the sources can be further improved by engineering the resonators~\cite{Vernon2016}, and the photon spectra can be topologically protected against fabrication errors~\cite{Mittal2018}. 
Together with the developed multiplexing technology~\cite{multiplexing}, our  sources would allow near-deterministic generation of single-photons. 
Moreover, silicon offers the unique capability of integrating electronics and photonics~\cite{Atabaki2018}, promising large-scale integration of quantum circuits~\cite{Wang16d} and their efficient quantum control. Simultaneously scaling up the number of photons and the dimensionality~\cite{Malik2016} would allow the opportunity to have exponentially larger Hilbert spaces with integrated optics. 
In general, our demonstrations pave the way for a complex integration of quantum nonlinear and linear optic devices in silicon, that may provide a scalable and versatile platform for the study of quantum photonic~\cite{Gu2018}, computational~\cite{One-way,RomeBS}, physical~\cite{Harris}, and biochemical~\cite{Sparrow2018} systems.  

\section*{Author contributions}
All authors contributed to the discussion and development of the project.  J.W. devised the concept of the experiment. Y.D. designed and fabricated the device. D.L., Y.D., I.F., J.W., S.P., D.B., and R.S. performed the experiment. J.W., D.L., Y.D., I.F., Y. Q., Y.L., Y.X., M.H., M.M., G.F.S., and X.Z. performed the theoretical analysis. K.R., J.L.O., J.G.R., Q.G., L.K.O., Y.D., J. W. and M.G.T. managed the project. The manuscript was written by J.W., D.L., Y.D. and I.F. with the input from all others.  

\section*{Acknowledgments} 
D.L, I.I.F., J.G.R. and M.G.T. acknowledge support from UK Quantum Technology Hub for Quantum Communication Technologies funded by the EPSRC:  EP/M013472/1; ; programme grant no EP/L024020/1.
Y.D. acknowledges support from Denmark SPOC (DNRF123) and VILLUM FONDEN, QUANPIC (00025298).
I.I.F. acknowledges FP7 Marie Curie Initial Training Network PICQUE (608062). 
We acknowledge support from the Natural Science Foundation of China (nos. 61975001, 61590933, 11527901, 11825402), National Key R\&D Program of China (nos. 2018YFB1107205), Beijing Natural Science Foundation (Z190005), Beijing Academy of Quantum Information Sciences (Y18G21) and Key R\&D Program of Guangdong Province (2018B030329001). 
M.H. acknowledges support from the Austrian Science Fund (FWF) through the START project Y879-N27 and the joint Czech-Austrian project MultiQUEST (I 3053-N27 and GF17-33780L). 
M.M. acknowledges support from the EPSRC (EP/P024114/1) and the QuantERA ERA-NET co-fund (FWF Project I3773-N36). 
Y.L. acknowledges the support from National Key R and D Program of China (grant nos. 2018YFB1107202, 2016YFA0301302) and NSFC (nos. 61590933, 11627803). 
Y.X. acknowledges the NSFC (grant nos. 11825402, 11654003). 
K.R. acknowledges the support from the QuantERA (SQUARE). 
J.L.O. acknowledges a Royal Society Wolfson Merit Award and a Royal Academy of Engineering Chair in Emerging Technologies.  
Q.G. acknowledges the National Key R and D Program of China (no.2013CB328704). 
We acknowledge the support from the Engineering and Physical Sciences Research Council (EPSRC), European Research Council (ERC), National Natural Science Foundation of China (NSFC), Denmark SPOC (ref DNRF123), Bristol NSQI.
M.G.T. acknowledges support from the ERC starter grant (ERC-2014-STG 640079) and and an EPSRC Early Career Fellowship (EP/K033085/1). 
We thank G. J. Mendoza and D. Bonneau for useful discussions. We thank W. A. Murray, M. Loutit, E. Johnston, J. W. Silverstone and L. Kling for experimental assistance.

\bibliographystyle{unsrtnat}

\bibliography{biblio}


\newpage 
\clearpage

\newcommand{\hbAppendixPrefix}{S}
\renewcommand{\thefigure}{\hbAppendixPrefix\arabic{figure}}
\setcounter{figure}{0} 

\renewcommand{\thetable}{\hbAppendixPrefix\arabic{table}} 
\setcounter{table}{0}
\renewcommand{\theequation}{\hbAppendixPrefix\arabic{equation}} 
\setcounter{equation}{0}

\pagenumbering{arabic}
\setcounter{page}{1}
\onecolumn

\section*{\centering\fontsize{15}{15}\selectfont 
Supplementary Information: \\
Chip-to-chip quantum teleportation and \\
\qquad\qquad\qquad\qquad multi-photon entanglement in silicon}

\section{Device Details and Experimental Setup }
\label{sec:DesignSetup}
In our experiment, we prepare high fidelity dual-rail qubits ($\alpha \ket{0}+\beta\ket{1}$) on-chip by confining single-photons between two optical modes $\ket{0}$ and $\ket{1}$, physically in two optical waveguides or paths. 
Measurements on each qubit in the computational basis (\{$\ket{0}$, $\ket{1}$\}) are achieved by detecting single-photons at each of the optical modes. Single qubit unitary transformations and arbitrary projective measurements are performed by controlling the relative phase between the two modes and the probability amplitude of detection events between each optical mode. 
These effects are realised on our device by integrating thermo-optic phase-shifters and optical MZIs. \\

On this device, up to four logical qubits are created and manipulated simultaneously. In order to entangle and process multiple qubit states, it is crucial to develop pure and identical photon sources (see Fig.~\ref{fig:SourcesResults} and section~\ref{sec:MRRsource}) and to develop high-fidelity multi-qubit operators (see Fig.~\ref{fig:OperatorResults} and section~\ref{sec:opeators}). 
Spectrally identical single photons are created by overlapping each of the micro-ring resonances. A high number purity is achieved by operating the device in a low-squeezing regime. 
After the single-photons pass through the asymmetric MZI, the photon states can be mapped into the orthonormal basis vectors given by the four logical states: $\ket 0 _{\lambda,k}$, and $\ket 1 _{\lambda,k}$, where $\lambda$ represent the photon wavelength, and $k$ is the qubit number. In this representation, one can map well-known quantum information protocols to single-photon state propagation on-chip. The four qubits are entangled and processed in a programmable linear-optic quantum circuit, able to implement Bell projection $\hat{O}_\text{Bell}$ and fusion operation $\hat{O}_\text{fusion}$. 
To certify the developed high-quantity multiphoton sources and multiqubit operators, we performed three main experiments: single-qubit teleportation, entanglement swapping (see Fig.~\ref{fig:TeleGHZResults} and section~\ref{sec:TELE}), and four-photon GHZ state generation (see Fig.~\ref{fig:TeleGHZResults} and section~\ref{sec:GHZ}). 

\subsection{Device Design} 
\label{subsec:Design}
\begin{figure*}[!b]
  \centering
  \includegraphics[width=0.65\textwidth]{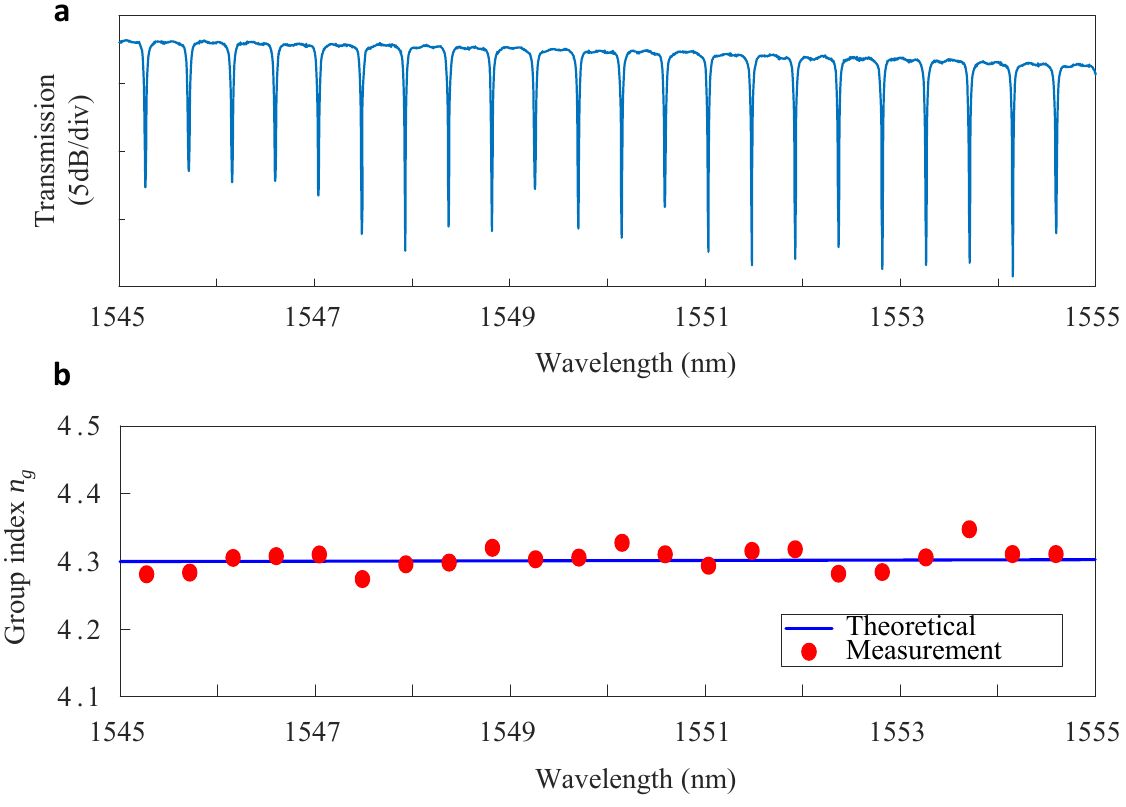}
  \caption{ \footnotesize	
Measurement of the group index $n_g$ of silicon waveguide. \textbf{a,} Measured transmission of a MRR with $R_{\text{test}}=200\mu m$. 
\textbf{b,} Theoretical calculated (blue line) and measured group index (red points) from the MRR, showing a good agreement and an average group index of 4.31. } 
\label{WGng}
\end{figure*}

\textbf{Group index:}  The standard single mode silicon waveguides of our silicon devices have geometry of 450~nm$\times$250~nm. The group index of the standard silicon waveguide is experimentally characterised by designing a test microring resonator (MRR) with radius $R_{\text{test}}$ of 200$\mu$m. The transmission of the test MRR is measured, as presented in Fig.~\ref{WGng}a, and the group index $n_g$ is given by 
\begin{equation}
 n_g = \frac{c_0}{\text{FSR}_{\text{test}} 2\pi R_{\text{test}}},
\end{equation}
where $c_0$ is the light speed in vacuum, and $\text{FSR}_{\text{test}}$ is the free spectral range (FSR) of the test MRR. An average group index $n_g$ of the silicon waveguide around 1550~nm is measured to be 4.31, which is in good agreement with theoretical calculation, as shown in Fig.~\ref{WGng}b.\\

\begin{figure*}[h!]
  \centering
  \includegraphics[width=0.65\textwidth]{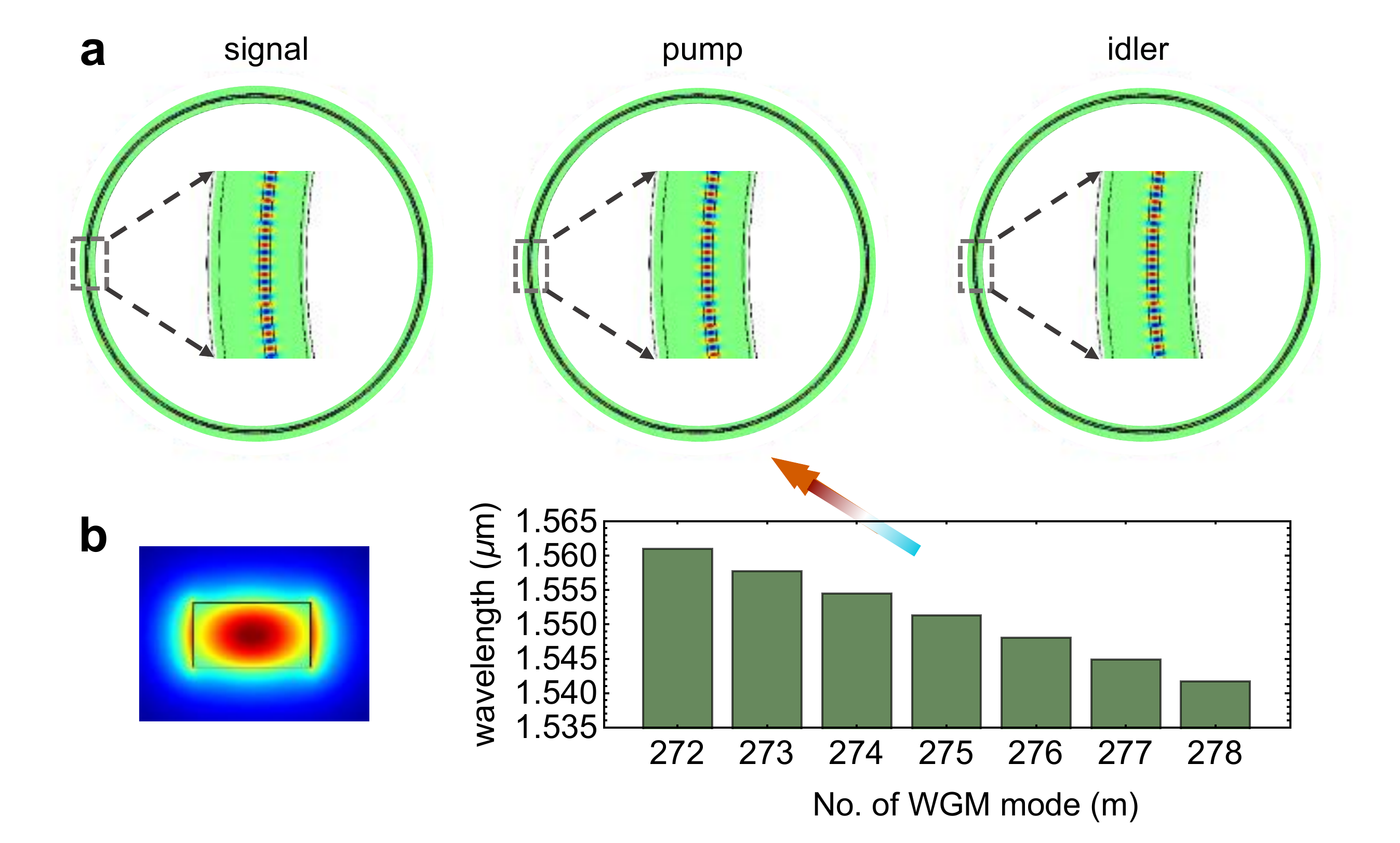}
  \caption{ \footnotesize	
Numerical simulation of a MRR source with $R=27.68\mu m$. \textbf{a,} Simulated electric field distributions at the resonant wavelengths for the pump ($m=275$), signal ($m=278$) and idler ($m=272$) photons, respectively. The integer $m$ refers to the  quantum number of WGM eigenmodes. 
\textbf{b,}  Simulated transverse electric TE mode in the MRR. } 
\label{SIMRRSim}
\end{figure*}

\noindent \textbf{MRR:} Optical microresonators can confine light by internally reflecting light at the edges of the resonator. 
In this work, we develop silicon MRRs for multiphoton generation. The FSRs of four MRRs are all designed to be 400 GHz, compatible with the channel spacing in the ITU (international telecommunication union) standardisation. This allows us to perform the quantum experiments using off-the-shelf telecommunication instruments. The 400GHz $\text{FSR}_\text{MRR} $ is corresponding to radius of $R=27.68\mu m$ in our design. 
Figure~\ref{SIMRRSim} shows the numerical simulation results for the designed MRR source. We used a mode-solution with 2D rotational symmetric structure and EWFD (Electromagnetic Waves, Frequency Domain) module in COMSOL. 
A series of standing-wave are existing at the resonance wavelengths of the MRR, given by: 
\begin{equation}
 m \lambda_m = 2\pi R n_{\text{eff}}, 
\end{equation}where $\lambda_m$ is the resonant wavelength in the $m$-th whispering-gallery mode (WGM) in MRR, and  $n_{\text{eff}}$ is the effective refractive index. Here we only excited the transverse electric (TE) mode. Fig.~\ref{SIMRRSim}b shows the intrinsic TE mode in the MRR. With an inclusion of material dispersion, the simulation results confirm the FSR of 400 GHz in our design. 
The simulated electric field distributions for the pump, signal and idler photons are respectively provided in Figs.~\ref{SIMRRSim}a, where enlarged images are shown in the insert. Figure 2a in the main text shows the experimentally measured $\text{FSR}_\text{MRR}$ which is about 406.25 GHz. \\

\noindent \textbf{AMZI:} Asymmetric Mach-Zehnder interferometers (AMZIs) having two unbalanced arms with a length difference $\Delta L= 217.372 ~\mu m$ (see Fig. 1a) gives a $\text{FSR}_\text{AMZI}$ of 320 GHz. That means the ratio of $\text{FSR}_\text{MRR}/\text{FSR}_\text{AMZI}$ eqauls 5/4.  In our experiment, in order to efficiently remove the pump photons from the generated single-photons with a high extinction ratio, the signal and idler photons are chosen at $\pm$ 3 $\text{FSR}_\text{MRR}$) away from the pump wavelength, i.e., $\lambda_s=\lambda_{p}- 3  ~ \text{FSR}_\text{MRR} $ and $\lambda_i=\lambda_{p}+3  ~\text{FSR}_\text{MRR} $. 
The choice of 320 GHz $\text{FSR}_\text{AMZI}$ thus allows us to separate the signal and idler photons on-chip and collect the two photons at different output ports. The measured $\text{FSR}_\text{AMZI}$ is about 321.25 GHz, and experimental value of $\text{FSR}_\text{MRR}/\text{FSR}_\text{AMZI}$ is about 1.26, consistent with the theoretical design. Despite the slight deviation of FSR arising from the device fabrication, the large bandwidth of AMZI still allows an efficient separation and routing of the signal and idler photons. We obtained a high extinction of more than 30dB separating the signal and idler photons. The full spectra for AMZIs and MRRs are provided in Fig.~\ref{fig:SourcesResults}a. \\

\noindent \textbf{SGC:} Sub-wavelength grating couplers (SGC) can directly couple light from silicon nano-waveguides to single-mode fibers, and vice verse. To increase the coupling efficiency, we designed SGCs with an apodized photonic crystal  structure to match optical modes in single-mode fiber and SGC~\cite{Ding2013:SGC}. A 100 nm aluminium (Al) mirror is introduced below the lower cladding to further increase the coupling efficiency~\cite{Ding2014:SGC}. In our experiments, the coupling angle $\theta$ is designed to be 15 degree, which is given by: 
\begin{equation}
 \lambda = l_i (n_{eff,i}-n_0 \text{sin}\theta), 
\end{equation}
where $\lambda$ is the peak wavelength targeted near 1550 nm, $n_0$ is the refractive index of the uppermost cladding (air). $n_{eff,i}$ and $l_i$ are the effective refractive index and length of the $i$-th scattering unit of the photonic crystal structure based scattering units along the grating. By optimising the geometry of each scattering unit, \{$n_{eff,i}$, $l_i$\}, we can tailor the output optical field and obtain a Gaussian output profile from the SGC. The width of the photonic crystal slots is  345 nm. In our devices, 1D SGCs are designed for the TE mode, and exploited for chip-fiber coupling. We also designed 2D SGCs, formed by superposing two 1D SGCs at a right angle, and exploited the 2D SGCs as path-polarisation converters for chip-to-chip  entanglement distribution and teleporation experiments (see details in section~\ref{subsec:CTCTELE}). 

\subsection{Device Fabrication}
\label{subsec:Fabrication}
In on-chip quantum experiments, decreasing optical losses, in particular coupling loss and insertion loss of quantum optical components, is critical. For this purpose, we achieve ultra-high efficiency grating couplers by preparing a sophisticated silicon-on-insulator platform with bonded Al mirror ~\cite{Ding2014:SGC}. It starts from a commercial silicon-on-insulator (SOI) wafer with top silicon thickness of 250\,nm and a buried oxide layer of 3\,$\mu$m. Firstly, 1.6\,$\mu$m thick SiO$_2$, which is an optimum thick SiO$_2$ for fully-etched grating coupler with Al mirror~\cite{Ding2014:SGC}, is deposited by the plasma-enhanced chemical vapour deposition (PECVD) process on the SOI wafer. After that, the Al mirror is deposited by electron-beam (ebeam) evaporator, and followed by another thin layer of SiO$_2$ deposition with thickness of 1\,$\mu$m. The wafer is flip-bonded to another silicon carrier wafer by Benzocyclobutene (BCB) bonding process. The final Al-introduced SOI wafer is consequently achieved by removing the substrate and buried oxide (BOX) layers of the original SOI wafer. The silicon photonic circuit with fully-etched apodized grating couplers using a photonic crystal~\cite{Ding2013:SGC} are fabricated by standard ebeam lithography (EBL) followed by Inductively Coupled Plasma (ICP) etching and ebeam resist stripping. After the photonic circuit part is fabricated, 1.3\,$\mu$m thick SiO$_2$ is deposited by PECVD, followed by chemical mechanical polishing (CMP) process to planarize the surface with approximately 300\,nm sacrifice, resulting in a final top SiO$_2$-cladding layer of 1\,$\mu$m. The micro-heaters are patterned afterwards by standard ultraviolet (UV) lithography process followed by 100\,nm titanium (Ti) deposition and liftoff process. The conducting wires and electrode pads are obtained by a second UV lithography followed by Au/Ti deposition and lift-off process.\\

\begin{figure*}[h!]
  \centering
  \includegraphics[width=0.75 \textwidth]{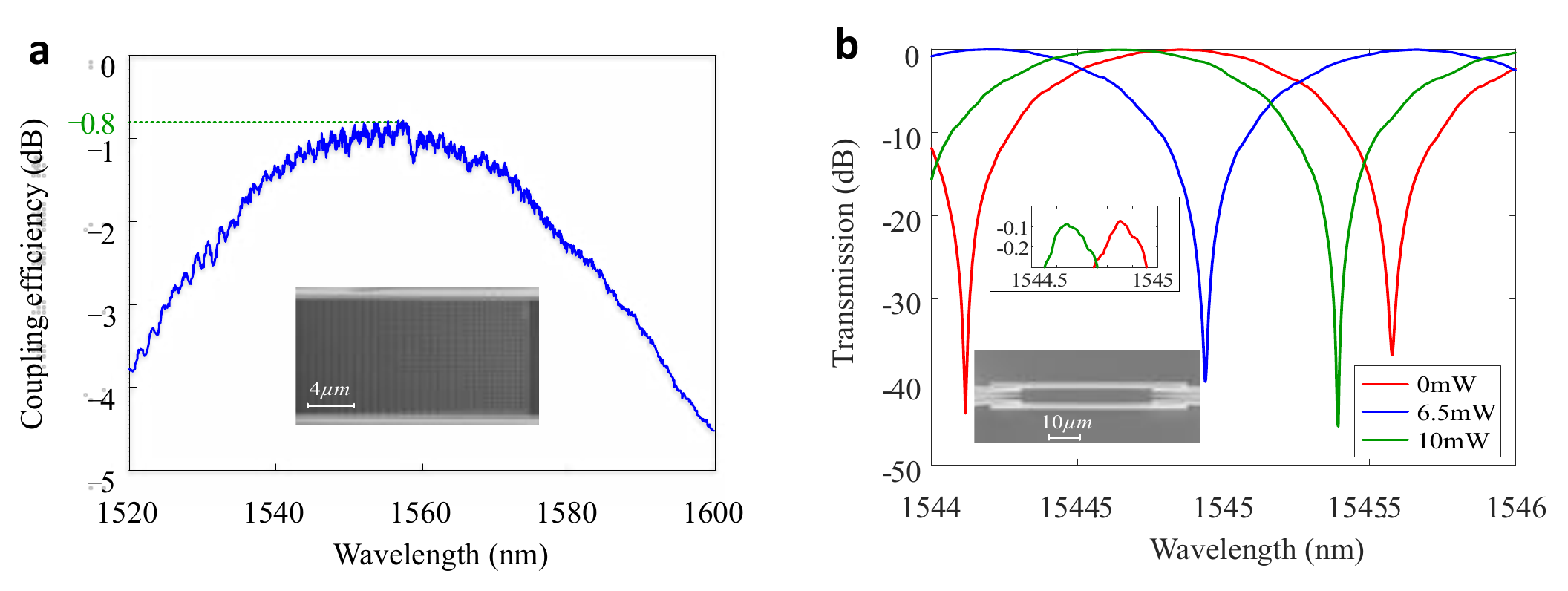}
  \caption{ \footnotesize	
Characterisations of the integrated optical components. \textbf{a,} measured spectrum for a fully etched 1d-dimensional grating coupler on the silicon-on-insulator platform with Al mirror. \textbf{b,} measured spectrum for a thermal-tunable AMZI used in the silicon circuits. The scanning electron microscope (SEM) images of the fabricated grating coupler with one-dimensional photonic crystal and 2$\times$2 MMI are presented in each figure. } 
\label{SIFig_components}
\end{figure*}

Our fabrication platform enables a propagation loss of $\sim$2\,dB/cm measured by the cut-back method for the standard fully-etched silicon waveguide with a geometry of 450\,nm$\times$250\,nm. Figure~\ref{SIFig_components}a shows the characterisation of 1d SGCs. Thanks to the the Al-mirror, a peak coupling efficiency of -0.8\,dB at 1555\,nm, with 1\,dB bandwidth of 40\,nm is achieved. The 2$\times$2 multimode interferometers (MMIs) are developed for 50:50 beamsplitters. In order to characterise the thermal tunability of the Ti heater and splitting ratio, insertion loss of the 2$\times$2 MMI structures, we implemented an AMZI filter with a Ti micro-heaters applied on one arm as phase-shifters. Note that this AMZI is only used for testing the performance of MMIs and phase-shifters, which have a different FSR as the ones in Fig.1a and section~\ref{subsec:Design}. 
In this situation, applying a heating power to the Ti micro-heater results in a change of the refractive index in the silicon waveguide, inducing a phase shift and thus transmission shift. As shown in Fig.~\ref{SIFig_components}b, 14.5~\,mW heating power results in a transmission shift of more than one FSR. The resistance of the Ti heaters is measured to be around 500 $\Omega$. 
Such efficient Ti heaters enable us to efficiently fully reconfigure the quantum circuit to prepare, operate and measure different quantum states, and also precisely align the four MRRs to obtain indistinguishable single photon generation. Moreover, the transmission presented in the inset of  Fig.~\ref{SIFig_components}b is less than 0.1\,dB, indicating a insertion loss less than 0.05\,dB for each 2$\times$2 MMI. The high extinction ratio in Fig.~\ref{SIFig_components}b also confirms the highly balanced splitting ratio in the MMIs. 

\subsection{Experimental Setup}
\label{subsec:Setup}
\begin{figure}[t!]
\centering
\includegraphics[width=1\textwidth]{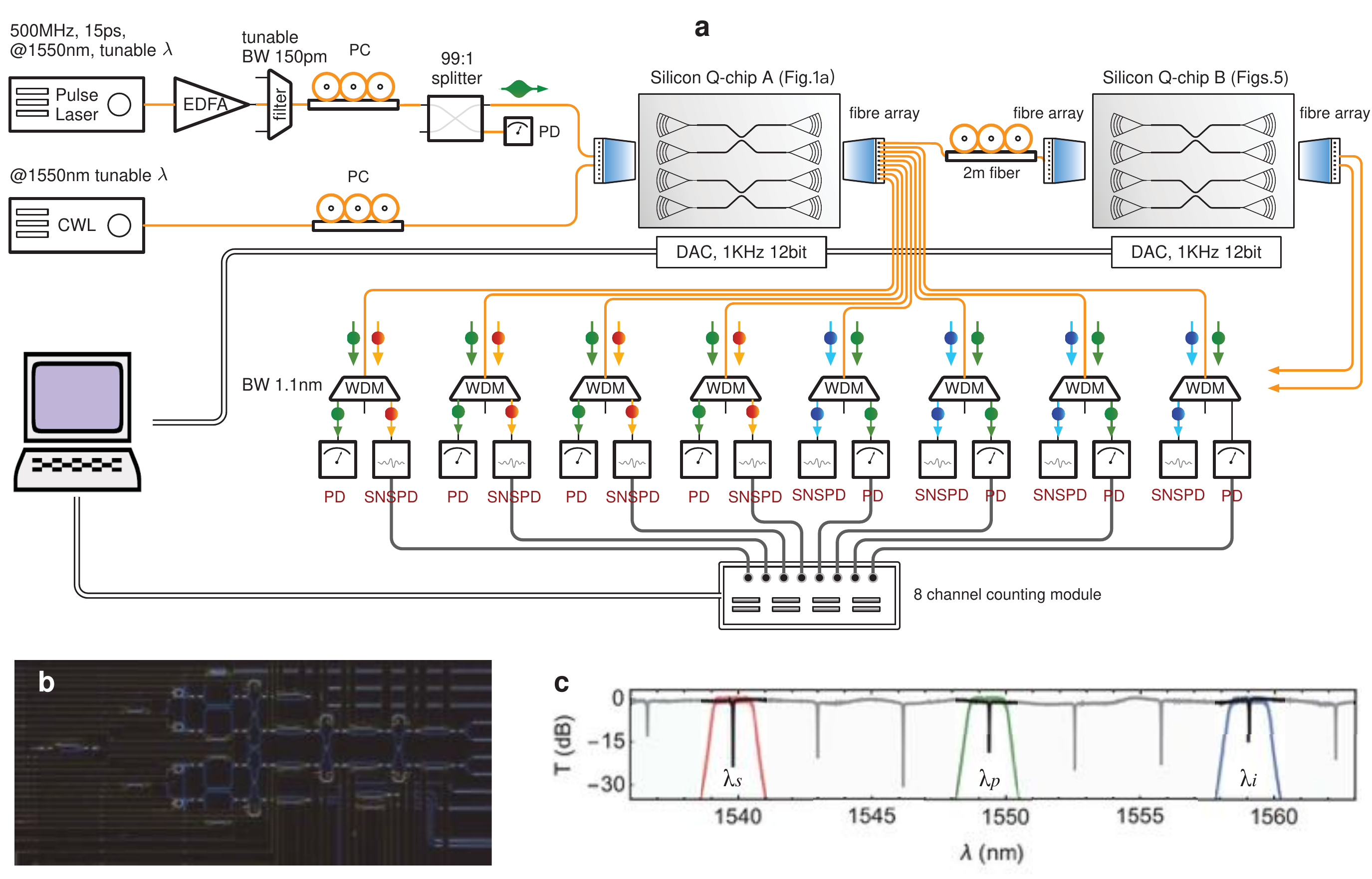}
\caption{
\textbf{a,} Schematic of the experimental setup. 
A pulse laser is used to pump the MRR single-photon sources. It has a repetition rate of 500 MHz, pulse width of 15 ps and tunable wavelength around 1550 nm. 
CWL: continuous-wave laser, which is tunable around 1550 nm; EDFA: erbium doped fiber amplifier;  WDM: wavelength-division (de)multiplexer; PD: photo-diode; SNSPD: superconducting nano-wire single-photon detectors; PC: fiber polarisation controller. DAC: digital-to-analog converter. Orange wires refer to single-mode optical fibers; back wires refer to electronic cables. 
The schematic for the chip A circuits is shown in Fig.1a in the main text, and the schematic for the chip B circuits is shown in Figs.~\ref{fig:CTC} in section~\ref{subsec:CTCTELE}. 
\textbf{b,} An optical microscope image of the silicon device in Fig.1a. The four MRR photon sources and MZIs can be observed. Blue lines are silicon waveguides, and gold wires are the phase-shifters and the conductive wires to access the phase-shifters. 
\textbf{c,} The measured spectra for the off-chip WDM filters with a bandwidth of 1.1nm and a high extinction ratio. These filters allow an efficient removal of residual pump photons (green) from the created signal photons (blue) and idler photons (red). 
The MRR spectrum is measured as shown in black and its resonance linewidth (30-40pm) is about 2-3 orders narrower. }
\label{fig:setup}
\end{figure}

A schematic of the experimental setup is shown in Fig.~\ref{fig:setup}. 
Two non-degenerated photon-pairs, in total 4 photons, are generated on-chip via the spontaneous four wave mixing (SFWM) process in the array of silicon MRRs.  
The MRR photon-pair sources are pumped by a pulse laser source amplified by a commercial erbium doped fiber amplifier (EDFA). 
The pulse laser has a 500MHz repetition rate and a 15ps pulse width. In our quantum experiments, the averaged power coupled into the chip was about 800 $\mu$W. 
The wavelength is tuned around 1550 nm. 
The spectrum of the amplified pump is filtered through a tunable filter with a {$0.12~\textrm{nm}$} narrow bandwidth, which is about 4 times wider than the linewidth of MRRs. 
The CW light is used for device characterisations and for monitoring the coupling and optimising input polarisation. The SFWM enhancement in Fig.~\ref{fig:SourcesResults}b was collected by pumping the MRRs with weak CW light  and scanning the wavelength over the resonance. 
The input light polarisation is optimised by a polarisation controller, then coupled into the chip through the 1d SGCs (Fig.1b and Fig.~\ref{SIFig_components}a). After the chip, photons are coupled into an array of optical fibers, then filtered through an array of wavelength division multiplexing (WDM) filters removing the residual pump photons from the generated single-photons. \\

In our experiments, we used pump light with wavelength of $\lambda_p=1549.35$ nm, and generated single photon-pairs at the resonances of the MRRs. 
We selected the signal photon at $\lambda_s=1539.758$ nm and idler photon at $\lambda_i=1559.015$ nm, by carefully designing the $\text{FSR}_\text{MRR}$ and $\text{FSR}_\text{MZI}$ (see Fig.~\ref{fig:SourcesResults}a and section~\ref{subsec:Design}), and by combing the off-chip WDMs (see Fig.~\ref{fig:setup}c). The >100dB off-chip WDM filters can efficiently remove the residual pump photons (green) from the created signal photons (blue) and idler photons (red). 
Since the bandwidth of the WDM filters  is about 1.1 nm, 2-3 orders wider than the linewidth ($37.7\pm 1.9$ pm) of MRR resonances (Fig.~\ref{fig:FigSMRRSources}), no spectral filtering was added in the created photon-pairs for improving the spectral purity of photons. This greatly improves the herald efficiency and the brightness of photons, which are essential to high-efficiency single-photon generation and any multiphoton quantum application. The single-photon pairs are detected through an array of 8 superconducting nano-wire single-photon detectors (SNSPDs) from PhotonSpot, with an average efficiency of $85\%$, $100$ Hz dark counts and $50$ ns dead-time. The 8 SNSPDs output electronic signals are analysed by a 8 channel logic counting module. The data was process in a classical computer. \\

In our chip-to-chip teleportion experiments, the quantum states were transmitted from chip A (transmitter, where the multiphoton states are created) to chip B (receiver, where the teleported states are reconstructed). The two chips are coherently linked by a single-mode optical fiber, using the path-polarisation conversion technique. The chip-to-chip experimental details and results are provided in section~\ref{subsec:CTCTELE}. \\

The devices are wire-bonded on PCB. All the thermo-optic phase-shifters can be individually controlled by computer-interfaced electronic controllers, with 12-bits resolution and micro-second speed. Overall all phase-shifters on the chip can be reconfigured at a kHz rate. Each phase-shifter is connected to two separate pads (one for signal and one for ground) to prevent electronic cross-talk. 
A Peltier-cell together with a thermistor and a proportional integrative derivative controller were used to keep the temperature of the photonic device stable. A standard water-cooling system was built to ensure an efficient dissipation of the power injected into the photonic chip. 

\subsection{Re-calibration of Single-photon Detectors }
In Figure~\ref{fig:setup},  eight SNSPDs are used in our experiment and 
multi-photon coincidence measurements are recorded in order to estimate the relative probability amplitudes of quantum states when projected into local bases. 
In order for the normalised counts to give a true estimate of these relative probabilities, we have to ensure that measurements corresponding to different eigenvalues have equal detection efficiency's. In reality, since single photons in our device travel different paths and are coupled to different optical fibres, the relative detection efficiency of photons in the $\ket{0}$ and $\ket{1}$ modes of each qubit will vary. \\

One solution is to collect photons only belonging to a single eigenstate, say $\ket{0000}$ (having a fixed heralding efficiency) and then use the four on-device projectors to rotate the measured basis accordingly. For example, rather than directly measuring coincidence counts corresponding to the eigenvector $\ket{1111}$ one could continue to measure $\ket{0000}$ but apply the transformation $\hat{\sigma}_x^{\otimes 4}$ along all of the qubits. In this scenario, the photons detected would see the identical losses (providing the setup is temporally stable) and so the relative coincidence counts would give an accurate description of the quantum mechanical probabilities of measuring each state. One drawback of this method is that the number of measurements required per measurement in a single basis (for N qubits) scales by a factor of $2^N$. In this case of our experiments, this gives a $2\times$ increase in measurements for teleportation, $4\times$ increase for entanglement swapping \& Bell projections and a $16\times$ increase in GHZ measurements. \\

An alternative approach is to keep track of the heralding efficiency's (they will vary each time the coupling between chip and fibre array changes) and correct ones counts in the following manner. Let the detection probability of a photon in the $\ket{0}$  ($\ket{1}$ ) mode of the $i^{\text{th}}$ qubit be written as $\eta_{i,0}$ $(\eta_{i,1})$. In this case the measured four-fold counts $cc_{\text{meas}}$ may be written in terms of the on-device four-fold counts $cc_{\text{true}}$ as 

\begin{equation}
cc_{\text{meas},ijkl}
= \eta_{1,i} \eta_{2,j} \eta_{3,k} \eta_{4,l} cc_{\text{true},ijkl}.
\end{equation}


In order to balance the losses across each of the measurements, we can manually correct the counts by normalising them relative to the $\ket{0000}$ modes 

\begin{equation}
\frac{cc_{\text{meas},ijkl}}{cc_{\text{meas},0000}}
=
\frac{\eta_{1,i} \eta_{2,j} \eta_{3,k} \eta_{4,l}}{\eta_{1,0} \eta_{2,0} \eta_{3,0} \eta_{4,0}}
\frac{cc_{\text{true},ijkl}}{cc_{\text{true},0000}}
\end{equation}
and so the desired corrected quantity, $cc_{\text{corr},ijkl}=cc_{\text{true},ijkl}/cc_{\text{true},0000}$, can be measured in terms of the measured quantities $cc'_{\text{meas},ijkl}=cc_{\text{meas},ijkl}/cc_{\text{meas},0000}$ the ratio of heralding efficiency's in the following way: 
\begin{align*}\label{}
cc_{\text{corr},0000}
&= 
1,
\\
cc_{\text{corr},0001}
&= 
\frac{\eta_{4,0}}{\eta_{4,1}}
cc'_{\text{meas},0001},
\\
cc_{\text{corr},0010}
&=
\frac{\eta_{3,0}}{\eta_{3,1}}
cc'_{\text{meas},0010},
\\
cc_{\text{corr},0011}
&= \frac{\eta_{3,0}}{\eta_{3,1}}  
\frac{\eta_{4,0}}{\eta_{4,1}}
cc'_{\text{meas},0011},
\\
cc_{\text{corr},0100}
&=  \frac{\eta_{2,0}}{\eta_{2,1}}
cc'_{\text{meas},0100},
\\
cc_{\text{corr},0101}
&=  \frac{\eta_{2,0}}{\eta_{2,1}}
\frac{\eta_{4,0}}{\eta_{4,1}}
cc'_{\text{meas},0101},
\\
cc_{\text{corr},0110}
&= \frac{\eta_{2,0}}{\eta_{2,1}}
\frac{\eta_{3,0}}{\eta_{3,1}}
cc'_{\text{meas},0110},
\\
cc_{\text{corr},0111}
&= \frac{\eta_{2,0}}{\eta_{2,1}}
\frac{\eta_{3,0}}{\eta_{3,1}}
\frac{\eta_{4,0}}{\eta_{4,1}}
cc'_{\text{meas},0111},
\\
cc_{\text{corr},1000}
&=  \frac{\eta_{1,0}}{\eta_{1,1}}
cc'_{\text{meas},1000},
\\
cc_{\text{corr},1001}
&= \frac{\eta_{1,0}}{\eta_{1,1}}
\frac{\eta_{4,0}}{\eta_{4,1}}
cc'_{\text{meas},1001},
\\
cc_{\text{corr},1010}
&= \frac{\eta_{1,0}}{\eta_{1,1}}
\frac{\eta_{3,0}}{\eta_{3,1}}
cc'_{\text{meas},1010},
\\
cc_{\text{corr},1011}
&= \frac{\eta_{1,0}}{\eta_{1,1}}
\frac{\eta_{3,0}}{\eta_{3,1}}
\frac{\eta_{4,0}}{\eta_{4,1}}
cc'_{\text{meas},1011},
\\
cc_{\text{corr},1100}
&=  \frac{\eta_{1,0}}{\eta_{1,1}}
\frac{\eta_{2,0}}{\eta_{2,1}}
cc'_{\text{meas},1100},
\\
cc_{\text{corr},1101}
&=  \frac{\eta_{1,0}}{\eta_{1,1}}
\frac{\eta_{2,0}}{\eta_{2,1}}
\frac{\eta_{4,0}}{\eta_{4,1}}
cc'_{\text{meas},1101},
\\
cc_{\text{corr},1110}
&=  \frac{\eta_{1,0}}{\eta_{1,1}}
\frac{\eta_{2,0}}{\eta_{2,1}}
\frac{\eta_{3,0}}{\eta_{3,1}}
cc'_{\text{meas},1110},
\\
cc_{\text{corr},1111}
&= \frac{\eta_{1,0}}{\eta_{1,1}}
\frac{\eta_{2,0}}{\eta_{2,1}}
\frac{\eta_{3,0}}{\eta_{3,1}}
\frac{\eta_{4,0}}{\eta_{4,1}}
cc'_{\text{meas},1111}.
\end{align*}

The four heralding efficiency ratios can be measured by comparing two fold coincidence counts from a single sources across both modes of each qubit. For example, in the regime where a single source is pumped, the single photons produced per pulse inside the device is the probability of generating a photon per pulse (at a given power) multiplied by the repetition rate, $Rp$. The coincidence counts across any two given modes are then $cc= \eta_1 \eta_2 R p$. The four correction ratios above can then be calculated by using an ancillary channel to compare the two-fold counts produced from the same source when swapping the path of single photons between the zero and one mode of a particular qubit. In the case of the first qubit, 

\begin{equation}
\frac{ cc_{0,\text{ancilla}} }{ cc_{1,\text{ancilla}} } = \frac{\eta_0}{\eta_1}.
\end{equation}

\section{Micro-ring Resonator Single-photon Sources}
\label{sec:MRRsource}
In silicon, single-photons are generated from heralded single-photon sources (HSPSs) such as waveguides or microresonators which have to fulfil some criteria to be scalable. In such a source, due to the third-order nonlinear optical process, spontaneous four-wave mixing, two pump photons are absorbed and in order to conserve the energy and momentum, a photon-pair, historically called signal-idler, is produced. Detecting the idler photon herald the presence of signal photon thus acting as a heralded single-photon source (HSPS). This heralding process projects the signal photon into specific spectro-temporal modes. For the heralded signal photons from multiple independent sources to be identical (i.e. spectrally pure), the signal photons need to have the same spectral information such as the spectral shape of the photons, and the joint spectral densities (JSD) of the signal and idler photons have to be separable (i.e. uncorrelated in the frequency domain). 

In reality, spectral purity has trade-offs with photon-number purity (ratio between single-photon state and multi-photon state), heralding efficiency and the brightness (amount of photons/s). Therefore, a high-quality HSPS which is a key for scalable quantum computing and quantum information processing, can be characterised by a set of metrics: spectral purity (P); photon-number purity; heralding efficiency; brightness and coincidence to accidental ratio. A perfect entangling operation among physical qubits requires these metrics to be high. Also, once we achieve a near ideal HSPS, we can use various multiplexing schemes to make a near ideal single-photon source.

All of these metrics can be captured by a single experiment of photon indistinguishability measurement (PIM). In a PIM with two HSPSs, we detect the idler photons from each HSPS to herald the signal photons, which then interferes in a HOM (Hong-Ou-Mandel) interferometer~\cite{HOM} or an MZI~\cite{Rarity1990}.  
The raw visibility of the interference indicates the indistinguishability of the heralded single-photons in the presence of the photon-number impurity and spectral impurity. The raw brightness of the four-fold coincidence indicates the heralding efficiency, loss of the device and the detection efficiency. Thus, visibility of the PIM and the total integration time to perform this measurement efficiently captures the metrics of the HSPSs and their possibility of scalability. 

\begin{figure*}[ht!] 
\centering
\includegraphics[width=1.0\textwidth]{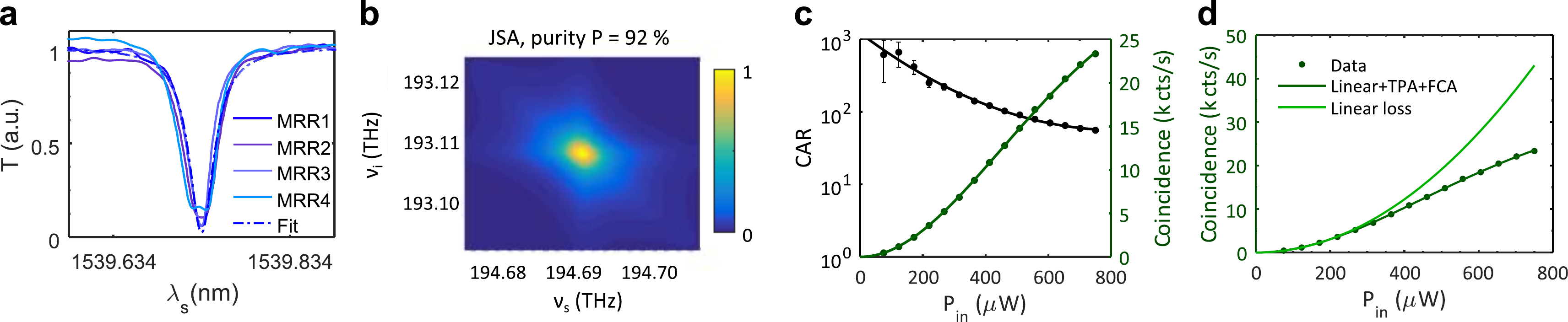}
\caption{Characterisations of the MRR source array: 
\textbf{a,} Near identical resonance spectra of the four MRRs near the signal-photon wavelength. 
\textbf{b,} Simulated spectral purity of the MRR sources.  It shows a theoretical limit of 92\% purity can be achieved for our design. \textbf{c,} Measured raw brightness and CAR as a function of input power to the 1st MRR. \textbf{d,} The reduction of brightness due to the non-linear losses in silicon.}
\label{fig:FigSMRRSources} 
\end{figure*}

\subsection{Resonant Enhancement of MRRs}
In SOI photonics, it has been shown that a MRR's resonant field enhancement allows high intrinsic brightness of photon-pair generation while at the same time keeping the photon-number purity close to unity \cite{Ma2017a}. The upper bound of the spectral purity is measured about 92 \% for a basic resonator design \cite{Helt2010, Grassani2016a} and newly proposed methods promise spectral purity near unity \cite{Vernon:17, Christensen2018}. Also, a theoretical investigation suggested that the heralding efficiency of a basic design can be boosted up to 80\% \cite{Vernon2016}. The values above are all in separately performed experiments, and the only PIM has shown 72\% raw visibility between two independent MRRs and assumes that the degradation from 92\% spectral purity is solely due to the multi-pair emission \cite{Faruque2018}.

We performed measurements on our MRRs as an array of HSPSs (Fig. \ref{fig:FigSMRRSources}) and tabulated the attributes in Tab. \ref{tab:MRR_attributes} to compare PIM. Figure \ref{fig:FigSMRRSources}a shows the resonance spectra of   all four MRRs near the signal-photon wavelength. The spectra are nearly identical suggesting a good degree of overlap. Also, they are fitted individually to the following equation to estimate the linewidths (Full Width Half Maximum---FWHM), quality factors (Q) and heralding efficiencies ($\eta_H$):
\begin{align}
    E_{\text{out}} = \frac{\tau - \alpha e^{i\theta}}{1-\alpha\tau^*e^{i\theta}}E_{\text{in}}
\end{align}
where, $E_{\text{in}}$ and $E_{\text{out}}$ are the input and output of the MRR respectively, $\kappa$ is the cross-coupling coefficient and $\tau$ is the self-coupling coefficient, $\theta$ is the round-trip phase and $\alpha$ is the round-trip transmission. If the resonance wavelength is measured as $\lambda_{\text{res}}$ and the FWHM as $\Delta\lambda_{\text{res}}$, then the quality factor is estimated using,
\begin{equation}
    Q = \frac{\lambda_{\text{res}}}{\Delta\lambda_{\text{res}}}
\end{equation}\\
Similarly, the heralding efficiency just after the MRRs are estimated by \cite{Vernon2016},
\begin{equation}
    (\eta_H)_{\text{corr}} = \frac{\tau}{\tau+\alpha}
\end{equation}

In our experiment, the corrected $(\eta_H)_{\text{corr}}$ just after the MRRs is measured to be $\sim$50\%. This value is intuitively understandable by realising that for our near critically coupled MRRs ($\tau \approx \alpha$), there is a 50\% chance that the signal photon of a photon-pair can escape from the MRR and detected.

Spectral purity is quantified by simulating the JSD using the above estimated parameters and the pump pulse configuration. The expression of JSD is found from the quantum mechanical description of the SFWM. Figure \ref{fig:FigSMRRSources}b shows the simulated JSD for one MRR source. Quantum mechanically, the interactions among the pump beam and the signal-idler photon-pairs can be described by an effective Hamiltonian $\hat{H}_{\text{int}}$ \cite{Christ2011},
\begin{equation}
\hat{H}_{\text{int}} = N \int \int d\omega_s d\omega_i f(\omega_s,\omega_i)~ \hat{a}^{\dagger}(\omega_s)~ \hat{a}^{\dagger}(\omega_i) + \text{H.c.} \label{eq:H_int}
\end{equation}
where the normalisation constant $N$ is related to the strength of the interaction, the bi-photon function $f(\omega_s,\omega_i) $ contains energy and momentum conservation of the interaction, $\text{H.c.}$ represents Hermitian conjugate, and $\hat{a}^{\dagger}(\omega_s),\ \hat{a}^{\dagger}(\omega_i)$ represent creation operators of signal and idler photons which act on the vacuum and generate signal and idler photons. 

\begin{equation}
	f(\omega_s,\omega_i) = \sum_{k = 1}^\infty \lambda_k h_k(\omega_s)\times g_k(\omega_i) \label{eq:Schmidt_decomposition}
\end{equation}
where $\lambda_k$ are normalised Schmidt coefficients. Hence, the purity ($P$) is quantified by the following two equations,
\begin{align}
	& P = tr(\hat{\rho}_s^2) = \sum \lambda_k^4 \label{eq:P_rho}\\
	& K_{\text{Schmidt}} = \frac{1}{P} = \frac{1}{\sum \lambda_k^4}
\end{align}
where, $K_{\text{Schmidt}}$ is called the Schmidt number. If there is only one element in the above expression ($k=1$ only), then it corresponds to a separable signal-idler spectra in the bi-photon function and $tr(\hat{\rho}_s^2) = tr(\hat{\rho}_s)$ will be true. In general, this statement is also used to specify that a quantum state is pure. Our JSD simulations gives us spectral purities (see Fig. \ref{fig:FigSMRRSources}) $P \approx$ 92\% of the basic resonator deign~\cite{Helt2010}.

We have also measured the intrinsic photon-pair generation efficiency $\gamma_{\text{eff}}$ in $cts/s/mW^2$ inside the MRR, collection efficiencies in the signal-idler channels $\eta_s$ \& $\eta_i$ and raw heralding efficiency $(\eta_H)_{\text{raw}}$. The signal-idler singles counts ($\text{C}(s)$, $\text{C}(i)$) and the coincidence counts $\text{CC}(s,i)$ generated from SFWM can be expressed as functions of the average pump power ($P$)~\cite{SiReview},
\begin{align}
&\text{C}(s) = (\eta_s\gamma_{\text{eff}}) P^2 + \beta_s P + \text{DC}_s\\
&\text{C}(i) = (\eta_i\gamma_{\text{eff}}) P^2 + \beta_i P + \text{DC}_i\\
&\text{CC}(s,i) = (\eta_i\eta_s\gamma_{\text{eff}}) P^2 + \text{ACC}\label{eq:cc}
\end{align}
Here, $\text{DC}$ represents dark counts and $\text{ACC}$ represents accidentals. The experimental data is first fitted with quadratic equations:
\begin{align}
	&\text{C}_k = a_k P^2 + b_k P + c_k\\
	&\text{CC} = a_{si} P^2 + \text{ACC}(P)
\end{align} 
where, $k=\{s,i\}$ represents signal or idler. The coefficients $a_k$ of the quadratic term is related to the detection efficiencies ($\eta$) and generation efficiency $\gamma_{\text{eff}}$ and can be extracted by using:
\begin{align}
	\gamma_{\text{eff}} = \frac{a_s a_i}{a_{si}}; \qquad \eta_s = \frac{a_{si}}{a_i}; \qquad \eta_i = \frac{a_{si}}{a_s}
\end{align}

The Coincidence to Accidental Ratio (CAR) is estimated as,
\begin{equation}
	\text{CAR} = \frac{\text{CC}}{\text{ACC}}
\end{equation}
\\
The raw heralding efficiency or the Klyshko efficiency is estimated by,
\begin{align}
    (\eta_H)_{\text{raw}} = \frac{\text{CC}(s,i)}{\text{C}(i)}
\end{align}

In practice, the above model is sufficient provided that we have photon number resolving (PNR) detectors and can always herald only single-photon states, or if the power is low enough that the multi-pair emission is negligible and the non-linear losses in the system are negligible. This model of brightness estimation needs modification, particularly, accounting for multi-pair emissions for measurements usually done with non-PNR detectors. A more accurate rate equation can be found in \cite{SiReview}, 
\begin{equation}
	\text{CC} = \frac{x\eta_i\eta_s (x^2(1-\eta_i)(1-\eta_s) - 1)}{(1 - x(1-\eta_i))(1 - x(1-\eta_s))(x(1-\eta_i)(1-\eta_s)-1)} \label{eq:cc_multi}
\end{equation}
\\
Here, the $\eta_i$ and $\eta_s$ has been modified to include the non-linear losses such as two-photon absorption (TPA) and free carrier absorption (FCA). Transmission efficiency after TPA and FCA are defined by the following equations:
\begin{align}
    & \eta_{TPA} = \frac{1}{1 + \beta_{TPA}L~f_E^2(P/R/\tau))^2/A_{\text{eff}}} \\
    & \eta_{FCA} = \frac{1}{\sqrt{1 + \sigma_C N_{C0}L~f_E^4 (P/R/\tau)^2}}
\end{align}
Where, $R$ is the repetition rate of the laser and $\tau$ is the pulse width, $L$ is the geometrical length of the MRR, $f_E$ is the field enhancement of the MRR, $\beta_{TPA}$ = $8\times10^{-12}$ m/W is the TPA coefficient, $A_{\text{eff}}$ = $500\times220\times10^{-18}$ $m^2$ is the effective occupied area of the guided mode, $\sigma_C$ = $1.45\times10^{-21}$ $1/m^2$ is the FCA coefficient, $N_{CO}$ is the free carrier density estimated using,
\begin{equation}
    N_{CO} = \frac{\beta_{TPA}\tau}{2h\nu A_{\text{eff}}^2}    
\end{equation}
where, $h$ is the Planck constant and $\nu$ = $194$ THz is the frequency of the photons. The total transmission in signal (similarly, idler) channel is then defined as,
\begin{equation}
    \eta_s = \eta_{s0} \times \eta_{TPA} \times \eta_{FCA}
\end{equation}
where $\eta_{s0}$ is the collection efficiency due to the linear losses. We have a few observations in fitting the brightness data using the above models. The values of $\gamma_{\text{eff}}$, $\eta_s$, $\eta_i$, $(\eta_H)_{\text{raw}}$ in Tab. \ref{tab:MRR_attributes} are estimated using the singles and coincidence counts at very low pump power such that the non-linear losses are negligible. Afterwards, the full non-linear model was used to fit the brightness data in Fig. \ref{fig:FigSMRRSources}d. We have found that the photon-pair generation rate inside the MRRs are very high ($\sim50~Mcts/s/mW^2$) but due to the channel losses only a handful of them are collected. Nevertheless, the raw coincidence counts are $\sim$ 20 kcts/s with a CAR above 50 for 800 $\mu$ W average pump power for 15 ps pulses and 500 MHz repetition rate. The measured data are reported in Fig.~\ref{fig:FigSMRRSources}c. 
Although with continuous wave laser a CAR above 600 is reported, the raw brightness was about 5 kcts/s at 60 $\mu$ W average power~\cite{Ma2017a}. 
Therefore, this is the first demonstration of an array of high raw brightness HSPSs with low CAR in silicon photonics which is also comparable with some spontaneous parametric down-conversion (SPDC) HSPSs~\cite{Weston2016}. Figure \ref{fig:FigSMRRSources}d also shows that the raw CC are almost halved due to the non-linear ls which can be mitigated in future designs.
{\renewcommand{\arraystretch}{1.8}
\begin{table}[htbp]
\centering
\begin{tabular}{P{3.3cm}|Q{1.5cm}|Q{1.5cm}|Q{1.5cm}|Q{1.5cm}}
\hline  \hline  
\centering Attributes & MRR1 & MRR2 & MRR3 & MRR4 \\
\hline
Q-factor  & 4.181$\times10^4$ & 3.531$\times10^4$ & 4.907$\times10^4$ & 3.151$\times10^4$ \\
\hline
FWHM (pm)  & 36.824 & 43.601 & 31.380 & 48.860 \\
\hline
Self-Coupling, $\tau$  & 0.9801 & 0.9767 & 0.9815 & 0.9740 \\
\hline
Round-trip T, $\alpha$  & 0.9854 & 0.9837 & 0.9875 & 0.9829 \\
\hline
$\gamma_{\text{eff}}$ ($M cts/s/mW^2$)  & $57.38$ & $48.94$ & $53.90$ & $44.14$ \\
\hline
signal collectn $\eta_s$ (\%)  & 3.232 & 2.507 & 2.603 & 2.369 \\
\hline
idler collectn $\eta_i$ (\%) & 4.127 & 3.625 & 2.589 & 2.529 \\
\hline
Heralding $\eta_H$ raw (\%) & 2.439 & 1.859 & 1.782 & 1.608 \\
\hline
Heralding $\eta_H$ corr. (\%) & 49.87 & 49.82 & 49.85 & 49.77 \\
\hline
\end{tabular}
\caption{\label{tab:MRR_attributes} Attributes of the MRR array.}
\end{table}}

The photon-number purity is measured using the Hanbury-Brown-Twis (HBT) optical interferometer~\cite{HBT1956} through heralded second order correlation ($g^{(2)}_H$). The outcome of $g^{(2)}_H$ is usually presented by a ratio of single-pair emission to multi-pair emission. Effectively, the ratio of two-fold events and three-fold events is plotted as a function of the input pump power (squeezing strength $x$):
\begin{equation}
	g^{(2)}_H(0) = \frac{D_{123}D_1}{D_{12}D_{13}} 
\end{equation}
where, $D_{123}$ is the three-fold coincidence event on detectors 1, 2 and 3, $D_{ij}$ is the two-fold events on detectors $i$ and $j$, and $D_1$ is the rate of the heralding idler photons. The experimental data can be fitted with the following equation which expresses that the multi-pair emission rate is a function of higher order terms of input pump power:
\begin{equation}
	g^{(2)}_H(0) = \frac{\Sigma_{k\geq2}~ a_k P^k}{1 + \Sigma_{k\geq2}~ a_k P^k}
\end{equation}
where $a_k$ are fitting parameters and $P$ is the optical power. This expression is similar to the expression used in \cite{Ma2017a}.

The mean photon number per pulse is expressed as,
\begin{equation}
    \bar{n} = \frac{\gamma_{\text{eff}} \times P^2}{R}
\end{equation}
If the input power to the chip is $P_{in}$ and the estimated loss from the input of the chip till the MRR is $-1.25$ dB, then $P = P_{\text{in}}\times10^{-1.25/10}$. For $P$ = 800 $\mu$W average pump power, $\bar{n}$ = 0.065.




\begin{figure*}[ht!] 
\centering
\includegraphics[width=1.0\textwidth]{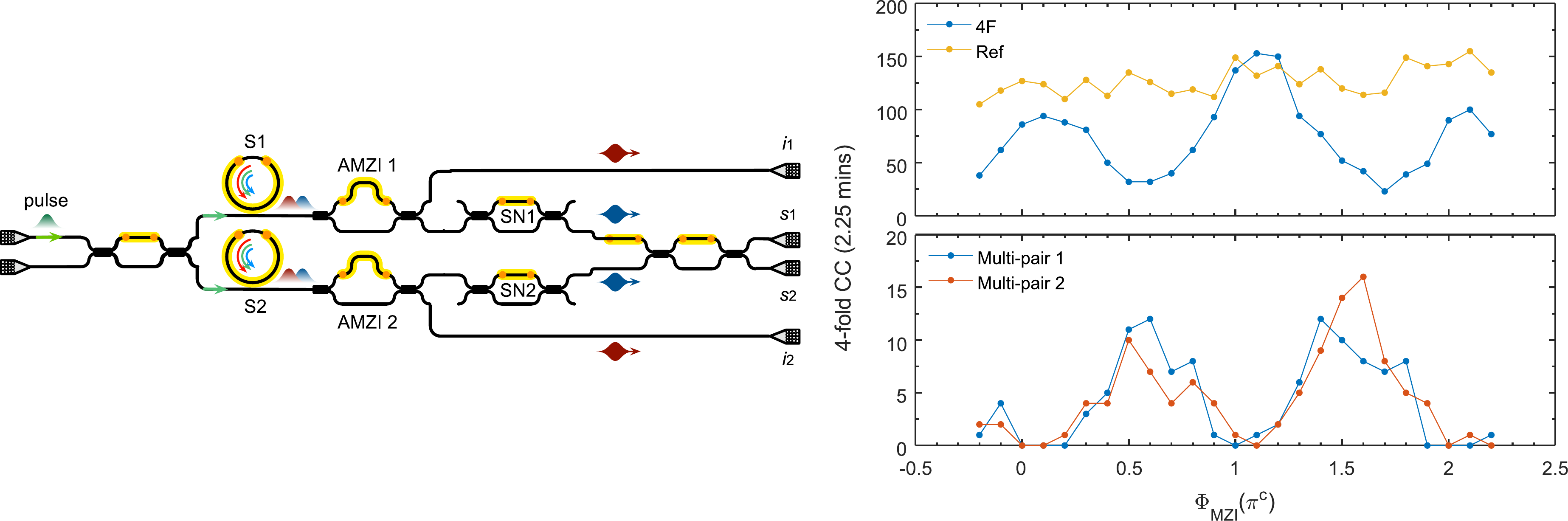}
\caption{PIM measurement. On the left is the equivalent circuit of the re-configured chip for the PIM and the multi-pair emission measurements from each source. On the right are the raw data showing the PIM, the steady coupling (indicated by Ref) and the multi-pair emission fringes. The multi-pair emission fringes are anti-phase with the original PIM. Here, the solid lines only connect the data points to guide the shapes of the data set.}
\label{fig:FigS_PIM} 
\end{figure*}    
\subsection{Photon Indistinguishability Measurement (PIM)}
The indistinguishability of two MRR sources are measured by interfering heralded single-photons generated by the two MRRs. The photonic circuit in Fig. 1 can be reconfigured to the circuit shown in Fig.~\ref{fig:FigS_PIM} for PIM between any two MRRs. For, example, in this figure, the signal-idler photons produced in the MRR 1 \& 2 are spectrally demultiplexed using the AMZI 1 \& 2. For the rest of the photonic circuit, idler photons 1 \& 2 pass through an effective straight waveguide (qubit operations are set to identity) 
and are coupled off-chip to herald the presence of the signal photons 1 \& 2. The signal photons can be routed through the device to interfere in one of two projective MZIs in $\hat{P}$. The resulting interference pattern is recorded as a four-fold coincidence fringe ($(CC)_{4F}$) as a function of the MZI phase. For the signal photons from any of these MRRs, the intermediate MZIs are used as switches (SN1 \& SN2) to block one of the input paths of the final MZI. In such cases, any measured four-fold coincidences arise due to the multi-pair emission from one of the sources. Therefore, the multi-pair emission for each of these PIM can be measured (see the data in  Fig.~\ref{fig:FigS_PIM}, and the four-fold fringe can be corrected as shown in Fig.~\ref{fig:SourcesResults}d, f. 
Systematic fluctuations in the interference fringe due to the device stability and coupling are compensated by regularly monitoring four-fold reference counts for a fixed MZI phase. 
The PIM is expressed by the visibility of the fringe as defined by the following equation~\cite{Rarity1990}, 
\begin{equation}
    V_{\text{MZI}} = \frac{(\text{CC}_{4F})_{\text{max}} - (\text{CC}_{4F})_{\text{min}}}{(\text{CC}_{4F})_{\text{max}} + (\text{CC}_{4F})_{\text{min}}}
\end{equation}

The fringe in Fig.~\ref{fig:SourcesResults}d is distorted due to the imperfect multi-mode interference couplers which have been used to form an MZI. If the MMIs are not perfectly balanced (50:50), only 100\% reflection can be achieved but not 100\% transmission through the MZI. Such Imperfect MMI does not reduce the raw MZI visibility though distorts the fringe ~\cite{Faruque2018}.

Even when the MRRs spectra overlaps perfectly, the simulation of the visibility of the interference in Fig.~\ref{fig:SourcesResults}d depends on two factors: spectral impurity and photon-number purity. The spectral purity determines the y-intercept or the maximum achievable visibility, and the photon-number impurity degrades the visibility from that value depending on the input pump power \cite{Faruque2018}. For our MRR sources, the photon-number impurity drastically reduces the visibility while the spectral purity remains high. Also, losses do not degrade the raw visibility but reduces the brightness, thus reducing the total time to record a PIM.




\subsection{Effect of Spectral Separability On Interference }
The performance of our device during each of the experiments is fundamentally limited by the ability to coherently control and interfere single photons created within one of the four MRR sources. In this section we discuss some of the major challenges involved in scaling the number of simultaneous photons produced on-device, and show that a limiting factor in integrated multi-photon quantum information processing is in the spectral correlations of single photons. \\

Single photons are created in our device due to the inherent $\chithree$ nonlinear process in silicon. The SFWM process converts pairs of photons from our pump into degenerate energy-conserved signal and idler photons, having the well-known interaction Hamiltonian 
\begin{equation}
\hat{\mathcal{H}}_{\text{Int}}
=
N\int_{\lambda_s,\lambda_i} d \lambda_s d\lambda_i  f(\lambda_s , \lambda_i) \hat{\lambda_s}^\dagger \hat{\lambda_i}^\dagger + \text{h.c,}
\end{equation}
where $f(\lambda_s , \lambda_i) $ describes the frequency correlations of the produced signal and idler photons (conserving energy and momentum), and $\hat{\lambda_s}^\dagger$, $ \hat{\lambda_i}^\dagger$ are the creation operators for the signal and idler photons at the wavelengths $\lambda_s $, $\lambda_i$. 

\subsubsection{Two-Photon Entanglement}
Generating Bell pairs in our device relies on the ability to interfere two photons generating in one of two MRR sources with high visibility. Without this quantum interference, the photons are found in a maximally mixed state as the coherence terms of the density matrix vanish. 
When attempting to generate the Bell state $\ket{\Phi}^+$, two sources are pumped in the low photon number regime. In the cases where only two photons are detected, the measured photons must arise from the same source with high probability. The full state can be written as a superposition of squeezed states in different spacial modes
\begin{equation}
\ket{\Psi}
= N
\left(
\int_{\lambda_s,\lambda_i} d \lambda_s d\lambda_i 
f(\lambda_s, \lambda_i) 
\hat{\lambda}_{s,\uparrow, 0}^\dagger \hat{\lambda}_{i,\downarrow, 0} ^\dagger 
+ 
\int_{\lambda'_s,\lambda'_i} d \lambda'_s d\lambda'_i 
f'(\lambda'_s, \lambda'_i) 
\hat{\lambda'}_{s,\uparrow, 1}^\dagger \hat{\lambda'}_{i,\downarrow, 1} ^\dagger 
\right)
\ket{\text{vac}},
\end{equation}
\\
where the subscripts $\{\uparrow, \downarrow\}$ reference photons in the upper/lower qubit and numbered subscripts refer to the particular waveguide mode.
Clearly, after the single photons pass through the integrated AMZI, they give rise to classical correlations in the $00$ and $11$ mode, corresponding to $\ket{00}\bra{00}$ and $\ket{11}\bra{11}$ terms in the reconstructed density matrix. Note that, here it is explicitly assumed that one may balance these contributions by altering the overall relative pumping strength provided to each source. In order to show that the generated state has the correct quantum properties, however, it is necessary to show the following
\begin{equation}
\begin{split}
\ket{\Phi^+}
&= 
(\ket{00}+\ket{11})/\sqrt{2}
\\
&=
(\ket{++}+\ket{--}) /\sqrt{2}
\\
&=
\hat{\sigma}_x \otimes \hat{\sigma}_x (\ket{00}+\ket{11})/\sqrt{2} .
\end{split}
\end{equation}
\\
The condition for our state to show the correct interference relies primarily on the ability to tune the spectrum of the single photons such that $f(\lambda_s, \lambda_i) =f'(\lambda_s, \lambda_i) $, requiring only that we are able to generate identical single photon sources. Once the photons are projected into the local bases $\hat{\sigma}_{x,\uparrow}\hat{\sigma}_{x,\downarrow}$, 
\begin{equation}
\begin{split}
\hat{\sigma}_{x,\uparrow}\hat{\sigma}_{x,\downarrow} 
\ket{\Psi}
= \frac{1}{2N}
\bigg(
\int_{\lambda_s,\lambda_i} d \lambda_s d\lambda_i 
&f(\lambda_s, \lambda_i) 
(\hat{\lambda}_{s,\uparrow, 0}^\dagger
+
\hat{\lambda}_{s,\uparrow, 1}^\dagger
)
(\hat{\lambda}_{i,\downarrow, 0}^\dagger 
+ 
\hat{\lambda}_{i,\downarrow, 1}^\dagger 
)
\\
&+
\int_{\lambda'_s,\lambda'_i} d \lambda'_s d \lambda'_i 
f(\lambda'_s, \lambda'_i) 
(\hat{\lambda'}_{s,\uparrow, 0}^\dagger
-
\hat{\lambda'}_{s,\uparrow, 1}^\dagger
)
(\hat{\lambda'}_{i,\downarrow, 0} ^\dagger 
-
\hat{\lambda'}_{i,\downarrow, 1} ^\dagger 
)
\bigg)\ket{\text{vac}}.
\end{split}
\end{equation}
The amount of quantum interference can now be seen by comparing the integrand for various values of $\lambda_s,\lambda_i$. In instances where $\lambda=\lambda'$, the state may be factorised in a clear way, giving a contribution to the total state 
$f(\lambda_s,\lambda_i)
(\hat{\lambda}_{i,\uparrow, 0}^\dagger 
\hat{\lambda}_{i,\downarrow, 0}^\dagger 
+
\hat{\lambda}_{i,\uparrow, 1}^\dagger 
\hat{\lambda}_{i,\downarrow, 1}^\dagger 
)$. In addition, due to the fact that $f=f'$ each integral in the general expression is symmetric around the case $\lambda = \lambda'$. For example, for every contribution to the overall state $$d \lambda_i d \lambda_s
f(\lambda_s, \lambda_i) 
(\hat{\lambda}_{s,\uparrow, 0}^\dagger
+
\hat{\lambda}_{s,\uparrow, 1}^\dagger
)
(\hat{\lambda}_{i,\downarrow, 0}^\dagger 
+ 
\hat{\lambda}_{i,\downarrow, 1}^\dagger 
)
+
d \lambda'_i d \lambda'_s
f(\lambda'_s, \lambda'_i) 
(\hat{\lambda'}_{s,\uparrow, 0}^\dagger
-
\hat{\lambda'}_{s,\uparrow, 1}^\dagger
)
(\hat{\lambda'}_{i,\downarrow, 0} ^\dagger 
-
\hat{\lambda'}_{i,\downarrow, 1} ^\dagger 
),$$
there is a symmetric contribution 
$$d \lambda'_i d \lambda'_s
f(\lambda'_s, \lambda'_i) 
(\hat{\lambda'}_{s,\uparrow, 0}^\dagger
+
\hat{\lambda'}_{s,\uparrow, 1}^\dagger
)
(\hat{\lambda'}_{i,\downarrow, 0}^\dagger 
+ 
\hat{\lambda'}_{i,\downarrow, 1}^\dagger 
)
+
d \lambda_i d \lambda_s
f(\lambda_s, \lambda_i) 
(\hat{\lambda}_{s,\uparrow, 0}^\dagger
-
\hat{\lambda}_{s,\uparrow, 1}^\dagger
)
(\hat{\lambda}_{i,\downarrow, 0} ^\dagger 
-
\hat{\lambda}_{i,\downarrow, 1} ^\dagger ),
$$
which provides the quantum interference. Summing these terms together one arrives at 
$$2 d \lambda'_i d \lambda'_s
f(\lambda'_s, \lambda'_i) 
(\hat{\lambda'}_{s,\uparrow, 0}^\dagger
\hat{\lambda'}_{s,\downarrow, 0}^\dagger
+
\hat{\lambda'}_{s,\uparrow, 1}^\dagger
\hat{\lambda'}_{s,\downarrow, 1}^\dagger
)
+
2 d \lambda_i d \lambda_s
f(\lambda_s, \lambda_i) 
(\hat{\lambda}_{s,\uparrow, 0}^\dagger
\hat{\lambda}_{s,\downarrow, 0}^\dagger
+
\hat{\lambda}_{s,\uparrow, 1}^\dagger
\hat{\lambda}_{s,\downarrow, 1}^\dagger
),
$$exploiting this symmetry for all values of $\lambda_i,\lambda_s$ the full quantum state may be written as
\begin{equation}
\begin{split}
\frac{1}{\sqrt{2}}
\int_{\lambda_s,\lambda_i} d \lambda_s d\lambda_i 
f(\lambda_s, \lambda_i) 
(\hat{\lambda}_{s,\uparrow, 0}^\dagger
\hat{\lambda}_{i,\downarrow, 0}^\dagger
+
\hat{\lambda}_{s,\uparrow, 1}^\dagger
\hat{\lambda}_{i,\downarrow, 1}^\dagger)
\ket{\text{vac}}.
\end{split}
\end{equation}
\\
The key point here is that no assumption is placed on the correlation functions $f(\lambda_s,\lambda_i)$ except for the fact we can generate an identical joint spectral function from each source. This is analogous in our experiment to the overlapping of microring resonances and explains the high fidelity two-photon entangled states generated in our device. The main source of noise in this experiment is due to the multiphoton terms which can be controlled by reducing the pump power. 

\subsubsection{Heralded Quantum Interference}
In the case where multiple photons are simultaneously produced from different single photon sources, perfect quantum interference possible only in the case where $f$ is spectrally pure, i.e. $f(\lambda_s,\lambda_i) = A(\lambda_s) B(\lambda_i)$. During the heralded interference, the initial state is
\begin{equation}
\ket{\Psi}
= N
\int_{\lambda_s,\lambda_i,\lambda'_s,\lambda'_i}
d \lambda_s d\lambda_i d \lambda'_s d\lambda'_i 
f(\lambda_s, \lambda_i)  
g(\lambda'_s, \lambda'_i) 
\hat{\lambda}_{s,\uparrow, 0}^\dagger 
\hat{\lambda'}_{s,\uparrow, 1}^\dagger 
\hat{\lambda}_{i,\downarrow, 0} ^\dagger 
\hat{\lambda'}_{i,\downarrow, 1}^\dagger 
\ket{\text{vac}}.
\end{equation}If each of the biphoton functions are separable, then the state can be written as
\begin{equation}
\ket{\Psi}
= N
\int_{\lambda_i}
A(\lambda_i) 
\hat{\lambda}_{i,\downarrow, 0} ^\dagger 
d\lambda_i
\int_{\lambda'_i}
B(\lambda'_i)
\hat{\lambda'}_i^\dagger
d\lambda'_i 
\int_{\lambda_s,\lambda'_s} 
C(\lambda_s)D(\lambda'_s)
\hat{\lambda}_{s,\uparrow, 0}^\dagger 
\hat{\lambda'}_{s,\uparrow, 1}^\dagger 
d \lambda_s d \lambda'_s
\ket{\text{vac}}.
\end{equation}
Hence in a similar way as above, when projecting the signal photons into $\hat{\sigma}_{x,\uparrow}^\dagger  \hat{\sigma}_{x,\downarrow}^\dagger$ we get complete quantum interference (no photons are simultaneously found ) when $C(\lambda)=D(\lambda)$, giving the final state
\begin{equation}
\hat{\sigma}_x \otimes \hat{\sigma}_x
\ket{\Psi}
= \frac{N}{2}
\int_{\lambda_i}
A(\lambda_i) 
\hat{\lambda}_{i,\downarrow, 0} ^\dagger 
d\lambda_i
\int_{\lambda'_i}
B(\lambda'_i)
\hat{\lambda'}_i^\dagger
d\lambda'_i 
\int_{\lambda_s,\lambda'_s \geq \lambda_s} 
C(\lambda_s)C(\lambda'_s) 
\big(
\hat{\lambda}_{s,\uparrow, 0}^\dagger 
\hat{\lambda'}_{s,\uparrow, 0}^\dagger 
+
\hat{\lambda'}_{s,\uparrow, 0}^\dagger 
\hat{\lambda}_{s,\uparrow, 0}^\dagger 
-
\hat{\lambda}_{s,\uparrow, 1}^\dagger 
\hat{\lambda'}_{s,\uparrow, 1}^\dagger 
-
\hat{\lambda'}_{s,\uparrow, 1}^\dagger 
\hat{\lambda}_{s,\uparrow, 1}^\dagger 
\big)
d \lambda_s d \lambda'_s 
\ket{\text{vac}}.
\end{equation}
\\
Finally consider the case where the joint spectra from each source is identical, however the signal and idler photons from each source are completely correlated (i.e. single source produces $ \approx \lambda_s \lambda_i + \lambda'_s \lambda'_i + \lambda''_s \lambda''_i +\ldots$). In this case, multiphoton terms (from different sources) where $\lambda_s=\lambda'_s$ still experience quantum interference, however the orthogonal terms (for D dimensions) in the total state dominate with a ratio given by $\lim_{D \to \infty} (D^2-D)/D$. In addition, the perfectly correlated signal and idler photons break the spectral symmetry seen in the previous two cases. For example consider the two orthogonal terms 
\begin{equation}
\frac{1}{2}
d \lambda_s d\lambda_i d \lambda'_s d\lambda'_i 
f(\lambda_s, \lambda_i)  
f(\lambda'_s, \lambda'_i) 
(\hat{\lambda}_{s,\uparrow, 0}^\dagger
+
\hat{\lambda}_{s,\uparrow, 1}^\dagger)
(\hat{\lambda'}_{s,\uparrow, 0}^\dagger
-
\hat{\lambda'}_{s,\uparrow, 1}^\dagger)
\hat{\lambda}_{i,\downarrow, 0} ^\dagger 
\hat{\lambda'}_{i,\downarrow, 1}^\dagger ,
\end{equation}and
\begin{equation}
\frac{1}{2}
d \lambda'_s d\lambda'_i d \lambda_s d\lambda_i 
f(\lambda'_s, \lambda'_i)  
f(\lambda_s, \lambda_i) 
(\hat{\lambda'}_{s,\uparrow, 0}^\dagger
+
\hat{\lambda'}_{s,\uparrow, 1}^\dagger)
(\hat{\lambda}_{s,\uparrow, 0}^\dagger
-
\hat{\lambda}_{s,\uparrow, 1}^\dagger)
\hat{\lambda'}_{i,\downarrow, 0} ^\dagger 
\hat{\lambda}_{i,\downarrow, 1}^\dagger .
\end{equation}
\\
Here the idler photons prevent factorisation of these terms and hence also prevent the quantum interference between signal photons. In addition, since these orthogonal frequency terms completely dominate the total wavefunction (unless $f$ is extremely narrow in line-width, forcing $\lambda=\lambda'$), the overall state will see no interference. This case results in a heralded MZI interference fringe with a visibility of $33\%$ (where 
$v=(cc_{\text{max}}-cc_{\text{min}})/(cc_{\text{max}}+cc_{\text{min}})$).

\subsection{Locking Micro-ring Sources}
Indistinguishable single-photons are generated in our device by precisely tuning the resonances of the MRRs as shown in Fig.~\ref{fig:SourcesResults} in main text and above subsections. 
Since each of the MRRs are precisely fabricated to have identical FSR, we may simultaneously match the overlap of signal and idler photons by scanning each of the rings at only one resonance. 
In practice, the cavity resonances are exceptionally sensitive to external temperature fluctuations. 
To minimise this effect, a thermistor is used to measure the surrounding temperature and a temperature controller stabilises the overall device at $22$ degrees Celsius in order to to minimise the relative frequency shift of overlapped resonances caused by temperature drift.\\

\begin{figure*}[ht!] 
\centering
\includegraphics[width=1.0\textwidth]{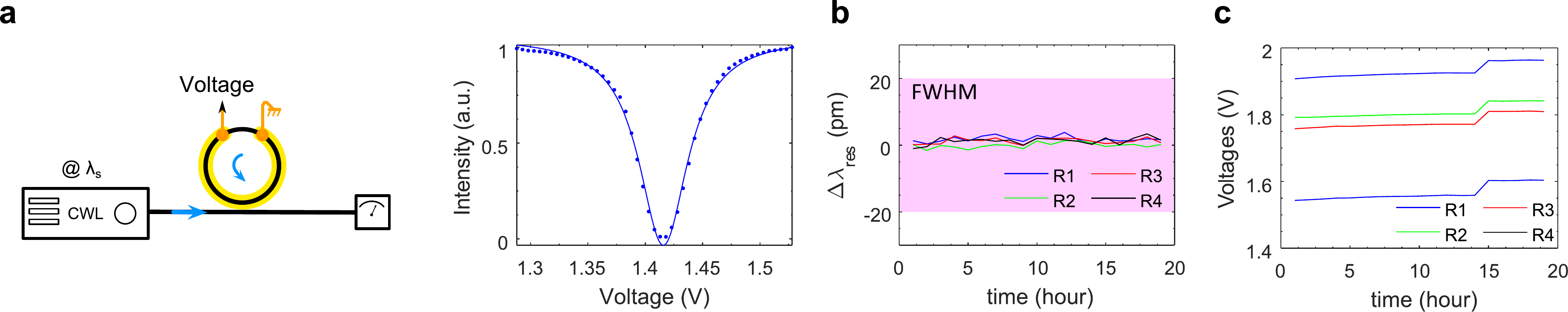}
\caption{Aligning and locking the MRR resonances. 
\textbf{a,} a weak CW light is injected into the MRR at the signal photon's resonance ($\lambda_s$: locking wavelength) and the output intensity is measured while varying the voltage on the thermo-optic phase-shifter on the MRR. The collected data is fitted with a Lorentzian line shape and the minima is the required voltage for the resonance alignment. 
\textbf{b,} The resonance positions of the all four MRRs are well-aligned over the whole time of the experiments using \textbf{c,} thermo-optic tuning by applying electrical voltage to the heaters on the MRRs. }
\label{fig:MRRLocking} 
\end{figure*}

To improve the stabilisation further we perform spectral resonance locking, which utilises a fixed frequency probe with weak power (CW laser fixed at desired frequency resonance) in order to correct for any unwanted drift. 
This is achieved by monitoring the power at separate outputs where each separate port contains light only passing through a single MRR. Figure~\ref{fig:MRRLocking}a reports the experimental setup for locking the MRRs. 
In this case the resonances can then be maintained by simultaneously scanning each of the ring heater voltages at values close to each resonance and measuring the optical power from each of the outputs, as shown in Fig.~\ref{fig:MRRLocking}b. 
The point where each of the rings are maximally overlapped corresponds to the minimisation of power in all of the separate powermeters. 
The resulting fringes are fit with a standard Lorenzian using a non-linear model fit, and the optimal voltages are then applied to correct the four resonances. 
Figure \ref{fig:MRRLocking}c shows the measured resonance position of each of the overlapped rings over a 20 hour period, showing how they remain centred with one-another within a few pm. And Figure \ref{fig:MRRLocking}d shows the corresponding voltage required to stabilise the resonance wavelength of the four MRRs. 

\section{Linear-optic  Multi-qubit Operations}
\label{sec:opeators}

\subsection{Single-qubit Operation and Projection }
The silicon device in Fig.1 enables us to prepare arbitrary single-qubit states and to perform arbitrary single-qubit projective measurements across four qubits. 
These operations and projections are broken down into the components $\hat{U}_{\text{Phase}}(\phi)$,  which controls the relative photon phase across adjacent modes, and $\hat{U}_{\text{MZI}}(\theta)$, which controls the relative probability amplitude of measuring a photon in the $\ket{0}$ or $\ket{1}$ mode, a combination of these affects gives $\hat{U} = \hat{U}_{\text{MZI}}(\theta)\hat{U}_{\text{Phase}}(\phi)$. A quantitative assessment of these operations can be given by summarising the linear optical transformations on each of the modes as follows \begin{equation}
\hat{U}_{\text{Phase}}(\theta) =  \begin{pmatrix}
    1 & 0  \\
    0 & e^{i\theta}
  \end{pmatrix}
  ,
  \hat{U}_{\text{MMI}}=  \frac{1}{\sqrt{2}} \begin{pmatrix}
    i & 1  \\
    1 & i
  \end{pmatrix} 
  ,\\
  \hat{U}_{\text{MZI}}(\theta)=  \hat{U}_{\text{MMI}} \hat{U}_{\text{Phase}}\hat{U}_{\text{MMI}} 
=
e^{i(\theta + \pi)/2} 
\begin{pmatrix}
    \sin{(\theta/2)} & \cos{(\theta/2)}  \\
    \cos{(\theta/2)} & - \sin{(\theta/2)}
  \end{pmatrix}.
 \end{equation}
 
Therefore the total unitary is written as 
\begin{equation}
\hat{U}=
e^{i(\theta + \pi)/2} 
\begin{pmatrix}
    \sin{(\theta/2)} & e^{i \phi} \cos{(\theta/2)} \\
    \cos{(\theta/2)} & -  e^{i \phi}\sin{(\theta/2)}
  \end{pmatrix}
  \text{,}
\end{equation}
  which transforms the logical basis states (up to a global phase) in the following way
  \begin{equation}
  \{ \ket{0}, \ket{1} \}
  \to
  \{ 
  \sin{(\theta/2)} \ket{0} 
  +
  e^{i \phi} \cos{(\theta/2)} \ket{1} 
  \text{, }
  \cos{(\theta/2)} \ket{0} 
  -
  e^{i \phi} \sin{(\theta/2)} \ket{1} 
  \},
  \end{equation}
  forming an arbitrary orthonormal basis set, spanning the two dimensional space. \\
  
Projective measurements should perform the adjoint transformation $\hat{U}^\dagger$, projecting arbitrary bases back into the computational basis such that photon detection events in each mode infer information about the probability distribution of measurements projected into that particular basis, requiring
\begin{equation}
\hat{U}^\dagger \{ 
  \sin{(\theta/2)} \ket{0} + e^{i \phi} \cos{(\theta/2)} \ket{1} 
  \text{, }
  \cos{(\theta/2)} \ket{0} 
  -
  e^{i \phi} \sin{(\theta/2)} \ket{1} 
  \}
  = \hat{U}^\dagger \hat{U}
\{ \ket{0}, \ket{1} \} = \{ \ket{0}, \ket{1} \} .
\end{equation}
Calculating this transformation is straightforward since \begin{equation}
\hat{U}^\dagger = (\hat{U}_{\text{MZI}}(\theta)\hat{U}_{\text{Phase}}(\phi))^\dagger
= 
\hat{U}_{\text{Phase}}^\dagger(\phi)
\hat{U}_{\text{MZI}}^\dagger(\theta) 
= e^{-i(\theta + \pi)} \hat{U}_{\text{Phase}}(-\phi)
\hat{U}_{\text{MZI}}(\theta) ,
\end{equation}
hence projecting back into the computational basis is a matter of reversing the operation order and reversing the sign of the phase applied to the $\ket{1}$ mode. Figure~\ref{fig:FigSMathforO}a,b show the linear-optic circuit diagram for the preparation arbitrary single-qubit state, and for the implementation of arbitrary single-qubit projective measurements. 

\subsection{Bosonic Bell Operation  }
Fundamentally, quantum information tasks involving photons rely on the ability to create and interfere pure identical single-photons. In section~\ref{sec:MRRsource}, we have discussed the major challenge in interfering multiple photons from separate sources. The best measure of our ability to do that is shown in the heralded two-photon interference pattern achieved using the MRR sources. However, in this section we add a layer of complexity to the previous section, and assess the ability to interfere multiple qubits consisting of identical photons in different spacial modes. The motivation for this is the following: \\

The heralded two-photon interference experiment gives us information primarily about the quality of MRR sources. The visibility of the interference pattern tells us about how spectrally identical the photons are, how low the average photon number is, as well as how low the spurious counts and background noise are. 
By correcting for multi-pair terms (or in the low photon number limit) the corrected visibility of the herald two-photon interference gives us a measure of the percentage of photons which interfere at the beam splitter, i.e. how identical they are. \\

Though we have shown the high-performance MRR sources in Fig.~\ref{fig:SourcesResults} and section~\ref{sec:MRRsource}, it is extremely important  to understand whether we can build a useful device, able to generate high-quality multiphotons and process the multiple qubits prepared in the multiphotons. This is key to any multiphoton quantum photonic implementation. 
In order to assess that question, we have to verify how the building blocks, i.e., $\hat{O}_{\text{Bell}}$ and $\hat{O}_{\text{fusion}}$, can operate multiple qubits states, which are defined in Fig.~\ref{fig:OperatorResults}  in main text. \\

\begin{figure*}[ht!] 
\centering
\includegraphics[width=0.66\textwidth]{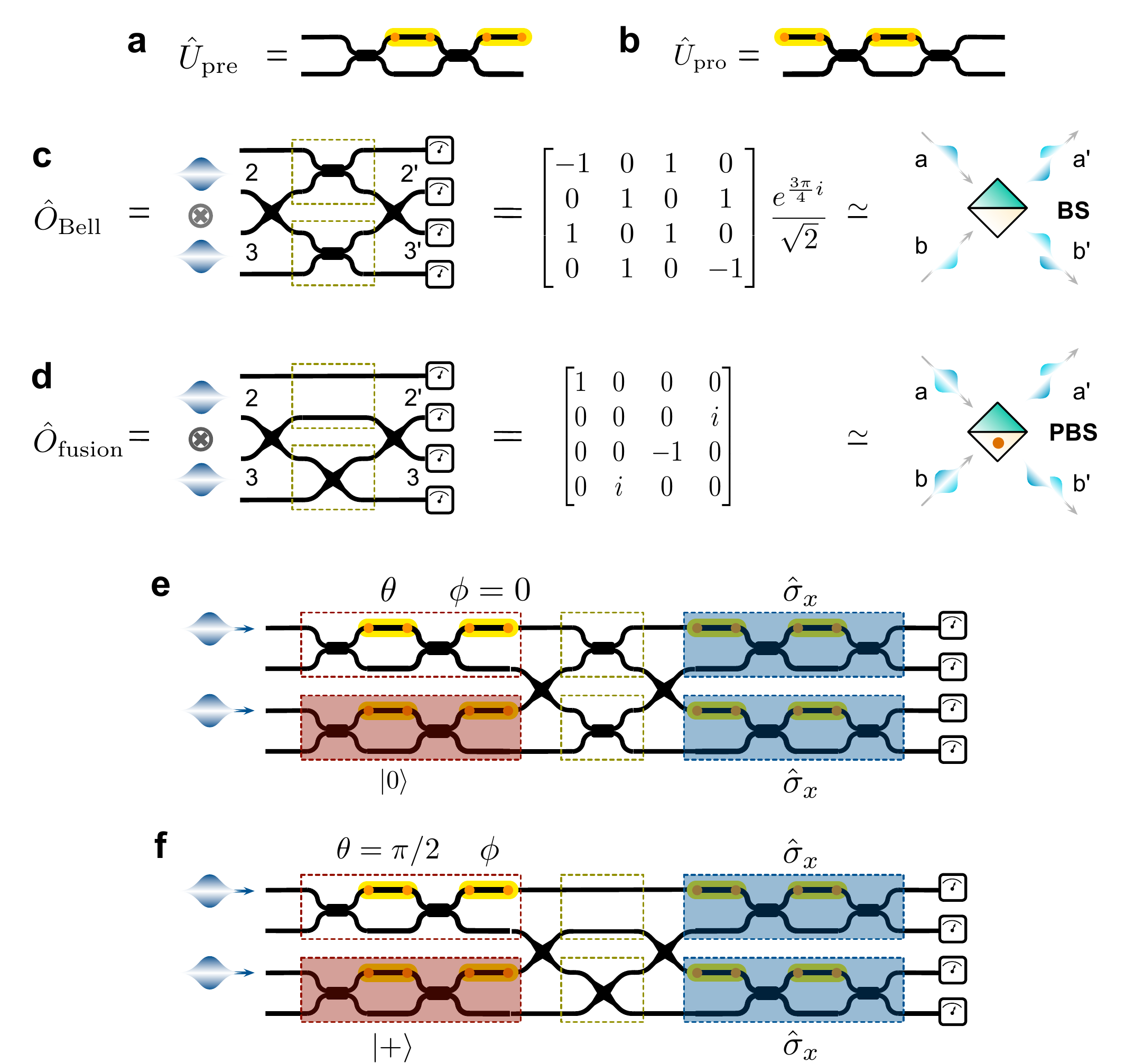} 
\caption{Linear-optic quantum components and circuits. 
\textbf{a}, a circuit diagram for the preparation arbitrary single-qubit state, and \textbf{b}, a circuit diagram for the implementation of arbitrary single-qubit projective measurements. 
\textbf{c}, a circuit diagram and matrix presentation of the $\hat{O}_\text{Bell}$ operator, and \textbf{d},  a circuit diagram and matrix presentation of the  $\hat{O}_\text{fusion}$ operator. 
BS: beam splitter for polarisation-encoded qubits; PBS: polarisation beam splitter for polarisation-encoded qubits. 
\textbf{e}, and \textbf{f},  Configurations of the circuits performing HOM interference and Bell generation at the $\hat{O}_\text{Bell}$ and $\hat{O}_\text{fusion}$. The HOM-like fringes in Fig.~\ref{fig:OperatorResults}h,i were observed by rotating the $\theta$ or $\phi$ in the $\hat{U}_{MZI}$ of 2-nd qubit and measuring the two qubits 2,3 both in the $\hat{\sigma}_x$ basis. 
}
\label{fig:FigSMathforO} 
\end{figure*}

The $\hat{O}_{\text{Bell}}$ circuit diagram is shown in Fig~\ref{fig:OperatorResults}f, and its matrix representation is shown in Fig.~\ref{fig:FigSMathforO}c. As an example, we prepared an initial state $\ket{1,0}_{2,3}$ (see the measured state density matrix $\rho_{2,3}$ in Fig.~\ref{fig:OperatorResults}e) and consider its evolution under the  $\hat{O}_{\text{Bell}}$transformation. 
The initial state can be written in terms of the creation operators for the signal photons on the vacuum as 
\begin{equation}
\Hat{S}_{\downarrow,2}^ \dagger  \Hat{S}_{\uparrow,3}^ \dagger \ket{\text{vac}}, 
\end{equation}
where the notation $\hat{S}^\dagger _{\uparrow ,i}$ ($\hat{S}^\dagger _{\downarrow, i}$)  describes a photon arising in the zero $\ket{0}$ (one $\ket{1}$) mode of the $i$-th qubit.  
The operation $\hat{O}_{\text{Bell}}$ can be decomposed into three time-ordered transformations
$\hat{O}_{\text{Ex}} \hat{O}_{\text{Int}} \hat{O}_{\text{Ex}}$, which matrix representation is given in (Fig.~\ref{fig:FigSMathforO}c), analogue to the beam splitter (BS) transformation on two polarisation-encoded qubits. In detail, $\hat{O}_{\text{Ex}}$ is an operation which exchanges photon modes and performs the mapping $\Hat{S}_{\downarrow,2}^ \dagger \leftrightarrow \Hat{S}_{\uparrow,3}^ \dagger$ in our circuit. 
$\hat{O}_{\text{Int}}$ allows photons from qubits 2 \& 3 to interact and interfere each other. The dashed box (Fig.~\ref{fig:FigSMathforO}c) refers to a Hadamard-like transformation up to a global $z$-rotation $e^{i \frac{3\pi}{4}}$, when the $\hat{U}_{\text{MZI}}(\theta)$) in the dashed box is set as $\theta=\pi/2$. 
The $\hat{O}_{\text{Int}}$ operation is described by applying the four local Hadamard-like transformations as follows
\begin{equation}
\begin{split}
\Hat{S}_{\uparrow,2}^ \dagger \rightarrow \frac{1}{\sqrt{2}}(\Hat{S}_{\uparrow,2}^ \dagger + \Hat{S}_{\downarrow,2}^ \dagger) 
\\
\Hat{S}_{\downarrow,2}^ \dagger \rightarrow \frac{1}{\sqrt{2}}(\Hat{S}_{\uparrow,2}^ \dagger - \Hat{S}_{\downarrow,2}^ \dagger) 
\\
\Hat{S}_{\uparrow,3}^ \dagger \rightarrow \frac{1}{\sqrt{2}}(\Hat{S}_{\uparrow,3}^ \dagger + \Hat{S}_{\downarrow,3}^ \dagger) 
\\
\Hat{S}_{\downarrow,3}^ \dagger \rightarrow \frac{1}{\sqrt{2}}(\Hat{S}_{\uparrow,3}^ \dagger - \Hat{S}_{\downarrow,3}^ \dagger).
\end{split}
\end{equation}
Applying these transformations $\hat{O}_{\text{Ex}} \hat{O}_{\text{Int}} \hat{O}_{\text{Ex}}$, the initial state evolves to become 
\begin{equation}
\frac{1}{2}(\Hat{S}_{\uparrow,2}^ \dagger \Hat{S}_{\downarrow,2}^ \dagger +
\Hat{S}_{\uparrow,2}^ \dagger \Hat{S}_{\downarrow,3}^ \dagger -
\Hat{S}_{\downarrow,2}^ \dagger \Hat{S}_{\uparrow,3}^ \dagger -
\Hat{S}_{\uparrow,3}^ \dagger \Hat{S}_{\downarrow,3}^ \dagger) 
\ket{\text{vac}}.
\end{equation}\\
Finally, four-fold coincidence measurements are recorded  only when each qubit has one photon, meaning outcomes where multiple signal photons are found in the same qubit are discarded. This relies on the bosonic nature of photons. 
Under these conditions the final two-qubit measurable state is the singlet entangled state 
\begin{equation}
\frac{1}{\sqrt{2}}(\Hat{S}_{\uparrow,2}^ \dagger \Hat{S}_{\downarrow,3}^ \dagger -
\Hat{S}_{\downarrow,2}^ \dagger \Hat{S}_{\uparrow,3}^ \dagger)  \ket{\text{vac}}
\rightarrow \ket{\Psi^-}.
\end{equation}
\\
In our device, the initial state is generated by simultaneously generating photons in rings 1 \& 3. By measuring four-fold coincidence counts at the output of our device, we naturally measure the eigenstate $\ket{0000}_{1,2,3,4}$ (when the  $\hat{O}$ operator is set as off, that is the configuration is set to identity). The first and last qubits are treated as ancillary qubits in order to herald the presence of the two qubit state $\ket{00}_{2,3}$. The second qubit is then rotated under an MZI via $\hat{U}_{\text{MZI,2}}(\theta)\ket{00}_{2,3}$ which evolved the state as $\ket{10}_{2,3}$ when $\theta=0$. \\

To test the performance of $\hat{O}_{\text{Bell}}$, we performed an experiment analogous to that of the qubit HOM fringe (Fig.~\ref{fig:OperatorResults}h in main text), interfering two qubits in a $\hat{O}_{\text{Bell}}$. 
Considering the bosonic nature of photons meeting at the  $\hat{O}_{\text{Bell}}$, we expect logical modes to bunch and orthogonal modes to antibunch at the 2' amd 3' ports (see Fig.~\ref{fig:FigSMathforO}c). 
 The visibility of the interference between the orthogonal and identical qubits reflects on how well we can prepare our qubits, and the fidelity of the $\hat{O}_{\text{Bell}}$ operation. 
By rotating the state $\hat{O}_{\text{Bell}}\hat{U}_{\text{MZI}}(\theta)\ket{00}_{2,3}$ around $\theta \in [0,\pi]$ --- evolving $\ket{10}_{2,3}$ when $\theta=0$,  to $\ket{00}_{2,3}$ when $\theta=\pi$ --- we see the interference fringe in Fig.~\ref{fig:OperatorResults}h with a visibility of $80 \pm 3 \%$. The qubits 2 and 3 were measured in the $\hat{\sigma}_x\hat{\sigma}_x$ basis, in which case we can observe the coherent interference fringe. Figure~\ref{fig:FigSMathforO}e shows the configuration of the circuits performing HOM interference and Bell generation at the $\hat{O}_\text{Bell}$. 
Moreover, we performed quantum state tomography to reconstruct the density matrix of $\ket{\Psi^-}$.  We thus confirm the Bell entangling operation using the circuit in Fig.~\ref{fig:OperatorResults}f. 

\subsection{Fusion Entangling Operation}
We next consider the state evolution under $\hat{O}_{\text{fusion}}$, as shown in Fig. \ref{fig:OperatorResults}g and its matrix representation is shown in Fig. ~\ref{fig:FigSMathforO}d. 
In this experiment we first configure the circuit to prepare the separable state 
\begin{equation}
\ket{0}_{1} \otimes \ket{+}_{2} \otimes (\ket{0}_{3}+e^{i\phi}\ket{1}_{3}) \otimes \ket{0}_{4} .
\end{equation} 
\\
In this scenario we prepare the idler photons, i.e., 1st and 4th, in the zero mode to act as ancillary qubits, allowing heralded four photon measurements. 
Turning off the $\hat{O}$, we can take the full state tomography of the qubits $ \rho_{2,3}$ state. Fig.~\ref{fig:OperatorResults}e shows the $\ket{++}_{2,3}$ state when $\phi=0$. 
When turned it on, we performed a fusion operation $\hat{O}_{\text{fusion}}$ on dual-rail qubits, which is mathematically analogous to polarisation qubits incident on a polarisation beamsplitter (PBS). That means $\hat{O}_{\text{fusion}}$ transmit photons in the $\ket{0}$ mode and swap photons in $\ket{1}$ mode. This is also confirmed by its unitary in Fig.~\ref{fig:FigSMathforO}d. 
Similarly, the operation $\hat{O}_{\text{fusion}}$ can be decomposed into three time-ordered transformations. In the middle part, the dashed boxes in Fig.~\ref{fig:FigSMathforO}d refer to an identity-like  transformation and $\hat{\sigma}_x$-like transformation 
up to global $z$-rotations, when the $\hat{U}_{\text{MZI}}(\theta)$) in the dashed box are set as $\theta=\pi$ in the top and $\theta=0$ in the bottom, respectively. \\

The initial states of the interacting qubits can be written as 
\begin{equation}
\frac{1}{2}(\Hat{S}_{\uparrow,2}^ \dagger + \Hat{S}_{\downarrow,2}^ \dagger) (\Hat{S}_{\uparrow,3}^ \dagger + e^{i \theta} \Hat{S}_{\downarrow,3}^ \dagger) \ket{\text{vac}},
\end{equation}which evolve under $\hat{O}_{\text{fusion}}$ as 
\begin{equation}
\frac{1}{2}(-\Hat{S}_{\uparrow,2}^ \dagger + i \Hat{S}_{\downarrow,3}^ \dagger) (\Hat{S}_{\uparrow,3}^ \dagger +i e^{i \phi} \Hat{S}_{\downarrow,2}^ \dagger) \ket{\text{vac}}.
\end{equation} 
In this state, only the terms where the physical meaning of the dual-rail qubit is preserved can lead to four-fold coincidence detection events, leading to the final measurable state 
\begin{equation}
-\frac{1}{\sqrt{2}}(\Hat{S}_{\uparrow,2}^ \dagger \Hat{S}_{\uparrow,3}^ \dagger  + e^{i \phi} \Hat{S}_{\downarrow,3}^ \dagger \Hat{S}_{\downarrow,2}^ \dagger) \ket{\text{vac}},
\end{equation} giving the maximally entangled states $\ket{\Phi^+}$ ($\ket{\Phi^-}$) when $\phi = 0$ ($\phi = \pi$), and in general $(\ket{00}+e^{i \phi}\ket{11})/\sqrt{2}$. As an example, we performed quantum state tomography to reconstruct the density matrix of $\ket{\Phi^+}$ and a state fidelity of $0.830\pm 0.032$ was obtained (see Fig.~\ref{fig:OperatorResults}k). Moreover, we also performed the qubits HOM-like interference fringe (Fig.~\ref{fig:OperatorResults}i in main text), interfering two qubits in the $\hat{O}_{\text{fusion}}$. 
The simultaneous clicks of detectors in the 2' and 3' ports projects the state into $(\ket{00}+e^{i \phi}\ket{11})/\sqrt{2}$. The state can be written as $(\ket{++}\text{cos}\phi + \ket{--}\text{cos}\phi+\ket{+-}\text{sin}\phi+\ket{-+}\text{sin}\phi)/2$. 
When detecting qubits 2 and 3 in the $\hat{\sigma}_x\hat{\sigma}_x$ basis, e.g., in $(\ket{++}$, and rotating the input state $\ket{+}_{2}(\ket{0}_3+e^{i\phi}\ket{1}_3)$ around $\phi \in [0,\pi]$, we observed the HOM-like interference fringe, as shown in Fig.~\ref{fig:OperatorResults}i. Figure~\ref{fig:FigSMathforO}f shows the configuration of the circuits performing HOM interference and Bell generation at the $\hat{O}_\text{fusion}$. 
A high visibility of $86 \pm 4.0\%$ was measured in our experiment. We thus confirm the entangling fusion operation using the circuit in Fig.~\ref{fig:OperatorResults}g.

\subsection{Bell-state Measurement}


A Bell measurement is a two-qubit measurement which can unambiguously determine each of the four Bell states. For example, the Hamamard gate followed by a CNOT gate is sufficient to project two-qubit pure states 
into the basis of Bell states given by $\lbrace \ket{\Phi^+}, \ket{\Phi^-}, \ket{\Psi^+}, \ket{\Psi^-} \rbrace$. Since these gates are both unitary and hermitian, reversing the order of operation performs a projective measurement from the Bell basis to the computational basis 
\begin{equation}
(\hat{U}_{\text{CNOT}} )
(\hat{U}_{\text{Had}} \otimes \hat{I})
\lbrace
\ket{\Phi^+},
\ket{\Phi^-},
\ket{\Psi^+},
\ket{\Psi^-}
\rbrace \to
\lbrace 
\ket{00},
\ket{01},
\ket{10},
\ket{11}
\rbrace.
\end{equation}

Though a deterministic Bell state analysis is not possible with linear optics, a partial measurement able to distinguish up to two of the Bell states is possible. Three types of linear-optical Bell analysers are possible: KLM CNOT gate (1/9), bosonic Bell projector (1/2), and fusion operator (1/2), where the values in () refer to the success probability of Bell measurement. 
The KLM CNOT gate and its Bell measurements have been reported in other materials systems~\cite{OxfordChip,JeremyPhotons}, which reply on multiphotons states generated by off-chip SPDC sources. 
In our device, we demonstrate that the latter two, i.e.,  bosonic Bell projector $\hat{O}_{\text{Bell}}$ and fusion operator $\hat{O}_{\text{fusion}}$, with higher success probability, are able to perform projective measurements on the Bell states. In this section, we discuss the Bell analyser using $\hat{O}_{\text{Bell}}$, and discuss the Bell analyser using $\hat{O}_{\text{fusion}}$ in section ~\ref{subsec: GHZgeneation}. \\

Consider first the evolution of the states $\ket{\Phi^{\pm}}$, which cannot be determined on chip. 
On our device, we perform the operator $\hat{O}_{\text{Bell}}$ on signal photons only (the operation happens on qubits 2 and 3). 
To keep the analysis simple, we first assume that the signal photons from either source are perfectly indistinguishable from one another. 
In this case, the state that is created on our device is 

\begin{equation}
\frac{1}{\sqrt{2}}(\hat{S}^\dagger _{\uparrow,2} \hat{S}^\dagger _{\uparrow, 3} \pm \hat{S}^\dagger _{\downarrow,2} \hat{S}^\dagger _{\downarrow , 3} ) \ket{\text{vac}}.
\end{equation}
Swapping waveguide modes applies the transformation $\hat{S}^\dagger _{\downarrow, 2} \longleftrightarrow \hat{S}^\dagger _{\uparrow ,3}$, which produces the state 
\begin{equation}
\frac{1}{\sqrt{2}}(\hat{S}^\dagger _{\uparrow,2} \hat{S}^\dagger _{\downarrow, 2} \pm \hat{S}^\dagger _{\uparrow,3} \hat{S}^\dagger _{\downarrow , 3} ) \ket{\text{vac}}
\end{equation}
Next, each qubit interferes on a Hadamard-like operator. 
Both terms of the evolved state will see HOM-like interference: 
\begin{equation}
\frac{1}{2\sqrt{2}} 
\big(
(\hat{S}^\dagger _{\uparrow, 2}  + \hat{S}^\dagger _{\downarrow,2})  
(\hat{S}^\dagger _{\uparrow, 2} - \hat{S}^\dagger _{\downarrow,2})  
\pm   
(\hat{S}^\dagger _{\uparrow, 3} + \hat{S}^\dagger _{\downarrow,3})    
(\hat{S}^\dagger _{\uparrow,3} - \hat{S}^\dagger _{\downarrow,3}) 
\big)
\ket{\text{vac}}. 
\end{equation}
which gives the state 
\begin{equation}
\frac{1}{2\sqrt{2}} 
\big(
\hat{S}^\dagger _{\uparrow,2} \hat{S}^\dagger _{\uparrow,2} 
-
\hat{S}^\dagger _{\downarrow,2} \hat{S}^\dagger _{\downarrow,2} 
\pm   
\hat{S}^\dagger _{\uparrow,3} \hat{S}^\dagger _{\uparrow,3} 
-
\hat{S}^\dagger _{\downarrow,3} \hat{S}^\dagger _{\downarrow,3} 
\big)
\ket{\text{vac}}.
\end{equation}
Finally, we reapply the waveguide swapping: 
\begin{equation}
\frac{1}{2\sqrt{2}} 
\big(
\hat{S}^\dagger _{\uparrow,2} \hat{S}^\dagger _{\uparrow,2} 
-
\hat{S}^\dagger _{\uparrow,3} \hat{S}^\dagger _{\uparrow,3} 
\pm   
\hat{S}^\dagger _{\downarrow,2} \hat{S}^\dagger _{\downarrow,2} 
-
\hat{S}^\dagger _{\downarrow,3} \hat{S}^\dagger _{\downarrow,3} 
\big)
\ket{\text{vac}}.
\end{equation}
\\
Importantly, since we collect only four-fold coincidence measurements and since photons are indistinguishable, after passing through $\hat{O}_{\text{Bell}}$ the single photons are completely bunched at the output port and should not contribute to coincidence measurements - and so we cannot detect them.\\

Next, consider the other two Bell states
\begin{equation}
\ket{\Psi^{\pm}} 
=
\frac{1}{\sqrt{2}}
(
\hat{S}^\dagger _{\uparrow, 2}  \hat{S}^\dagger _{\downarrow,3}  
\pm
\hat{S}^\dagger _{\downarrow,2}  \hat{S}^\dagger _{\uparrow,3}  
)
\ket{\text{vac}}
\end{equation}
Under the same transformations the two states become $$
\ket{\Psi^{\pm}} 
=
\frac{1}{\sqrt{2}}
(
\hat{S}^\dagger _{\uparrow,2}  \hat{S}^\dagger _{\downarrow,3}  
\pm
 \hat{S}^\dagger _{\uparrow,3}    \hat{S}^\dagger _{\downarrow,2} 
)
\ket{\text{vac}} 
$$
$$
\Rightarrow
\frac{1}{2\sqrt{2}} 
\big(
(\hat{S}^\dagger _{\uparrow,2}  + \hat{S}^\dagger _{\downarrow,2})  
(\hat{S}^\dagger _{\uparrow,3}  - \hat{S}^\dagger _{\downarrow,3})  
\pm   
(\hat{S}^\dagger _{\uparrow ,3} + \hat{S}^\dagger _{\downarrow ,3})    
(\hat{S}^\dagger _{\uparrow ,2} - \hat{S}^\dagger _{\downarrow ,2}) 
\big)
\ket{\text{vac}}
$$
\begin{align}
\begin{split}
\Rightarrow
\frac{1}{\sqrt{2}} 
(
\hat{S}^\dagger _{\uparrow,2}  \hat{S}^\dagger _{\uparrow,3}  
-
\hat{S}^\dagger _{\downarrow,2}  \hat{S}^\dagger _{\downarrow,3}  
) 
\ket{\text{vac}} 
~[\text{for} \ket{\Psi^+}]
\\
\frac{1}{\sqrt{2}} (
\hat{S}^\dagger _{\downarrow,2}  \hat{S}^\dagger _{\uparrow,3}  
-
\hat{S}^\dagger _{\uparrow,2}  \hat{S}^\dagger _{\downarrow,3}  
)
\ket{\text{vac}} 
~ [\text{for}  \ket{\Psi^-}]\\
\Rightarrow
\frac{1}{\sqrt{2}} 
(
\hat{S}^\dagger _{\uparrow,2}  \hat{S}^\dagger _{\downarrow,2}  
-
\hat{S}^\dagger _{\uparrow,3}  \hat{S}^\dagger _{\downarrow,3}  
)
\ket{\text{vac}} 
 ~[\text{for}  \ket{\Psi^+}]
\\
\frac{1}{\sqrt{2}} (
\hat{S}^\dagger _{\downarrow,2}  \hat{S}^\dagger _{\uparrow,3}  
-
\hat{S}^\dagger _{\uparrow,2}  \hat{S}^\dagger _{\downarrow,3}  
)
\ket{\text{vac}} 
~ [\text{for}  \ket{\Psi^-}]
\end{split}
\end{align}

Hence the input state $\ket{\Psi^+}$ leads to coincidence measurements in the same qubit but opposite modes, while the asymmetric input state $\ket{\Psi^-}$ leads to coincidences in the opposite qubit and opposite modes. 
Therefore, given that the state must be in one of the four Bell states (as in the teleportation and entanglement swapping protocols), we can determine which one by looking at the coincidence events corresponding to the correct modes.

\section{Teleportation and Entanglement Swapping }
\label{sec:TELE}
\subsection{Quantum Teleportation of Single-qubit States}
Quantum teleportation has the surprising property whereby a transmitter (Bob) is able to transfer information to a receiver (Daniel) at another location, without the need to transmit the physical system in which information is encoded. 
This is counter-intuitive, since information is physical and should therefore be stored in a physical system, whereby information can be extracted by performing measurements on that system. 
In fact, quantum teleportation is made possible by the use of pre-shared entanglement between the sender and receiver. \
Since this quantum channel shares non-local correlations, it is possible to reconstruct the qubit state remotely. 
Daniel is also required to take part in the reconstruction of Bob's qubit, since the teleportation can only be achieved up to a local rotation \{$\hat{\sigma}_{x}, \hat{\sigma}_{y},\hat{\sigma}_{z}$\}, where $\hat{\sigma}_{x}, \hat{\sigma}_{y},\hat{\sigma}_{z}$ represent the three Pauli operators. 
Moreover, due to the no-cloning theorem, a single unknown quantum state cannot be perfectly replicated or precisely measured, however, quantum teleportation does allow the remote preparation of arbitrary quantum bits. \\

We define the four involved parties in the teleportation as following: Alice (qubit 1), Bob (qubit 2), Charlie (qubit 3) and Daniel (qubit 4). 
In our silicon device, the quantum state in Bob to be teleported is prepared in the logical qubit $\psi_2$, which are initially 
created by SFWM inside MRR 1, resulting in the state $\ket 0_{i,1}  \ket 0_{s,2} $. 
In this experiment, the role of the Alice is to heralded the presence of Bob. 
The $\ket 0_{s,2}$ state is then rotated via a local unitary transformation, $\hat U$, such that (for clarity, we only denote the number of qubits in subscript) 
\begin{equation}
\hat U \ket 0_{s,2} = \alpha \ket 0_{2} + \beta \ket 1_{2} \coloneqq \ket{\psi}_2 
\end{equation}where $\alpha$ and $\beta$ are complex coefficients having$|\alpha|^2+|\beta|^2=1$.  
In order to perform the teleportation protocol, it is required that the sender and the receiver each share a quantum channel where both parties (Charlie and Daniel) obtain a particle who's state is maximally entangled with one another $\ket {\Phi^+} = \frac{1}{\sqrt{2}}(\ket{00} + \ket {11})$. 
This state is generated on-chip by simultaneously and coherently pumping MRR sources 3 and 4. 
By overlapping the MRRs, tuning the relative source pumping strength and compensating the relative photon phase, we prepare the entangled state shared between Charlie and Daniel $\ket {\Phi^+}_{3,4} = \frac{1}{\sqrt{2}}(\ket 0_{3}  \ket 0_{4}+\ket 1_{3}  \ket 1_{4}) $. The reconstructed density matrix result for the $\ket {\Phi^+}_{3,4}$ as well as the full set of Bell states is provided in Figs.~\ref{fig:FullBell}a-d. 
When simultaneously measuring four photons from each of the qubits (MRR2 in this case is switched off), the initial state becomes $\ket 0_{1} \otimes \ket{\psi}_2 \otimes \ket {\Phi^+}_{3,4} $, which can be written in the basis of Bell states for particles 2 and 3 as 
\begin{equation}
\ket 0_{1} \otimes \frac{1}{2}(  \ket{\Phi^+}_{2,3} \otimes \ket {\psi}_{4} + \ket{\Phi^-}_{2,3} \otimes \hat{\sigma}_{z} \ket {\psi}_{4} +  \ket{\Psi^+}_{2,3} \otimes \hat{\sigma}_{x} \ket {\psi}_{4} +  \ket{\Psi^-}_{2,3} \otimes \hat{\sigma}_{y}  \ket {\psi}_{4}),
\end{equation}
where $ \ket{\psi}_4 = \alpha \ket 0_{4} + \beta \ket 1_{4} $. 
Bob performs the teleportation of qubit state $\psi_2$ by implementing the Bell measurement $\hat{O}_\text{Bell}$ between Bob's qubit  $\psi_2$ and Charlies' qubit 3 that is entangled with Daniel's qubit 4 (see discussions on Bell measurement in section~\ref{sec:opeators}). 
The success of $\hat{O}_\text{Bell}$ allows the projection of the two-qubit state into $\ket {\Psi^+}_{2,3}$, which in turn transmits Bob's $\ket{\psi}_2$ state to Dan who obtain the state $\ket{\psi}_4$ up to a local rotation, $\hat{\sigma}_{x}$. 
Alternatively, projecting the two-qubit state into $\ket {\Psi^-}_{2,3}$ transmits Bob's $\ket{\psi}_2$ to Dan who will in this case obtain the state $\ket{\psi}_4$ up to a local rotation, $\hat{\sigma}_{y}$. \\

In our experiment the ideal transmitted state is $\ket{\psi}_2$, while the received state is $\rho_4$.  To assess the effectiveness of the teleportation protocol, we measure the fidelity between the tomographically estimated state, $\rho_{4}$, and ideal state, $\ket{\psi_{\text{2}}}$, defined as the overlap between two states $F=\Tr (\rho_{4} \ketbra{ \psi_{\text{2}} }{ \psi_{\text{2}} })$. 
The QST estimation was achieved by collecting four-fold coincidence counts while projecting the state into the observable $\ketbra{0}{0}_1 \otimes \ketbra{\Psi^+}{\Psi^+}_{2,3} \otimes \hat{U}_4 \ketbra{0}{0}_{4}  \hat{U}_4^\dagger$. 
The coincidences are used in order to estimate the probability amplitudes of the state under local projective measurements and maximum likelihood methods estimate the overall density matrix. We have prepared the six states $\ket{\psi}_2=$ \{$\ket{0}$,$\ket{1}$,$\ket{+}$,$\ket{-}$,$\ket{+ i}$,$\ket{- i}$\}. 
By implementing QSTs, we obtain the reconstructed $\rho_{4}$ for the six states. The outcomes of these measurements are summarised in Fig.~\ref{fig:TeleGHZResults}d, where we report an average fidelity of $ 91\%$. The measured fidelity values are provided in Table~\ref{table_allFid}.

\begin{figure*}[ht!] 
\centering
\includegraphics[width=1\textwidth]{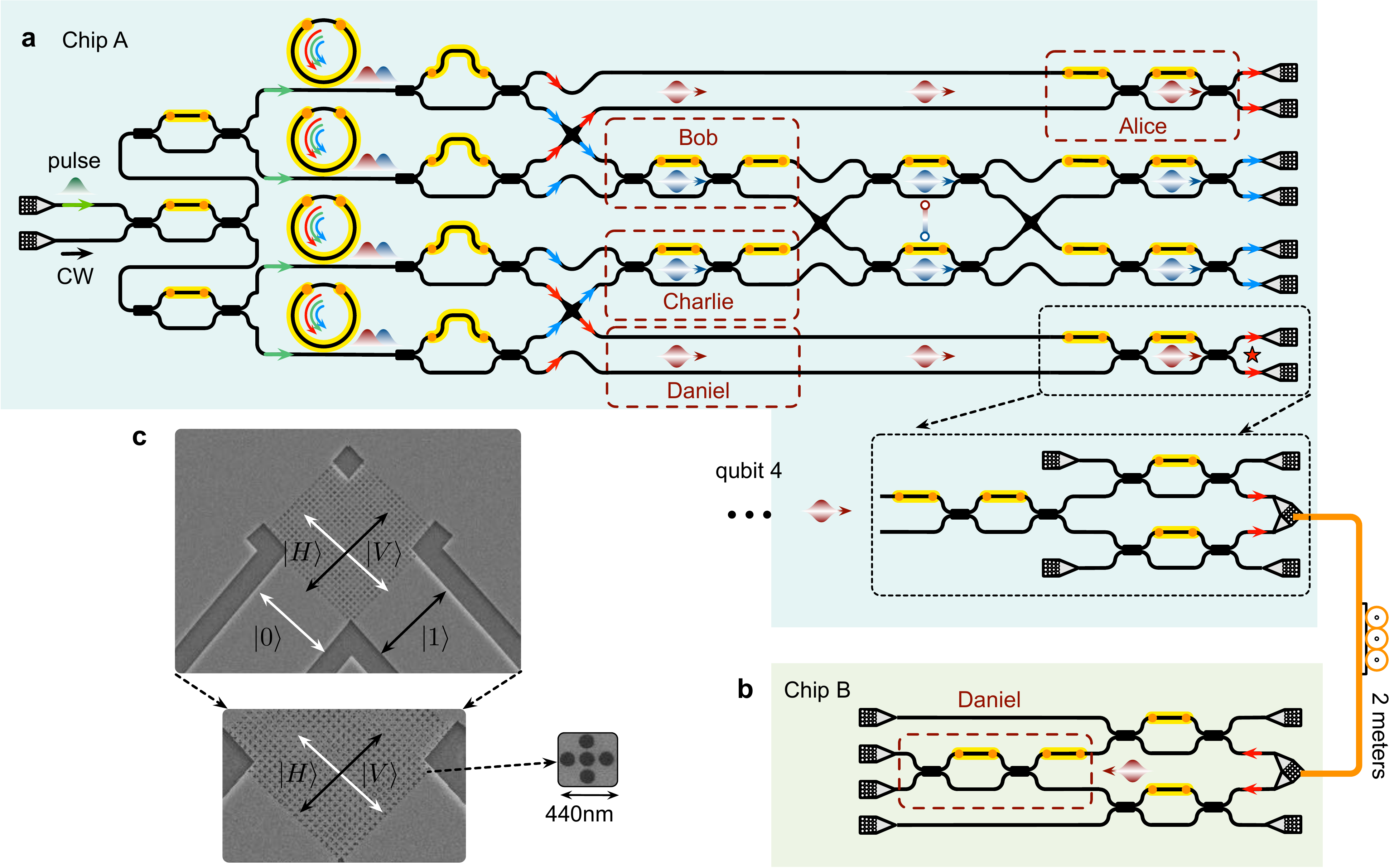} 
\caption{Schematic of the chip-to-chip entanglement distribution and teleportation of single qubits using the path-polarisation conversion technique. 
\textbf{a,} Schematic of chip A that includes a switchable router on qubit 4. A pair of MZIs choose whether the qubit is encoded into dual-rail and output via two 1d SGCs, or converted to polarisation qubits and output via the 2d SGC. The former enables the implementation of arbitrary single-qubit measurement in chip A, while the latter enables the coherent distribution of qubit 4 from chip A to chip B. 
\textbf{b,} Schematic of Bob chip, which is able to reconvert polarisation-encoded qubits to path-encoded qubits, in order to perform reconstructive projective measurements on Bob.  A pair of MZIs are used for the ease of calibrating the components in chip B. \textbf{c,} SEM images of the 2D SGC structure fabricated on chip A and chip B. It can coherently convert the two on-chip path-encoded \{$|0\rangle$,$|1\rangle$\} states to two orthogonal polarisation modes \{$|H\rangle$,$|V\rangle$\} in fiber, and vise verse. } 
\label{fig:CTC} 
\end{figure*}    

\subsection{Chip-to-chip Entanglement Distribution and Quantum Teleportation }
\label{subsec:CTCTELE}
Silicon is a compelling platform for classical optical telecommunications as well as quantum communications, both through optical fibers. In general, the silicon-based integrated quantum transceivers could provide possible low-cost and high-performance secure communication networks.  
A reliable transfer of single photon qubits from one silicon device to another has already been shown, and the distribution of entangled states has been verified through the violation of Bell inequalities~\cite{Wang2016}. Here, we first distribute the full set of entangled Bell states between two devices, and then demonstrate key missing capabilities so far, i.e., the chip-to-chip teleportation. \\

\begin{figure*}[ht!] 
\centering
\includegraphics[width=0.6\textwidth]{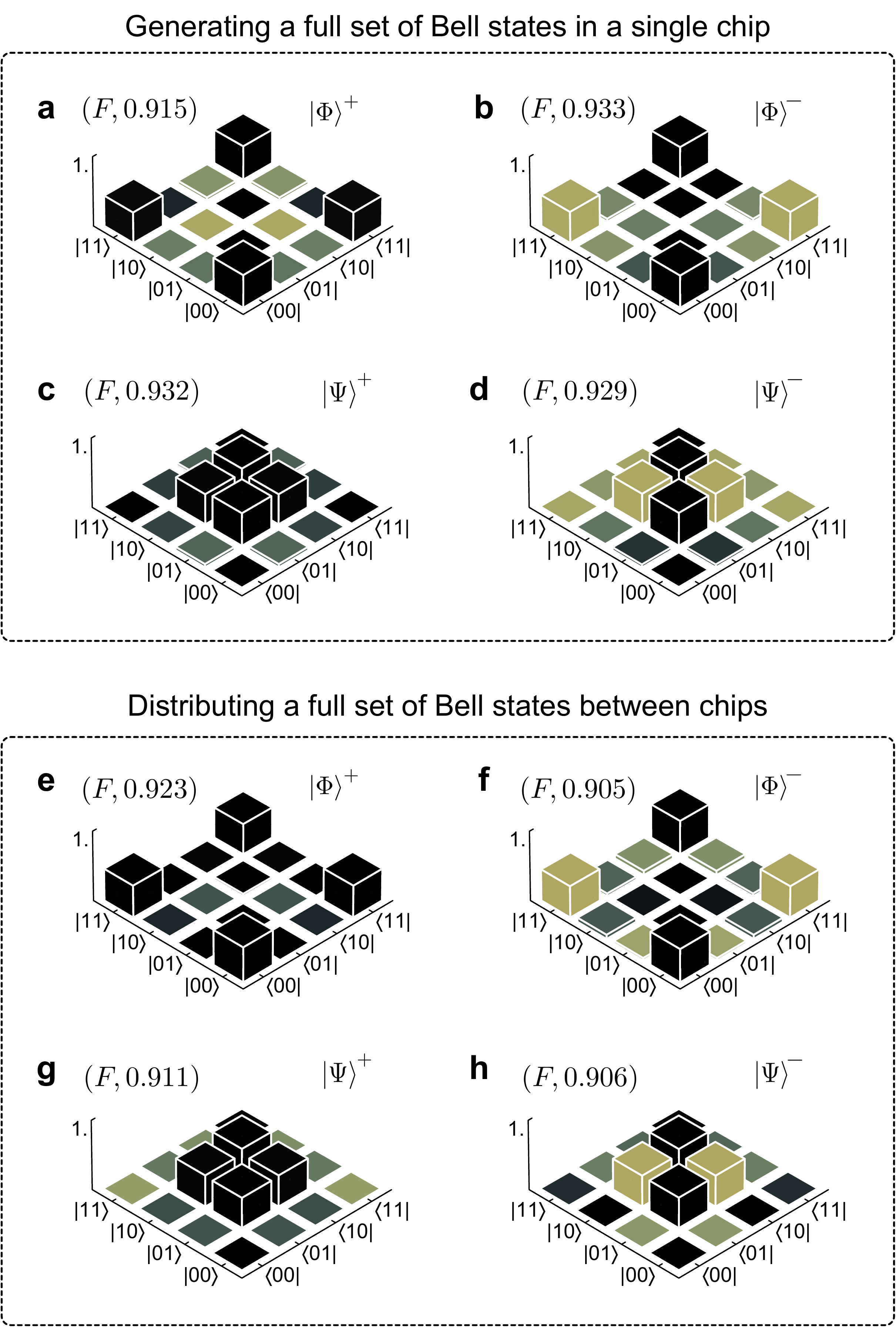} 
\caption{On-chip generation and inter-chip distribution of a compete set of four Bell states.  
Reconstructed density matrices for the Bell states $\ket{\Phi^+}$ (\textbf{a}), $\ket{\Phi^-}$ (\textbf{b}), $\ket{\Psi^+}$ (\textbf{c}), and $\ket{\Psi^-}$ (\textbf{d}), which are all generated and measured on a single chip A. The four entangled states are then distributed across the two chips, keeping qubit 3 in chip A and distributing qubit 4 to chip B. 
After the entanglement distribution, reconstructed density matrices for the Bell states $\ket{\Phi^+}$ (\textbf{e}), $\ket{\Phi^-}$ (\textbf{f}), $\ket{\Psi^+}$ (\textbf{g}), and $\ket{\Psi^-}$ (\textbf{h}).  Two local measurements are respectively performed on chip A and Chip B for quantum state tomography.}
\label{fig:FullBell} 
\end{figure*}    

We first create a complete set of four Bell entangled states, i.e.,$\ket{\Phi^ \pm}$ and $\ket{\Psi^ \pm}$. 
The Bell pairs are generated in our device by simultaneously, equally and coherently pumping two MMR sources. Since the top (bottom) source creates signal and idler photons prepared in the $\ket{00}$ ($\ket{11}$) mode. In the event where only two photons are generated (from only one source) the superposition state $\ket{\Phi^+}$ can be initially generated. 
The other three Bell states are created by performing local rotations on one of the Bell qubits in the following way $\ket{\Phi^-}=\hat{\sigma}_z\ket{\Phi^+}$, $\ket{\Psi^+}=\hat{\sigma}_x\ket{\Phi^+}$,$\ket{\Psi^-}=\hat{\sigma}_y\ket{\Phi^+}$, where $\hat{\sigma}_x$, $\hat{\sigma}_y $, $\hat{\sigma}_z$ are the Pauli operators. Figures~\ref{fig:FullBell} a-d  shows the reconstructed density matrices for the $\ket{\Phi^ \pm}$ (a, b) and $\ket{\Psi^ \pm}$ (c, d) states. Two copies of the full set of Bell states can be produced in the device, \{$\ket{\Phi^ \pm}_{1,2}$, $\ket{\Psi^ \pm}_{1,2}$ \} and \{$\ket{\Phi^ \pm}_{3,4}$, $\ket{\Psi^ \pm}_{3,4}$ \}, and further allow the teleportation, entanglement swapping and GHZ entanglement. \\

To coherently distribute and teleport states from one chip (chip A, Fig.~\ref{fig:CTC}a) to another (chip B, Fig.~\ref{fig:CTC}b), we exploited the path-polarisation conversion technique~\cite{Wang2016}. The path-encoded states are very reliable in integrated optical chips, while polarization-encoded states are robust in free space and in optical fiber. Figure 1 and Fig.~\ref{fig:CTC}c shows the SEM images of 2d SGC, which is formed by superposing two 1d SWGs at a right angle, can coherently convert the two orthogonal polarisation modes in fiber \{$|H\rangle$,$|V\rangle$\} into two on-chip paths, say \{$|0\rangle$,$|1\rangle$\} states, both in the TE polarised mode. This results in a  coherent  mapping of the polarisation-encoded $\alpha |H\rangle +\beta|V\rangle$ state in fiber and path-encoded $\alpha |0\rangle +\beta|1\rangle$ state on chip. We distributed the  $\ket{\Phi^ \pm}$ and $\ket{\Psi^ \pm}$ entangled states over a 2-meter single-mode fiber, connected by the two path-polarisation converters, one on each chip. A fiber polarisation controller was used to compensate the random polarisation rotation in the fiber. 
Figures~\ref{fig:FullBell} e-h  show the reconstructed density matrices for the $\ket{\Phi^ \pm}$ (a, b) and $\ket{\Psi^ \pm}$ (c, d) states, after their entanglement distribution over the two chips. High fidelities are observed for all Bell states, certifying the high-fidelity quantum states distribution between the two chips. \\

\begin{figure*}[ht!] 
\centering
\includegraphics[width=0.5\textwidth]{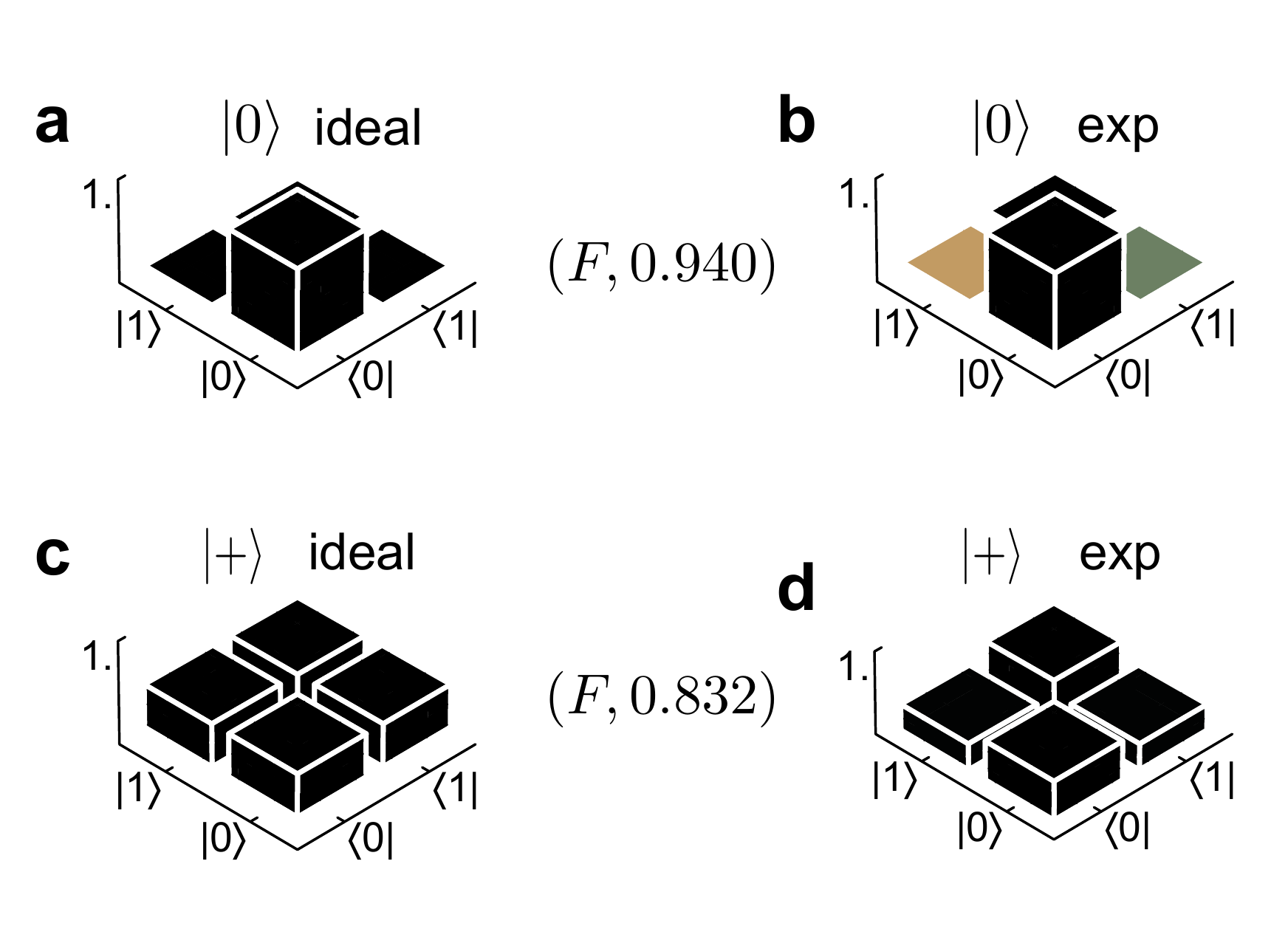} 
\caption{Experimental results of chip-to-chip teleportation of single-qubit states. Reconstructed density matrices for \textbf{a}, ideal state $\ket{0}$, \textbf{b}, teleported state $\ket{0}$, \textbf{c}, ideal state $\ket{+}$, and \textbf{d}, teleported state $\ket{+}$.  
Six measurements are performed on chip B to achieve QST of each state. The $F$ refers to the measured fidelity $\bra{ \psi_{2}}\rho_4\ket{\psi_{2}} $, where $\rho_{4}$ is the teleported state and $\ket{\psi_{2}}$ is the ideal state. }
\label{fig:CTCteledata} 
\end{figure*}

Furthermore, we show a proof-of-principle demonstration of quantum teleportation between the sender and receiver circuits, necessary for quantum networks, quantum cryptography and distributed quantum computing. 
The $\ket{\psi}_{2}\otimes \ket{\Phi^+_{3,4}}$ state is created in the chip A (photon 1 is heralded) and the 4-th qubit is distributed to chip B, and then the Bell measurements $\hat{O}_\text{Bell}$ are performed on the qubits 2,3 in the chip A. The success of joint clicks in the Bell basis in chip A enables the teleporation of state $|\psi_2\rangle$ from chip A to chip B. As examples, we prepared two states $|\psi_2\rangle =\{ |0_2\rangle, |+_2\rangle\}$, and implemented the chip-to-chip teleporation. The teleported states $\ket{\phi_4} $ are reconstructed in the chip B by performing QST (see Fig.~\ref{fig:CTC}). 
Figure~\ref{fig:CTCteledata} shows the reconstructed density matrices for the teleported $\ket{0}$ and $\ket{1}$ states. High state fidelities of $0.940\pm 0.041$ and $0.832\pm 0.048$ have been demonstrated. Compared to the single-chip results, we remark on the fact that the distribution of entangled states and teleported qubits does not have a major impact in the resulting state fidelity. This arises from the high-fidelity of chip-to-chip interconnection. Our results confirm the success of chip-to-chip quantum teleporation of single-qubit states.

\subsection{Entanglement Swapping (Teleportation of Entanglement)}
The distribution and swapping of entanglement are of fundamental interest in quantum communications, having direct applications in device-independent quantum key distribution, quantum secret sharing and quantum networks. 
Optical loss in quantum channel limits the distance to which these quantum protocols can operate. 
A proposed solution is that of the quantum repeater, which can in principle achieve long distances by utilising entanglement swapping and quantum memory.  

Entanglement swapping is a quantum information protocol whereby sets of maximally entangled qubits, say between (Alice\& Bob) and (Charlie \& Daniel), are swapped with one another by conducting a Bell measurement on two unentangled qubits. 
For instance, projecting Bob \& Charlie into a Bell state would also entangle Alice \& Daniel.  
This is of particular interest since there is no requirement for Alice and Daniel to have ever interacted with one another, yet they now experience nonlocal quantum correlations, and thus allowing the extension of entanglement for long-distance quantum communication and network. \\

In our integrated device, two sets of entangled particles in the state $\ket{\Phi^+}_{1,2} \otimes \ket{\Phi^+}_{3,4}$ are generated when detecting fourfold coincidences across each qubit in the regime where all four MRR sources are spectrally overlapped and simultaneously \& coherently pumped. 
The density matrices for  $\ket{\Phi^+}_{1,2}$ and $\ket{\Phi^+}_{3,4}$ are experimentally reconstructed with a state fidelity of $0.996\pm0.002$ and $0.915\pm0.003$ respectively. 
In the basis of Bell states for qubits 2 and 3, the  state can be rewritten as 

\begin{equation}
\frac{1}{2}(\ket{\Phi^+}_{1,4} \otimes \ket{\Phi^+}_{2,3} + \ket{\Phi^-}_{1,4} \otimes \ket{\Phi^-}_{2,3} + \ket{\Psi^+}_{1,4} \otimes \ket{\Psi^+}_{2,3} + \ket{\Psi^-}_{1,4} \otimes \ket{\Psi^-}_{2,3}).
\end{equation}

Once more, we perform a Bell measurement between the signal qubits by switching on $\hat{O}_{\text{Bell}}$, which collapses the remaining two-photon idler state to $\ket{\Psi^+}_{1,4}$. We perform a full tomography on this state by heralding the signal photons whilst performing local projective measurements on the idler qubits, $[(\hat{U}_1 \otimes \hat{U}_4) \ketbra{\Psi^+}{\Psi^+}_{1,4} (\hat{U}_1^\dagger \otimes \hat{U}_4^\dagger)] \otimes \ketbra{\Psi^+}{\Psi^+}_{2,3}$. The reconstructed density matrix is shown in Fig. \ref{fig:TeleGHZResults}e.  

\section{Generation, Certification, and Quantification of Genuine Multipartite GHZ Entangled States}
\label{sec:GHZ}

\subsection{Generation of Four-photon and Three-photon GHZ States}
\label{subsec: GHZgeneation}
We generate the on-chip entangled GHZ states by applying the two-qubit entangling fusion operator $\hat{O}_{\text{fusion}}$ (see section \ref{sec:opeators}) onto two signal photons 2,3, which initially each form part of the Bell state $ \ket{\Phi^+}_{1,2}$ and $ \ket{\Phi^+}_{3,4}$. We represent the state evolution as following. 
\begin{equation}
\begin{split}
\hat{O}_{\text{fusion,2,3}}
(\ket{\Phi^+}_{1,2} \otimes \ket{\Phi^+}_{3,4})
=
\hat{O}_{\text{fusion,2,3}} ~
\frac{1}{2}
(
\hat{S}_{\uparrow, 1}^\dagger 
\hat{S}_{\uparrow, 2}^\dagger
+
\hat{S}_{\downarrow, 1}^\dagger 
\hat{S}_{\downarrow, 2}^\dagger
)
\otimes
(
\hat{S}_{\uparrow, 3}^\dagger 
\hat{S}_{\uparrow, 4}^\dagger
+
\hat{S}_{\downarrow, 3}^\dagger 
\hat{S}_{\downarrow, 4}^\dagger
)
\ket{\text{vac}},
\end{split}
\end{equation}
where the notation $ \hat{S}_{\uparrow, i}^\dagger $ and  ($\hat{S}_{\downarrow, i}^\dagger$ describe a photon arising in the zero $\ket{0}$ mode and one $\ket{1}$ mode of the $i$-th qubit or photon. As we have discussed in Fig.~\ref{fig:OperatorResults} in the main text and section~\ref{sec:opeators}, the $\hat{O}_{\text{fusion}}$ on dual-rail qubits can transmit photons in the $\ket{0}$ mode and swap photons in $\ket{1}$ mode. Applying  the $\hat{O}_{\text{fusion}}$ evolves the state as 
\begin{equation}
\begin{split}
& \Rightarrow 
\frac{1}{2}
(
\hat{S}_{\uparrow, 1}^\dagger 
\hat{S}_{\uparrow, 2}^\dagger
+
\hat{S}_{\downarrow, 1}^\dagger 
\hat{S}_{\downarrow, 3}^\dagger
)
\otimes
(
\hat{S}_{\uparrow, 3}^\dagger 
\hat{S}_{\uparrow, 4}^\dagger
+
\hat{S}_{\downarrow, 2}^\dagger 
\hat{S}_{\downarrow, 4}^\dagger
)
\ket{\text{vac}}
\\
& \Rightarrow
\frac{1}{2}
(
\hat{S}_{\uparrow,1}^\dagger 
\hat{S}_{\uparrow,2}^\dagger
\hat{S}_{\uparrow,3}^\dagger
\hat{S}_{\uparrow,4}^\dagger 
+
\underbrace{
\hat{S}_{\uparrow,1}^\dagger 
\hat{S}_{\uparrow,2}^\dagger
\hat{S}_{\downarrow,2}^\dagger
\hat{S}_{\downarrow,4}^\dagger 
+
\hat{S}_{\uparrow,1}^\dagger 
\hat{S}_{\uparrow,3}^\dagger
\hat{S}_{\downarrow,3}^\dagger
\hat{S}_{\downarrow,4}^\dagger 
}_{\text{Not measured}}
+
\hat{S}_{\downarrow,1}^\dagger 
\hat{S}_{\downarrow,2}^\dagger
\hat{S}_{\downarrow,3}^\dagger
\hat{S}_{\downarrow,4}^\dagger 
)
\ket{\text{vac}}
\end{split}
\end{equation}

When detecting only one photon in each of the detector combinations \{D1,D2\}, \{D3,D4\}, \{D5,D6\} and \{D7,D8\}, the middle two items are not measured and thus disregarded via post-selection. The observation of the four-fold coincidence events thus results in the four-photon entangled GHZ state (after re-normalisation) as below: 
\begin{equation}
\begin{split}
& \Rightarrow 
\frac{1}{\sqrt{2}}
(
\hat{S}_{\uparrow,1}^\dagger 
\hat{S}_{\uparrow,2}^\dagger
\hat{S}_{\uparrow,3}^\dagger
\hat{S}_{\uparrow,4}^\dagger 
+
\hat{S}_{\downarrow,1}^\dagger 
\hat{S}_{\downarrow,2}^\dagger
\hat{S}_{\downarrow,3}^\dagger
\hat{S}_{\downarrow,4}^\dagger 
)
\ket{\text{vac}}
\\
& \Rightarrow 
\ket{\text{GHZ}}_4 =
\frac{1}{\sqrt{2}}
(
\ket{0000}+\ket{1111}
). 
\end{split}
\end{equation}

\textbf{Three-photon GHZ: } To create the three-photon GHZ entangled state, we locally measure qubit 4 by performing projection in the $\hat{\sigma}_x$ basis. 
Three photon GHZ states are generated by measuring the remaining photons in the diagonal basis, for example: 
\begin{equation}
\begin{split}
 \ket{\text{GHZ}}_4 
&= 
\frac{1}{2} (
\ket{000}(\ket{+}+\ket{-})
+
\ket{111}(\ket{+}-\ket{-})
).
\end{split}
\end{equation}
Hence when measuring a positive 
eigenvalue the remaining photon state is collapsed to $ \ket{\text{GHZ}}_3=(\ket{000}+\ket{111})/\sqrt{2}$ . \\

\textbf{Two-photon GHZ or entanglement swapping: } Likewise, we locally measure qubits 2, 3, in the $\hat{\sigma}_x \hat{\sigma}_x$ basis, and obtain the two-photon GHZ entangled state. 
When measuring photons 2 and 3 in this basis, the remaining state is a Bell state between the qubit 1 and 4,  by swapping entanglement $ \ket{\Phi^+}_{1,2}$ and $ \ket{\Phi^+}_{3,4}$. Figure~\ref{fig:TeleGHZResults}e shows the reconstructed density matrix of the swapped entangled state $\ket{\Phi^+}_{1,4}$ by performing a full QST. In the other words, the $\hat{O}_\text{fusion}$ works as the Bell analyser here, projecting the qubits 2,3 onto the Bell basis.

\subsection{Certification of GME by Entanglement Witness}
To certify genuine multipartite GHZ entanglement, we measured a suitable entanglement witness. The multipartite entanglement witness $\hat{W}$ is an operator which, when measured, gives values greater than 0 for all biseparable states and a value less than 0 for all genuine multipartite entangled states. Such an operator is trivial to construct, yet not always straightforward to evaluate. For example, assume our system produces a statistical mixture of states $\rho$ whose target state is a pure genuine multipartite entangled state $\ket{\psi}_{\text{GME}}$. The goal is to measure an observable that when measured gives an expectation value of $F_\alpha - F_\rho$, where $F_\rho$ is the fidelity between our generated state and target state, while $F_\alpha$ is the \textit{maximum} fidelity between the target state and all possible biseparable pure states. As a result
\begin{equation}
\begin{split}
\Tr(\rho\hat{W}) &=F_\alpha - F_\rho
\\ &=\max\limits_{\phi \in B} |\braket{\phi|\psi}_{\text{GME}}|^2 - \bra{\psi}\rho \ket{\psi}_{\text{GME}}
\end{split}
\end{equation}
giving,
\begin{equation}
\begin{split}
\hat{W} =\alpha \hat{I} - \ket{\psi} \bra{\psi}_{\text{GME}},
\end{split}
\end{equation}
where 
$\alpha = \max\limits_{\phi \in B} |\braket{\phi|\psi}|^2 .$ 
For the GHZ states considered in this experiment, the value of $\alpha$ is $1/2$, meaning that any any overlap between our state $\rho$ and the target GHZ state will certify genuine multipartite entanglement for measured fidelities greater than 50\%. In order to measure the fidelity in each case, the measureable $\ket{\text{GHZ}} \bra{\text{GHZ}}$ is decomposed into local measureable observables which can be easily verified on our device. In practice, this is achieved by measuring the average of two expectation values
\begin{equation}
F_{\text{GHZ}} = \frac{\langle \hat{A} \rangle + \langle \hat{B} \rangle}{2}
\end{equation}
where $\hat{A}$ is the population term given by
\begin{equation}
\hat{A}_N  = \ket{0} \bra{0} ^{\otimes N} +\ket{1} \bra{1} ^{\otimes N} ,
\end{equation}
and $\hat{B}$ is the coherence term, given by
\begin{equation}
\hat{B}_N= \sum_{k=0}^{N-1}(-1)^k \langle \hat{\Omega}_{k \pi /N}^{\otimes N} \rangle,
\end{equation}
where 
\begin{equation}
\hat{\Omega}_{\theta} = \cos \theta \hat{\sigma}_x + \sin{\theta} \hat{\sigma}_y.
\end{equation}

In our experiment, both the population term $\hat{A}$ and coherence term $\hat{B}$ can be measured and the results are respectively provided in Fig.~\ref{fig:TeleGHZResults}g and Fig.~\ref{fig:TeleGHZResults}i. 
In this manner, the state fidelity $F_{\text{GHZ}}  $ and entanglement witness operator $\langle \hat{W} \rangle$ can be estimated.  These were obtained by controlling the four on-chip projectors which allow arbitrary projective measurement on single-qubit states. The measurement of this GME witness requires $n+1$ global measurement settings, where $n$ is the number of particles in the GHZ state. As the number of particles $n$ grows, the measurement of this witness also becomes impractical. However, a recently introduced technique allows one to lower bound the state fidelity via only two global measurement settings, independent of the number of particles \cite{OAMNatPhys}. In the next section, we discuss the application of this new witness to our experimental data.

\subsection{Quantification of GME by Bounding GME-concurrence with Measurements in Two Bases}
The GME-concurrence is a measure of multipartite entanglement that is obtained by extending the familiar bipartite entanglement quantifier, concurrence, in the following manner. The concurrence of a pure state $\ket{\psi}$ is defined as $C(\ket{\psi}) = \sqrt{2(1-\textrm{Tr}\rho_A^2)}$, where $\rho_A = \textrm{Tr}_B\rho$ is the reduced density matrix of $\rho = \ket{\psi}\bra{\psi}$. The concurrence can be generalised for mixed states by the convex roof construction $C(\rho) = \textrm{inf}_{\{p_i,\ket{\psi_i}\} } \sum_i p_i C(\ket{\psi_i})$ where the infimum is taken over all possible pure state decompositions of $\rho$. For separable states, the concurrence gives a value of $0$, while for maximally entangled states such as a Bell state, it reaches its maximum value of $1$. Intermediate values of concurrence quantify the amount of entanglement present in a given bipartite state.

For pure multipartite states, a similar measure can be obtained by looking at all possible bipartitions of a multipartite state and calculating its concurrence \cite{GMEconcurrence}. In this manner, the GME-concurrence can be defined as $C_{GME}\coloneqq \min\limits_{\gamma_i\in\gamma} \sqrt{2[1-\textrm{Tr}(\rho_{A_{\gamma_i}}^2)]}$ where $\gamma = \{\gamma_i\}$ represents the set of all possible bipartitions $\{A_i | B_i\}$ of $\{1, 2,...,n\}.$ The $C_{GME}$ can be generalised for mixed multipartite states by making a convex roof construction in a manner similar to above. In a set of recent works, it was shown how the $C_{GME}$ can be obtained from measurements in two global product bases, drastically reducing the number of measurements required for estimating it, especially in the high-dimensional multipartite scenario \cite{OAMNatPhys,Erker:2017cb}. Here we apply this technique to calculate the GME-concurrence for qubit GHZ states of three and four photons.

\textbf{Three Photons: }The GME-concurrence for a tripartite qubit GHZ state can be lower bounded by the following expression that involves measurements in two global product bases, $\hat{\sigma}_x^{\otimes3}$ and $\hat{\sigma}_z^{\otimes3}$ \cite{Erker:2017cb}:

\begin{equation}
C_{GME} \geq C_{3,2} - 4\,\bigg(\sqrt{\bra{001}\rho\ket{001}\bra{110}\rho\ket{110}}
+ \sqrt{\bra{010}\rho\ket{010}\bra{101}\rho\ket{101}} + \sqrt{\bra{011}\rho\ket{011}\bra{100}\rho\ket{100}}\bigg),
\end{equation}

\noindent where the term $C_{3,2}$ is obtained in the following way from diagonal measurements in the first mutually unbiased basis $\hat{\sigma}_x=\{+,-\}$:

\begin{equation}
\begin{split}
C_{3,2} &= \bra{+++}\rho\ket{+++} + \bra{+--}\rho\ket{+--} + \bra{-+-}\rho\ket{-+-} + \bra{--+}\rho\ket{--+}\\
&- \bra{++-}\rho\ket{++-} - \bra{+-+}\rho\ket{+-+} - \bra{-++}\rho\ket{-++} - \bra{---}\rho\ket{---}.
\end{split}
\end{equation}

\noindent By measuring four-fold coincidence counts (tracing out photon 4) in the $\hat{\sigma}_z=\{0,1\}$ (see Fig.~\ref{fig:TeleGHZResults}g) and $\hat{\sigma}_x=\{+,-\}$ (see Fig.~\ref{fig:TeleGHZResults}h)  bases and calculating the 14 diagonal density matrix elements above, we obtain a value of $C_{GME} \geq 0.390\pm0.040$, which certifies that we are multipartite entangled with $n=3$ by at least 9 standard deviations. Errors are calculated via a Monte Carlo simulation of the experiment assuming Poissonian statistics.

\textbf{Four Photons: }The calculation for four photons proceeds in a manner similar to above. The GME-concurrence for a four particle qubit GHZ state can be lower bounded by the expression

\begin{equation}
\begin{split}
C_{GME} \geq C_{4,2} &- 4\,\bigg(\sqrt{\bra{0001}\rho\ket{0001}\bra{1110}\rho\ket{1110}} + \sqrt{\bra{0010}\rho\ket{0010}\bra{1101}\rho\ket{1101}} + \sqrt{\bra{0011}\rho\ket{0011}\bra{1100}\rho\ket{1100}}\\ 
&+ \sqrt{\bra{0100}\rho\ket{0100}\bra{1011}\rho\ket{1011}} 
+ \sqrt{\bra{0101}\rho\ket{0101}\bra{1010}\rho\ket{1010}} + \sqrt{\bra{0110}\rho\ket{0110}\bra{1001}\rho\ket{1001}}\\
&+ \sqrt{\bra{0111}\rho\ket{0111}\bra{1000}\rho\ket{1000}} \bigg),
\end{split}
\end{equation}

\noindent where the term $C_{4,2}$ is obtained in the following way from diagonal measurements in the first mutually unbiased basis $\hat{\sigma}_x=\{+,-\}$:

\begin{equation}
\begin{split} 
C_{4,2} &= \bra{++++}\rho\ket{++++} + \bra{++--}\rho\ket{++--} + \bra{+-+-}\rho\ket{+-+-} + \bra{+--+}\rho\ket{+--+}\\ 
&+ \bra{-++-}\rho\ket{-++-} + \bra{-+-+}\rho\ket{-+-+}
+ \bra{--++}\rho\ket{--++} + \bra{----}\rho\ket{----}\\ 
&- \bra{+++-}\rho\ket{+++-} - \bra{++-+}\rho\ket{++-+} - \bra{+-++}\rho\ket{+-++} - \bra{+---}\rho\ket{+---}\\
&-  \bra{-+++}\rho\ket{-+++} - \bra{-+--}\rho\ket{-+--} - \bra{--+-}\rho\ket{--+-} - \bra{---+}\rho\ket{---+}.
\end{split}
\end{equation}

\noindent By measuring four-fold coincidence counts in the $\hat{\sigma}_z=\{0,1\}$ (see Fig.~\ref{fig:TeleGHZResults}g) and $\hat{\sigma}_x=\{+,-\}$ (see Fig.~\ref{fig:TeleGHZResults}h) bases and calculating the 30 diagonal density matrix elements above, we obtain a value of $C_{GME} \geq 0.192\pm0.039$, which certifies that we are multipartite entangled with $n=4$ by at least 4 standard deviations. Errors are calculated via a Monte Carlo simulation of the experiment assuming Poissonian statistics.

\subsection{Certification of GME with Measurements in Two Bases}
Using the same figures of merit also allows for bounding respective state fidelities. For three qubits we find that
\begin{align}
   F_{GHZ}\geq\frac{1}{2} C_{3,2} -\,\bigg(\sqrt{\bra{001}\rho\ket{001}\bra{110}\rho\ket{110}}
+ \sqrt{\bra{010}\rho\ket{010}\bra{101}\rho\ket{101}} + \sqrt{\bra{011}\rho\ket{011}\bra{100}\rho\ket{100}}\bigg)\nonumber\\
+\frac{1}{2}(\bra{000}\rho\ket{000}+\bra{111}\rho\ket{111}),
\end{align}
as well as for four qubits
\begin{align}
   F_{GHZ}\geq\frac{1}{2}C_{4,2} &- \,\bigg(\sqrt{\bra{0001}\rho\ket{0001}\bra{1110}\rho\ket{1110}} + \sqrt{\bra{0010}\rho\ket{0010}\bra{1101}\rho\ket{1101}} + \sqrt{\bra{0011}\rho\ket{0011}\bra{1100}\rho\ket{1100}}\nonumber\\ 
&+ \sqrt{\bra{0100}\rho\ket{0100}\bra{1011}\rho\ket{1011}} 
+ \sqrt{\bra{0101}\rho\ket{0101}\bra{1010}\rho\ket{1010}} + \sqrt{\bra{0110}\rho\ket{0110}\bra{1001}\rho\ket{1001}}\nonumber\\
&+ \sqrt{\bra{0111}\rho\ket{0111}\bra{1000}\rho\ket{1000}} \bigg)+\frac{1}{2}(\bra{0000}\rho\ket{0000}+\bra{1111}\rho\ket{1111}),
\end{align}
\newpage
{\renewcommand{\arraystretch}{1.8}
\begin{table}[ht!]
\centering
\begin{tabular}{ c    | l   | c    | c| c }
\hline \hline  
\multicolumn{1}{ c| }{}
&\multicolumn{1}{ c| }{quantum state}
&\multicolumn{1}{ c| }{intra-chip state fidelity}
& \multicolumn{1}{c|}{inter-chip state fidelity}
&\multicolumn{1}{ c}{verification method}
\\ \hline  
1  &$\ket{\Phi}_\text{Bell}^+$ (Bell entangled, qubits 3,4) & $0.915\pm0.003$ & $0.923\pm0.027$  & QST [9] \\  
2 &$\ket{\Phi}_\text{Bell}^-$ (Bell entangled, qubits 3,4) & $0.933\pm0.002$& $0.905\pm0.015$ & QST [9]  \\  
3  &$\ket{\Psi}_\text{Bell}^+$ (Bell entangled, qubits 3,4) & $0.932\pm0.002$ & $0.911\pm0.019$ & QST [9]  \\  
4 &$\ket{\Psi}_\text{Bell}^-$ (Bell entangled, qubits 3,4) & $0.929\pm0.002$& $0.906\pm0.014$ & QST [9]  \\   \hline  
5  &$\ket{10}$ (separated, qubits 2,3) & $0.964\pm0.072$&  - & QST [9] \\  
6  &$\ket{++}$ (separated, qubits 2,3)  & $0.966\pm0.002$& -& QST [9] \\  \hline  
7  &$\ket{\Psi}_\text{Bell}^+$ (heralded Bell , qubits 2,3) & $0.851\pm0.040$ & - & QST [9]  \\  
8  &$\ket{\Phi}_\text{Bell}^+$ (heralded Bell , qubits 2,3)  & $0.830\pm0.032$ & -& QST [9]  \\  \hline  
9  &$\ket{0}$ (teleportation, qubit 2 $\Rightarrow $ 4)  & $0.957\pm0.020$  & $0.940\pm0.041$ & QST [3] \\
10  &$\ket{1}$ (teleportation, qubit 2 $\Rightarrow $ 4) & $0.976\pm0.026$ & -& QST [3] \\
11  &$\ket{+}$ (teleportation, qubit 2 $\Rightarrow $ 4) & $0.857\pm0.034$ & $0.832\pm0.048$ & QST [3]   \\
12 &$\ket{-}$ (teleportation, qubit 2 $\Rightarrow $ 4) & $0.863\pm0.039$  & -& QST [3] \\
13  &$\ket{+i}$ (teleportation, qubit 2 $\Rightarrow $ 4) & $0.893\pm0.040$ & - & QST [3] \\
14  &$\ket{-i}$ (teleportation, qubit 2 $\Rightarrow $ 4) & $0.889\pm0.044$  & - & QST [3]  \\\hline  
15  &$\ket{\Psi}_\text{Bell}^+$ (swapping, qubits 1,4) &$0.776\pm0.019$& -& QST  [9] \\
16  &$\ket{\Phi}_\text{Bell}^+$ (swapping,  qubits 1,4) & $0.737\pm0.019$ & -& QST  [9] \\\hline  
17  &$\ket{\Phi}^4_\text{GHZ}$ (GHZ entangled, qubits 1,2,3,4) &$0.683\pm0.014$ & - & EW   [5] \\
18  &$\ket{\Phi}^3_\text{GHZ}$ (GHZ entangled, qubits 1,2,3) &$0.735\pm0.017$& - & EW   [4] \\
19  &$\ket{\Phi}^2_\text{GHZ}=\ket{\Phi}_\text{Bell}^+$  (qubits 1,4) &$0.786\pm 0.019$ & -& EW   [3] \\\hline  
20  &$\ket{\Phi}^4_\text{GHZ}$ (GHZ entangled, qubits 1,2,3,4) &  $0.593\pm0.019$ & -& TBM  [2] \\
21  &$\ket{\Phi}^3_\text{GHZ}$ (GHZ entangled, qubits 1,2,3) & $0.693\pm0.020$& - & TBM  [2] \\
22 &$\ket{\Phi}^2_\text{GHZ}=\ket{\Phi}_\text{Bell}^+$  (qubits 1,4) & $0.689 \pm 0.017$ & - & TBM  [2] \\
\hline
\end{tabular}
\caption{Experimental fidelities for the multiphoton quantum states. These states are generated, controlled and measured either intra-chip or inter-chip. The verification and quantification methods, QST: quantum state tomography; EW: entanglement witness; TBM: two-basis measurement. In the [*] it shows the required number of global measurement setting, each having $2^n$ $n$-fold coincidence events in the experiment ($n$ is the number of qubits). 
The Bell states for the qubits 1, 2 are not listed, but similar fidelities were observed as the qubits 3, 4.  
Note that the "heralded Bell" (no.7 \& 8) refers to the probabilistic generation of qubits 2, 3 entangled states in the presence of joint clicks in photons 1 \& 4. 
For the 4-photon and 3-photon GHZ states, we implemented EW by 5 measurements (no.17) and by 4 measurements (no.18), respectively, and we also implemented TBM by 2 measurements (no.20 \& 21). The latter gains an efficient verification of GME with a slight scarification of fidelity, due to the less number of measurements. For the 2-photon entangled state $\ket{\Phi}^2_\text{GHZ}=\ket{\Phi}_\text{Bell}^+$, the full QST requires 9 global measurement settings (no.16) while TBM only requires 2 measurement setting (no.22). For larger GHZ entangled states, TBM provides a much efficient approach for GME verification.  
Up to eight SPSNDs are used to collect the data. All the error bars are calculated via a Monte Carlo simulation of the experiment assuming Poissonian statistics of photons. 
\label{table_allFid}}
\end{table}

\cleardoublepage
\newpage

\bibliography{biblio}

\end{document}